\documentclass[pra,aps,singlecolumn,10pt,superscriptaddress,notitlepage,longbibliography]{revtex4-2}
\usepackage[utf8]{inputenc}

\usepackage[english]{babel}

\usepackage{tikz}
\usetikzlibrary{positioning}

\usepackage{graphicx}
\usepackage{float}
\usepackage{color}
\usepackage{epstopdf}
\usepackage{mathtools}
\usepackage{amsmath}
\usepackage{amssymb}
\usepackage{float}
\usepackage{comment}

\usepackage{physics} 

\usepackage{amsfonts}
\usepackage{bm}
\usepackage{bbold}

\usepackage{color}

\graphicspath{{./figs/}}
\usepackage{gensymb}
\usepackage[colorinlistoftodos,prependcaption]{todonotes}
\usepackage{xargs}  
\newcommandx{\unsure}[2][1=]{\todo[linecolor=red,backgroundcolor=red!25,bordercolor=red,#1]{#2}}
\newcommandx{\change}[2][1=]{\todo[linecolor=blue,backgroundcolor=blue!25,bordercolor=blue,#1]{#2}}
\newcommandx{\info}[2][1=]{\todo[linecolor=OliveGreen,backgroundcolor=OliveGreen!25,bordercolor=OliveGreen,#1]{#2}}
\newcommandx{\improvement}[2][1=]{\todo[linecolor=Plum,backgroundcolor=Plum!25,bordercolor=Plum,#1]{#2}}
\newcommandx{\thiswillnotshow}[2][1=]{\todo[disable,#1]{#2}}
\newcommandx{\greencom}[2][1=]
{\todo[inline, color=green!40,#1]{#2}}
\newcommandx{\bluecom}[2][1=]
{\todo[inline, color=blue!40,#1]{#2}}

\usepackage[colorlinks]{hyperref}
\definecolor{winered}{rgb}{0.5,0,0}
\hypersetup
{
colorlinks=true,
linkcolor=winered,
urlcolor={winered},
filecolor={winered},
citecolor={winered},
allcolors={winered}
}

\usepackage{blindtext}
\usepackage{setspace}
\usepackage[margin=0.6in]{geometry}

\usepackage{letltxmacro}
\LetLtxMacro{\ORIGselectlanguage}{\selectlanguage}
\makeatletter
\DeclareRobustCommand{\selectlanguage}[1]{%
  \@ifundefined{alias@\string#1}
    {\ORIGselectlanguage{#1}}
    {\begingroup\edef\x{\endgroup
       \noexpand\ORIGselectlanguage{\@nameuse{alias@#1}}}\x}%
}
\newcommand{\definelanguagealias}[2]{%
  \@namedef{alias@#1}{#2}%
}
\makeatother

\definelanguagealias{en}{english}
\definelanguagealias{EN}{english}

\graphicspath{{figure/}}

\usepackage{graphicx}

\usepackage[caption=false]{subfig}

\begin{document}
\title{Gain-modified emission dynamics between two quantum emitters in a plasmonic gain cavity system}
\author{Becca VanDrunen}
\affiliation{\hspace{0pt}Department of Physics, Engineering Physics, and Astronomy, Queen's University, Kingston, Ontario K7L 3N6, Canada\hspace{0pt}}
\author{Juanjuan Ren}
\affiliation{\hspace{0pt}Department of Physics, Engineering Physics, and Astronomy, Queen's University, Kingston, Ontario K7L 3N6, Canada\hspace{0pt}}
\author{
Sebastian Franke}
\affiliation{Technische Universit\"at Berlin, Institut f\"ur Theoretische Physik,
Nichtlineare Optik und Quantenelektronik, Hardenbergstra{\ss}e 36, 10623 Berlin, Germany}
\affiliation{\hspace{0pt}Department of Physics, Engineering Physics, and Astronomy, Queen's University, Kingston, Ontario K7L 3N6, Canada\hspace{0pt}}
 \author{Stephen Hughes}
\affiliation{\hspace{0pt}Department of Physics, Engineering Physics, and Astronomy, Queen's University, Kingston, Ontario K7L 3N6, Canada\hspace{0pt}}

\def\bra#1{\mathinner{\langle{#1}|}}
\def\ket#1{\mathinner{|{#1}\rangle}}
\def\braket#1{\mathinner{\langle{#1}\rangle}}
\def\Bra#1{\left\langle#1\right|}
\def\Ket#1{\left|#1\right\rangle}

\begin{abstract}
We present a general theory of 
gain-modified emission dynamics between two quantum emitters (two level systems) in a linear gain medium, demonstrating how gain modifies the usual radiative decay rates and inter-emitter
coupling and decay rates, that are well known from purely lossy systems. 
We  derive a Born-Markov master equation  that shows explicitly how gain modifies the 
usual 
rates appearing for two emitters in a loss only medium,
and introduces new gain terms. We then present Bloch equations in the bare state basis as well as
a dressed-state basis (using superradiant and subradiant states). 
As an application of the theory, 
we show calculations for 
a fully three-dimensional metal dimer in a gain-compensated medium, containing two quantum emitters, and show explicit solutions in terms of the quasinormal modes, which easily allows one to obtain all the relevant decay rates.
We study the modified decay dynamics, 
gain excited pumping, as well as the entanglement entropy and spectral emission. We also demonstrate how a breakdown of the weak excitation approximation in the presence of gain requires a careful treatment beyond a linear response, which yields non-trivial incoherent coupling terms that typically are not included in dipole-dipole coupling models
or heuristic gain pumping models. 

\end{abstract}

\maketitle

\section{Introduction}

The study of radiative dynamics between coupled quantum emitters in photonic media has attracted significant attention 
due to its fundamental importance in understanding quantum optics, nanophotonics, and quantum information science. When two or more quantum emitters interact with one another through their 
medium-assisted electromagnetic fields, they can exhibit collective effects, such as superradiance (enhanced emission) and subradiance (suppressed emission)~\cite{crubellier1985superradiance,rehler1971superradiance,sheremet2023waveguide}. Understanding these phenomena is crucial for advancing technologies in quantum communication, light-matter interactions, and the development of novel quantum devices~\cite{PhysRevLett.129.120502}.

In the simple case of coupled quantum emitters in free space (or a homogeneous medium), interactions give rise to F\"orster coupling, a key mechanism in energy transfer processes between coupled dipoles and quantum emitters~\cite{selig2019theory,theuerholz2013influence,chen2021surface}. 
The radiative decay dynamics between two or more quantum emitters have also been well studied
in a variety of nanophotonic systems, including
multiple atoms in inhomogeneous dielectrics~\cite{PhysRevA.70.053823},
quantum dots in photonic
crystals~\cite{kim_2018_nanolett_QDs,PhysRevB.83.075305},
graphene~\cite{PhysRevB.85.155438,PhysRevB.90.085414}, and 
localized plasmons~\cite{PRB_Nerkararyan_2015_entanglement,PhysRevB.92.205420}.
The applications for coupled quantum emitters are numerous, including nanolasing~\cite{vyshnevyy_gain-dependent_2022,Liu:13} 
as well as creating entangled
stated of light and matter~\cite{PhysRevLett.106.020501,PhysRevB.110.075432}.

Recent studies have shown that even in the presence of open lossy systems, entanglement can be achieved and even sustained between coupled emitters. For example, this has been observed in experiments utilizing atomic arrays in optical lattices, where collective coupling to radiation modes can preserve coherence over extended time scales~\cite{PhysRevResearch.5.033108,schine2022long,jenkins2022ytterbium}.
The quantification of entanglement in these systems is often done using measures such as concurrence and negativity. Concurrence is a widely used measure for two-qubit systems, providing a general criterion for entanglement~\cite{Wootters_1998_entanglement,PhysRevA.91.051803,Gangaraj:15}. Negativity, on the other hand, is based on the partial transpose of the density matrix and serves as an entanglement monotone, particularly useful for detecting entanglement in mixed states~\cite{Plenio_2005_log_EN,Sang_2021_EN,Vidal_2002,Frank_Verstraete_2001}. These metrics are important in analyzing entanglement dynamics in various quantum systems~\cite{Adam_Miranowicz_2004,aolita2015open,PhysRevA.110.023720,barreiro2010experimental}.

Notably, in the context of complex photonic environments, much of the research in this field has been conducted in lossy media, which play a crucial role in shaping emission dynamics. 
Thus, the majority of theoretical and experimental work has focused on understanding how dissipation influences collective radiative effects, with particular emphasis on how loss modifies entanglement and coherence in these systems. 
When gain is present in the system, one can no longer apply the weak-excitation approximation (WEA), meaning that careful treatment of the master equation is required to derive the correct Bloch equations~\cite{PhysRevA.105.023702,franke_fermis_2021}, and many assumptions used for purely lossy media become invalid.
Indeed, even concepts such as the usual Fermi's ``golden rule'' for spontaneous emission (SE), which scales with the (projected)
local density of states (LDOS), breaks down
when gain is introduced into the system~\cite{franke_fermis_2021}, which so far has only been  derived for single quantum emitters. 

 Studies of gain-modified collective effects serve as a precursor to more quantitative laser theory, offering deeper insights into the fundamental mechanisms governing light amplification and coherence. The principles of superradiance and collective emission are central to understanding the behavior of lasers, particularly in systems where coherence emerges from many-body interactions~\cite{Glicenstein:22,PRXQuantum.5.010344}. Experimental evidence has also confirmed the presence of superradiance in semiconductor quantum dot lasers~\cite{PhysRevApplied.4.044018,jahnke2016giant}, where collective emission effects lead to enhanced output intensity and coherence properties. Moreover, recent work~\cite{vyshnevyy_gain-dependent_2022}  has investigated nanolasing in the presence of a linear gain medium,
 predicting 
a gain-dependent enhancement of the Purcell factor, which can also be explained classically~\cite{ren_classical_2023}, 
as well as the influence on 
 lasing and superradiant thresholds. 

From a theoretical viewpoint,
we have recently modified the theory for enhanced SE for a single emitter, treated as a two level system (TLS), in a cavity system with linear gain media~\cite{franke_fermis_2021,ren_classical_2023,VanDrunen:24}.
In this current work, we extend this work and provide a general theory of gain-modified emission dynamics from two (or more) coupled quantum emitters (TLSs) in 
any arbitrary system with both lossy and gain media; we 
also present a master equation solution in the Born-Markov approximation. We then use this to derive intuitive Bloch equations in both the bare state basis and the dressed state basis (using the subradiant and superradiant states), which help to  show the gain modifications to the usual coupled-emitter Bloch equations well known for lossy systems. We stress that our solution is completely general, with all coupling rates defined in terms of the Green's function of the medium, valid for any linear medium.

The rest of our paper is organized as follows:
in Sec.~\ref{sec: theory}, we first introduce the general Hamiltonian and quantum Helmholtz equation, which we use to derive the master equation with loss and gain terms, generalized to $n$ TLSs; we then restrict our main study to two TLSs for simplicity, which will be the focus of this paper. With the inclusion of new gain terms, we then derive the optical Bloch equations in both the bare state basis and dressed state basis, and discuss the breakdown of the WEA, for any finite gain values, and highlight the impact of gain on superradiance and subradiance. This renders the problem implicitly nonlinear, and thus there is a breakdown of simple classical theory often invoked to describe subradiant and superradiance in photonic media. We define several observables of interest including
entanglement negativity as well as the emitted spectrum, both of which can 
easily be computed from our master equation.

All coupled rates, emission rates, and gain rates are expressed in terms of the photonic Green function.
To compute the Green function of interest in a practical way, we also show how the parameters for the theory can be obtained from 
solutions or a linear gain cavity system, with solutions in terms of quasinormal modes (QNMs), which are the modes for open systems~\cite{kristensen_modes_2014}. Conveniently the QNMs can be used to rigorously obtain the Green function at various space points, and including gain presents no difficulty in the theory or calculation of such modes~\cite{ren_quasinormal_2021}.

Subsequently, in Sec.~\ref{sec: results}, we present calculations for a plasmonic resonator in a linear gain medium~\cite{VanDrunen:24} with two coupled quantum emitters, using  QNM theory, which we  justify using numerical full-dipole (non-modal) Maxwell calculations. We show that the gain medium influences the dynamics of the emitter populations and coherence,
as well as the 
 entanglement. We also study how the gain medium acts as a pump to populate excited states, including the two quanta state, which causes the superradiant dark state to become visible.
Subsequently, we investigate the role of gain in the emission spectra of the coupled emitter system, showing that gain media allows one to observe both the subradiant and superradiance resonances in the emitted power spectrum. We also identify and highlight the role of a 
cross-emitter pump term, similar to incoherent photon exchange, that appears in our gain model.
Finally, we summarize our key results and present our conclusions in Sec.~\ref{sec: conclusions}.

\section{Theory}
\label{sec: theory}

In this theory section, we first present the field quantization for $n$ TLSs (quantum emitters) within some arbitrary medium with both gain and lossy regions, which expands on previous work  for a single quantum emitter~\cite{PhysRevA.105.023702,franke_fermis_2021}. We then derive the reduced master equation within a Born-Markov approximation in the interaction picture, highlighting the role of new gain scattering terms. Subsequently, we restrict the solution to only consider two TLSs, and show the equations of motion for the population densities and coherence in the bare state basis. We next perform a basis change to the dressed basis (using the subradiant and superradiant states), where we show the dressed optical Bloch equations and the steady-state solutions. We then describe how to calculate emission spectra and logarithmic entanglement negativity, which will be used to investigate the impacts of gain on the coupled quantum emitter system. Finally, in this theory section, as a part of the specific example case we will show in this work, we then show how the decay rates required for the quantum approach can be analytically calculated from the medium Green's functions using QNMs (which we will use later for our applications).

\subsection{Field quantization for an arbitrary medium with loss and gain parts with two level dipole emitters}

To begin, we first extend previous works that focused on a single quantum emitter (TLS) in the presence of gain media~\cite{PhysRevA.105.023702,franke_fermis_2021}, by now including $n$ emitters, 
and field-mediated coupling.
We  consider 
the general Hamiltonian $H=H_{\rm atoms}
+H_B + H_I$, accounting for a system of TLSs interacting with the electromagnetic field in a  lossy~\cite{Dung,dung2000spontaneous} and amplifying media~\cite{raabe2008qed}, with
\begin{align}
    H_{\rm atoms}&=\hbar
    \sum_{\alpha=i,j} \omega_{ \alpha}\sigma^+_{\alpha}\sigma^-_{\alpha},\\
    H_B &= \hbar\int{\rm d}\mathbf{r}~{\rm sgn}(\epsilon_I)\int_0^\infty{\rm d}\omega~\omega \mathbf{b}^\dagger(\mathbf{r},\omega)\cdot\mathbf{b}(\mathbf{r},\omega),\\
    H_I &= -\sum_{\alpha=i,j}\left[\sigma^+_\alpha\int_0^\infty{\rm d}\omega \mathbf{d}_\alpha\cdot\hat{\mathbf{E}}(\mathbf{r}_{\alpha},\omega)+{\rm H.a.}\right],
\end{align}
where $\sigma^{+(-)}_\alpha$ are the raising (lowering) operator, $\mathbf{d}_\alpha$ are the dipole moments and $\mathbf{r}_{\alpha}$ are the positions of the emitters, e.g., for two  emitters,
we could have $a$ and $b$, respectively; 
in addition,  the spatial integral is over all 
space,  ${\rm sgn}$ is the sign function, $\epsilon_{I}$ is the imaginary part of the permittivity, 
  and
$\mathbf{b}^{(\dagger)}(\mathbf{r},\omega)$ are the bosonic annihilation (creation) operators of the medium and the electromagnetic degrees of freedom. For the case of
two quantum emitters,
$a$ and $b$, then $\alpha=a,b$.

\begin{figure}[h]
    \subfloat{\includegraphics[width=0.6\linewidth]{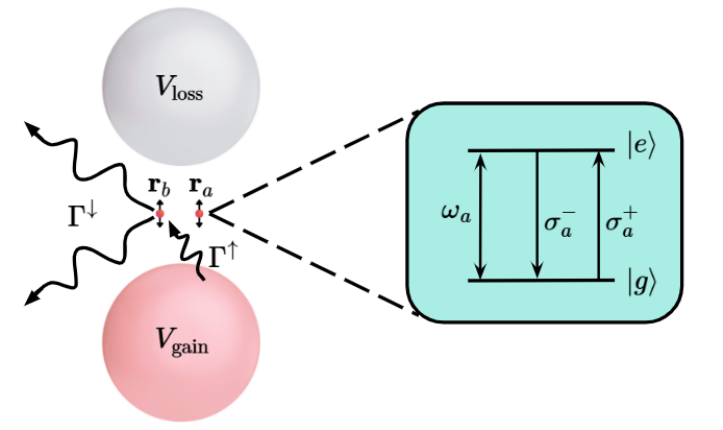}} \\
    \caption{
    Simple schematic of two quantum emitters,  located at positions $\mathbf{r}_a$ and $\mathbf{r}_b$, each modeled as TLSs in a dipole approximation, interacting with a lossy region (grey area, $\epsilon_I(\mathbf{r},\omega) > 0$) and a gain region (pink area, $\epsilon_I(\mathbf{r},\omega) < 0$). The gain-modified SE rate per emitter is $\Gamma^{\downarrow}$, and the emitters are pumped via the gain medium with $\Gamma^{\uparrow}$. For simplicity we only show these rates for emitter $b$, but similar rates occur for atom
    $a$, and there are also coupling rates between the atoms mediated by the fields ($\Gamma_{ab}$ and $\delta_{ab}$, derived and defined later). 
    At the right, we show the two level
    energy level diagram, for atom
    $a$ (bare state basis, and uncoupled); a similar bare-state energy level diagram exists from atom $b$.
    }
    \label{fig: TLS_loss_gain_schematic}
\end{figure}

The medium-assisted electric field operator, $\hat{\mathbf{E}}(\mathbf{r}_{ \alpha},\omega)$
satisfies the quantum Helmholtz equation,
    \begin{equation}
\left[\boldsymbol{\nabla}\times\boldsymbol{\nabla}\times-\epsilon(\mathbf{r},\omega)
\frac{\omega^2}{c^2}\right]\hat{\mathbf{E}}(\mathbf{r},\omega)=i\omega\mu_0\hat{\mathbf{j}}_{\rm N}(\mathbf{r},\omega),
\end{equation}
where $ \hat{\mathbf{j}}_{\rm N}(\mathbf{r},\omega)$ is the current noise operator, which satisfies the  QED commutation relation for arbitrary media~\cite{franke2020fluctuation}.
For any general media, 
with loss and amplifying parts~\cite{raabe2008qed}, then 
\begin{equation}
\hat{\mathbf{j}}_{\rm N}(\mathbf{r},\omega)=\omega\sqrt{
\frac{\hbar\epsilon_0|\epsilon_I(\mathbf{r},\omega)|}{\pi}}\left [\Theta(\epsilon_I)\hat{\mathbf{b}}(\mathbf{r},\omega)+\Theta(-\epsilon_I)\hat{\mathbf{b}}^\dagger(\mathbf{r},\omega)\right],
\end{equation}
where $\Theta[\epsilon_I]$ ($\Theta[-\epsilon_I]$) is the Heaviside function with respect to the spatial region, $\mathbb{R}^3-V_{\rm gain}$ ($V_{\rm gain}$), with passive (active) dielectric permittivity $\epsilon_I(\mathbf{r},\omega)>0$ ($\epsilon_I(\mathbf{r},\omega)<0$).
 The introduction of a phenomenological noise operator in  purely lossy (as well as in lossy and amplifying) media 
is rigorously justified 
 using a  microscopic oscillator model~\cite{Suttorp,philbin2010canonical,PhysRevA.84.013806}. 
 See 
 Fig.~\ref{fig: TLS_loss_gain_schematic} for a schematic example with two emitters, coupled to regions of loss and gain.
 The up-arrow rate represents excitation and the 
 down-arrow represents radiative decay, both of which 
are influenced by the gain part of the medium.

The quantum Helmholtz equation has the 
source-field 
solution 
\begin{equation}
\hat{\mathbf{E}}(\mathbf{r},\omega)=
\frac{i}{\omega\epsilon_0}\int{\mathrm d}{\bf s} \mathbf{G}(\mathbf{r},\mathbf{s},\omega){\cdot} \hat{\mathbf{j}}_{\rm N}(\mathbf{s},\omega)
\end{equation}
where $\mathbf{G}(\mathbf{r},\mathbf{s},\omega)$ is the
Green function of the medium, which 
satisfies
\begin{equation}
\left[\boldsymbol{\nabla}{\times}\boldsymbol{\nabla}{\times}
-\epsilon(\mathbf{r},\omega)\frac{\omega^2}{c^2}\right ]\mathbf{G}(\mathbf{r},\mathbf{s},\omega)=\frac{\omega^2}{c^2}\mathbb{1}\delta(\mathbf{{r}-\mathbf{s}}),
\end{equation}
~with suitable radiation boundary conditions. 

\subsection{Reduced master equation in the interaction picture for two emitters in a combined lossy-gain medium}

Following previous work
in gain media, where a single TLS was considered~\cite{franke_fermis_2021},
we extend that work to 
derive the reduced master equation for two
 emitters. The extension to more emitters is straightforward and our final solution can easily be applied to systems with more than two dipole emitters.

To derive a master equation, we 
start by defining the
Liouville von-Neumann equation,
\begin{equation}
    \partial_t\rho = -\frac{i}{\hbar}\left[H_0,\rho\right]-\frac{i}{\hbar}\left[H_I,\rho\right],
\end{equation}
with the bare Hamiltonian, $H_0=H_B + H_{\rm atoms}$, and the light-matter interaction Hamiltonian $H_I$, 
as defined above.
We next go to interaction picture and define
\begin{equation}
    \tilde{\rho}=e^{iH_0t/\hbar}\rho e^{-iH_0t/\hbar},
\end{equation}
such that
\begin{equation}
    \partial_t\tilde{\rho}(t)=-\frac{i}{\hbar}[\tilde{H}_I(t),\tilde{\rho}(t)]\label{eq: MasterInt1},
\end{equation}
with 
\begin{equation}
    \tilde{H}_I(t)=e^{iH_0t/\hbar}H_I e^{-iH_0t/\hbar}=-\sum_{\alpha=a,b}\left[\tilde{\sigma}^+_\alpha(t)\int_0^\infty{\rm d}\omega \mathbf{d}_\alpha\cdot\hat{\tilde{\mathbf{E}}}(\mathbf{r}_{\alpha},\omega,t)+{\rm H.a.}\right],
\end{equation}
and
\begin{align}
    \tilde{\sigma}^+_\alpha(t)&=e^{-i\omega_{\alpha}t}\sigma^+_\alpha,\\
    \hat{\tilde{\mathbf{E}}}(\mathbf{r}_{\alpha},\omega,t)&=e^{i\omega t}\hat{\mathbf{E}}(\mathbf{r}_{\alpha},\omega).
\end{align}

A formal solution of Eq.~\eqref{eq: MasterInt1} is given by 
\begin{equation}
    \tilde{\rho}(t)=\tilde{\rho}(0)-\frac{i}{\hbar}\int_0^t {\rm d}\tau [\tilde{H}_I(\tau),\tilde{\rho}(\tau)],
\end{equation}
and inserting this back into Eq.~\eqref{eq: MasterInt1}, we obtain 
\begin{equation}
    \partial_t\tilde{\rho}(t)=-\frac{i}{\hbar}[\tilde{H}_I(t),\tilde{\rho}(0)]-\frac{1}{\hbar^2}\int_0^t{\rm d}\tau[\tilde{H}_I(t),[\tilde{H}_I(\tau),\tilde{\rho}(\tau)]]. \label{eq: MasterInt2}
\end{equation}
For the (reduced) emitter density matrix $\tilde{\rho}_{\rm m}={\rm tr}_B\tilde{\rho}$ (`m' = matter), we trace over the
photonic bath (or reservoir) and obtain
\begin{equation}
    \partial_t\tilde{\rho}_{\rm m}(t)=-\frac{i}{\hbar}{\rm tr}_B[\tilde{H}_I(t),\tilde{\rho}(0)]-\frac{1}{\hbar^2}\int_0^t{\rm tr}_B[\tilde{H}_I(t),[\tilde{H}_I(\tau),\tilde{\rho}(\tau)]]{\rm d}\tau\label{eq: MasterInt3}.
\end{equation}

Applying a change of variables $\tau\rightarrow t-\tau\equiv \tau$ in the integral, leads to 
\begin{equation}
    \partial_t\tilde{\rho}_{\rm m}(t)=
    -\frac{1}{\hbar^2}\int_0^t{\rm tr}_B[\tilde{H}_I(t),[\tilde{H}_I(t-\tau),\tilde{\rho}(t-\tau)]]{\rm d}\tau\label{eq: MasterInt4},
\end{equation}
where we have neglected the first term. Furthermore, by assuming that we can separate $\tilde{\rho}(t)=\tilde{\rho}_{\rm m}(t)\tilde{\rho}_{\rm B}(t)$ (for all times) and that $\tilde{\rho}_{\rm B}(t)=\tilde{\rho}_{\rm B}(0)\equiv\rho_{\rm B}$, 
we arrive at \begin{equation}
     \partial_t\tilde{\rho}_{\rm m}(t)=-\frac{1}{\hbar^2}\int_0^t{\rm tr}_B[\tilde{H}_I(t),[\tilde{H}_I(t-\tau),\tilde{\rho}_{\rm m}(t-\tau)\rho_B]]{\rm d}\tau. \label{eq: MasterInt5}
\end{equation}
As a first Markov approximation, we assume $\tilde{\rho}_{\rm m}(t-\tau)\approx\tilde{\rho}_{\rm m}(t)$ under the temporal integral, leading to the time-local master equation
\begin{equation}
     \partial_t\tilde{\rho}_{\rm m}(t)=-\frac{1}{\hbar^2}\int_0^t{\rm tr}_B[\tilde{H}_I(t),[\tilde{H}_I(t-\tau),\tilde{\rho}_{\rm m}(t)\rho_B]]{\rm d}\tau.\label{eq: MasterInt6}
\end{equation}

\subsection{Evaluation of the nested commutator}
Next, we evaluate the nested commutator expressions:
\begin{align}
    -\frac{1}{\hbar^2}{\rm tr}_B[\tilde{H}_I(t),[\tilde{H}_I(t-\tau),\tilde{\rho}_{\rm m}(t)\rho_B]]\equiv -\frac{1}{\hbar^2}\sum_{i,j}\sum_{\alpha,\beta = a,b}\int_0^\infty{\rm d}\omega\int_0^\infty{\rm d}\omega'  \hat{C}_{\alpha i,\beta j}(\omega,\omega',t,\tau)]\label{eq: NestedCommute},
\end{align}
where 
\begin{align}
  \hat{C}_{\alpha i, \beta j}(\omega,\omega',t,\tau)=&  d_{\alpha,i} d_{\beta,j}{\rm tr}_B[\sigma^+_\alpha \hat{E}_i(\mathbf{r}_{\rm \alpha},\omega),[\sigma^+_\beta \hat{E}_j(\mathbf{r}_{\beta},\omega'),\tilde{\rho}_{\rm m}\rho_B]]e^{i(\omega-\omega_{\alpha})t}e^{i(\omega'-\omega_{\beta})(t-\tau)}\nonumber\\
  &+d_{\alpha,i} d_{\beta,j}^*{\rm tr}_B[\sigma^+_\alpha \hat{E}_i(\mathbf{r}_{\alpha},\omega),[\sigma^-_\beta\hat{E}_j^\dagger(\mathbf{r}_{\beta},\omega'),\tilde{\rho}_{\rm m}\rho_B]]e^{i(\omega-\omega_{\alpha})t}e^{-i(\omega'-\omega_{\beta})(t-\tau)}\nonumber\\
  &+d_{\alpha,i}^* d_{\beta,j}{\rm tr}_B[\sigma^-_\alpha \hat{E}_i^\dagger(\mathbf{r}_{\alpha},\omega),[\sigma^+_\beta \hat{E}_j(\mathbf{r}_{\beta},\omega'),\tilde{\rho}_{\rm m}\rho_B]]e^{-i(\omega-\omega_{\alpha})t}e^{i(\omega'-\omega_{\beta})(t-\tau)}\nonumber\\
  &+d_{\alpha,i}^* d_{\beta,j}^*{\rm tr}_B[\sigma^-_\alpha \hat{E}_i^\dagger(\mathbf{r}_{\alpha},\omega),[\sigma^-_\beta \hat{E}_j^\dagger(\mathbf{r}_{\beta},\omega'),\tilde{\rho}_{\rm m}\rho_B]]e^{-i(\omega-\omega_{\alpha})t}e^{-i(\omega'-\omega_{\beta})(t-\tau)}\label{eq: COp}.
\end{align}

To evaluate the trace terms we need to know the 
correlation functions between the various 
combinations of $b$, $b^\dagger$
operators with respect to the bath. 
To compute these, we 
adopt the following model:
\begin{align}
   {\rm tr}_B[b_i^{\dagger}(\mathbf{r},\omega)b_j(\mathbf{r}',\omega')\rho_B]&={\rm tr}_B[b_i(\mathbf{r},\omega)b_j(\mathbf{r}',\omega')\rho_B]={\rm tr}_B[b_i^{\dagger}(\mathbf{r},\omega)b_j^\dagger(\mathbf{r}',\omega')\rho_B]=0,\\
   {\rm tr}_B[b_i(\mathbf{r},\omega)b_j^\dagger(\mathbf{r}',\omega')\rho_B]&=\delta_{ij}\delta(\mathbf{r}-\mathbf{r}')\delta(\omega-\omega'),
\end{align}
which effectively treats the field-medium as a reservoir with $T=0\,{\rm K}$. 
To make this clearer, at
some finite temperature,
$T$, 
with $\rho_{B}=
\rho_{B}^{\rm amp} \rho_{B}^{\rm loss}$ (since the subspaces are independent without interactions), we would have
\begin{equation}
{\rm tr}_B[b_i^{\dagger}(\mathbf{r},\omega)b_j(\mathbf{r}',\omega')\rho_B]
= \left \{ 
\Theta(\epsilon_I)
n(\omega,T)
+ 
\Theta(-\epsilon_I)
n(\omega,|T|)
\right \}
\delta_{ij}\delta(\mathbf{r}-\mathbf{r}')\delta(\omega-\omega'),
\end{equation}
where for the amplifier
(inverted oscillator, which behaves as a thermal state with an effective negative temperature~\cite{PhysRevA.84.013806,franke_fermis_2021}), we have
\begin{equation}
n(\omega,|T|)=
\frac{N_L}{N_U-N_L},
\end{equation}
and for the loss parts, we have
\begin{equation}
n(\omega,T)=
\frac{N_U}{N_L-N_U},
\end{equation}
where $N_U$ and $N_L$
are the population densities of the upper and lower levels.
For transitions with frequencies
$\hbar\omega \gg k_B T$, then
we can
basically assume 
$N_L=0\, (N_U=1)$ for the gain part
and $N_U=0\, (N_L=1)$ for the loss part. 
For further details and a rigorous justification for this thermal 
oscillator model, see Supplementary Information of
Ref.~\onlinecite{franke_fermis_2021}.

Thus,  the first and fourth term on the RHS of Eq.~\eqref{eq: COp} vanish.
Using properties of the partial trace, and writing out the nested commutators, we obtain the remaining terms
\begin{align}
  \hat{C}_{\alpha i, \beta j}&(\omega,\omega',t,\tau)\nonumber\\
  =&d_{\alpha,i} d_{\beta,j}^*\left(\{\sigma^+_\alpha\sigma^-_\beta\tilde{\rho}_{\rm m}-\sigma^-_\beta\tilde{\rho}_{\rm m}\sigma^+_\alpha\}K_{\alpha i,\beta j}^{\rm AN}(\omega,\omega')+\{\tilde{\rho}_{\rm m}\sigma^-_\beta\sigma^+_\alpha-\sigma^+_\alpha\tilde{\rho}_{\rm m}\sigma^-_\beta\}K_{\beta j,\alpha i}^{\rm N}(\omega',\omega)\right) e^{i(\omega-\omega_{\alpha})t}e^{-i(\omega'-\omega_{\beta})(t-\tau)}\nonumber\\
  &+d_{\alpha,i}^* d_{\beta,j}\left(\{\sigma^-_\alpha\sigma^+_\beta\tilde{\rho}_{\rm m}-\sigma^+_\beta\tilde{\rho}_{\rm m}\sigma^-_\alpha\}K_{\alpha i,\beta j}^{\rm N}(\omega,\omega')+\{\tilde{\rho}_{\rm m}\sigma^+_\beta\sigma^-_\alpha-\sigma^-_\alpha\tilde{\rho}_{\rm m}\sigma^+_\beta\}K_{\beta j,\alpha i}^{\rm AN}(\omega',\omega)\right) e^{-i(\omega-\omega_{\alpha})t}e^{i(\omega'-\omega_{\beta})(t-\tau)}\label{eq: COp1_1},
\end{align}
with 
\begin{align}
    K_{\alpha i,\beta j}^{\rm N}(\omega,\omega')&={\rm tr}_B[\hat{E}_i^\dagger(\mathbf{r}_{\alpha},\omega)\hat{E}_j(\mathbf{r}_{\beta},\omega')\rho_B]=\langle 0|\hat{E}_i^\dagger(\mathbf{r}_{\alpha},\omega)\hat{E}_j(\mathbf{r}_{\beta},\omega')|0\rangle\\
    K_{\alpha i,\beta j}^{\rm AN}(\omega,\omega')&={\rm tr}_B[\hat{E}_i(\mathbf{r}_{\alpha},\omega)\hat{E}_j^\dagger(\mathbf{r}_{\beta},\omega')\rho_B]=\langle 0|\hat{E}_i(\mathbf{r}_{\alpha},\omega)\hat{E}_j^\dagger(\mathbf{r}_{\beta},\omega')|0\rangle, 
\end{align}
as the normal and anti-normal ordered vacuum  correlation functions of the electric field.

Using the explicit form of the medium-assisted field, 
\begin{equation}
\hat{E}_i(\mathbf{r},\omega)=i\sqrt{\frac{\hbar}{\pi\epsilon_0}}\sum_k\int{\mathrm d}\mathbf{r}' G_{ik}(\mathbf{r},\mathbf{r}',\omega)\cdot \left[\Theta(\epsilon_I)\sqrt{\epsilon_I(\mathbf{r}',\omega)}\hat{b}_k(\mathbf{r}',\omega)+\Theta(-\epsilon_I)\sqrt{|\epsilon_I(\mathbf{r}',\omega)|}\hat{b}^\dagger_k(\mathbf{r}',\omega)\right],\label{eq: SourceEField3}
\end{equation}
then yields 
\begin{align}
    K_{\alpha i,\beta j}^{\rm AN}(\omega,\omega')&=\sum_k\frac{\hbar}{\pi\epsilon_0}\int_{\mathbb{R}^3}{\rm d}\mathbf{r}\Theta[\epsilon_i]\epsilon_I(\mathbf{r},\omega)G_{ik}(\mathbf{r}_{\alpha},\mathbf{r},\omega) G_{jk}^*(\mathbf{r}_{\beta},\mathbf{r},\omega) \delta(\omega-\omega'),\\
    K_{\alpha i,\beta j}^{\rm N}(\omega,\omega')&=\sum_k\frac{\hbar}{\pi\epsilon_0}\int_{\mathbb{R}^3}{\rm d}\mathbf{r}\Theta[-\epsilon_i]|\epsilon_I(\mathbf{r},\omega)|G_{ik}^*(\mathbf{r}_{\alpha},\mathbf{r},\omega) G_{jk}(\mathbf{r}_{\beta},\mathbf{r},\omega) \delta(\omega-\omega').
\end{align}

Inserting these back into Eq.~\eqref{eq: NestedCommute},  gives 
\begin{align}
     -\frac{1}{\hbar^2}{\rm tr}_B[\tilde{H}_I(t),[\tilde{H}_I(t-\tau),\tilde{\rho}_{\rm m}(t)\rho_B]]=&\sum_{\alpha,\beta=a,b}e^{-i\Delta_{\alpha\beta}t}J_{\alpha\beta,\rm loss}(\tau)\left[ \sigma^-_\beta\tilde{\rho}_{\rm m}\sigma^+_\alpha - \sigma^+_\alpha\sigma^-_\beta\tilde{\rho}_{\rm m} \right]\nonumber\\
     &+e^{i\Delta_{\alpha\beta}t}J_{\alpha\beta,\rm loss}^*(\tau)\left[ \sigma^-_\alpha\tilde{\rho}_{\rm m}\sigma^+_\beta - \tilde{\rho}_{\rm m}\sigma^+_\beta\sigma^-_\alpha \right]\nonumber\\
     &+e^{i\Delta_{\alpha\beta}t}J_{\alpha\beta,\rm gain}(\tau)\left[ \sigma^+_\beta\tilde{\rho}_{\rm m}\sigma^-_\alpha - \sigma^-_\alpha\sigma^+_\beta\tilde{\rho}_{\rm m} \right]\nonumber\\
     &+e^{-i\Delta_{\alpha\beta}t}J_{\alpha\beta,\rm gain}^*(\tau)\left[ \sigma^+_\alpha\tilde{\rho}_{\rm m}\sigma^-_\beta - \tilde{\rho}_{\rm m}\sigma^-_\beta\sigma^+_\alpha \right],
\end{align}
with
\begin{align}
    J_{\alpha\beta,\rm loss}(\tau) &= \frac{1}{\hbar\pi\epsilon_0}\int_0^\infty{\rm d}\omega e^{-i(\omega-\omega_{\beta})\tau}\sum_{ijk}\int_{\mathbb{R}^3-V_{\rm gain}}{\rm d}\mathbf{r}\epsilon_I(\mathbf{r},\omega)d_{\alpha,i} G_{ik}(\mathbf{r}_{\alpha},\mathbf{r},\omega) G_{jk}^*(\mathbf{r}_{\beta},\mathbf{r},\omega) d_{\beta,j}^*\\
    J_{\alpha\beta,\rm gain}(\tau) &= \frac{1}{\hbar\pi\epsilon_0}\int_0^\infty{\rm d}\omega e^{i(\omega-\omega_{\beta})\tau}\sum_{ijk}\int_{V_{\rm gain}}{\rm d}\mathbf{r}|\epsilon_I(\mathbf{r},\omega)|d_{\alpha,i}^* G_{ik}^*(\mathbf{r}_{\alpha},\mathbf{r},\omega) G_{jk}(\mathbf{r}_{\beta},\mathbf{r},\omega) d_{\beta,j},
\end{align}
where $\Delta_{\alpha\beta} = \omega_\alpha - \omega_\beta$.

Importantly, $J_{\alpha\beta,\rm loss}(\tau)$ can be rewritten in terms the imaginary part of the Green's function and $J_{\alpha\beta,\rm gain}(\tau)$, so that 
\begin{equation}
    J_{\alpha\beta,\rm loss}(\tau) = \frac{1}{\hbar\pi\epsilon_0}\int_0^\infty{\rm d}\omega e^{i(\omega-\omega_{\beta})\tau}\sum_{ij}d_{i} {\rm Im}[G_{ij}(\mathbf{r}_{\alpha},\mathbf{r}_{\beta},\omega)] d_{j}^* + J_{\alpha\beta,\rm gain}^*(\tau),
\end{equation}
where we made use of the Green's identity, 
\begin{equation}
    \sum_k\int_{\mathbb{R}^3}{\rm d}\mathbf{r}\epsilon_I(\mathbf{r},\omega)G_{ik}(\mathbf{r}_{\alpha},\mathbf{r},\omega) G_{jk}^*(\mathbf{r}_{\beta},\mathbf{r},\omega) = {\rm Im}[G_{ij}(\mathbf{r}_{\alpha},\mathbf{r}_{\beta},\omega)]. 
\end{equation}
In the limit of one TLS, then we recover previous work~\cite{franke_fermis_2021}, where the usual SE rate is corrected
as being the LDOS contribution 
($\propto d_i{\rm Im}[G_{ii}({\bf r}_i,{\bf r}_i,\omega)] d_i$) plus a gain term.

\subsection{Second Markov approximation
and derivation of the Born-Markov master equation}

The full medium-dependent master equation, for two emitters, can now be written as 
\begin{align}
     \partial_t\tilde{\rho}_{\rm m}(t)=&\sum_{\alpha,\beta=a,b}\int_0^t{\rm d}\tau  \{ e^{-i\Delta_{\alpha\beta}t} J_{\alpha\beta,\rm loss}(\tau) [\sigma^-_\beta\tilde{\rho}_{\rm m}\sigma^+_\alpha - \sigma^+_\alpha\sigma^-_\beta\tilde{\rho}_{\rm m}] +{\rm H.a.}\}\nonumber\\
     &+\int_0^t{\rm d}\tau \{ e^{i\Delta_{\alpha\beta}t}J_{\alpha\beta,\rm gain}(\tau)[\sigma^+_\beta\tilde{\rho}_{\rm a}\sigma^-_\alpha - \sigma^-_\alpha\sigma^+_\beta\tilde{\rho}_{\rm a}] +{\rm H.a.}\}\label{eq: MasterInt7}.
\end{align}
In the Markov limit, we extend the upper limit of the integral $t\rightarrow\infty$, to obtain 
\begin{align}
    \int_0^\infty{\rm d}\tau J_{\alpha\beta,\rm loss}(\tau)=&\frac{1}{\hbar\epsilon_0}\sum_{ij}d_{\alpha,i} {\rm Im}[G_{ij}(\mathbf{r}_{\alpha},\mathbf{r}_{\beta},\omega_{\beta})] d_{\beta,j}^*\nonumber\\
    &-\frac{i}{\hbar\pi\epsilon_0}\mathcal{P}\int_0^\infty{\rm d}\omega \frac{1}{\omega-\omega_{\beta}}\sum_{ij}d_{\alpha,i} {\rm Im}[G_{ij}(\mathbf{r}_{\alpha},\mathbf{r}_{\beta},\omega)] d_{\beta,j}^*\nonumber\\
    &+\frac{1}{\hbar\epsilon_0}\sum_{ijk}\int_{V_{\rm gain}}{\rm d}\mathbf{r}|\epsilon_I(\mathbf{r},\omega_{\beta})|d_{i} G_{ik}(\mathbf{r}_{\alpha},\mathbf{r},\omega_{\beta}) G_{jk}^*(\mathbf{r}_{\beta},\mathbf{r},\omega_{\beta}) d_{\beta,j}^*\nonumber\\
    &-\frac{i}{\hbar\pi\epsilon_0}\mathcal{P}\int_0^\infty{\rm d}\omega \frac{1}{\omega-\omega_{\beta}}\sum_{ijk}\int_{V_{\rm gain}}{\rm d}\mathbf{r}|\epsilon_I(\mathbf{r},\omega)|d_{\alpha,i} G_{ik}(\mathbf{r}_{\alpha},\mathbf{r},\omega) G_{jk}^*(\mathbf{r}_{\beta},\mathbf{r},\omega) d_{\beta,j}^*,
    \label{eq:J_loss}
\end{align}
and 
\begin{align}
    \int_0^\infty{\rm d}\tau J_{\alpha\beta,\rm gain}(\tau)=& \frac{1}{\hbar\epsilon_0}\sum_{ijk}\int_{V_{\rm gain}}{\rm d}\mathbf{r}|\epsilon_I(\mathbf{r},\omega_{\rm a})|d_{\alpha,i}^* G_{ik}^*(\mathbf{r}_{\alpha},\mathbf{r},\omega_{\beta}) G_{jk}(\mathbf{r}_{\beta},\mathbf{r},\omega_{\beta})d_{\beta,j}\nonumber\\
    &+\frac{i}{\hbar\pi\epsilon_0}\mathcal{P}\int_0^\infty{\rm d}\omega \frac{1}{\omega-\omega_{\beta}}\sum_{ijk}\int_{V_{\rm gain}}{\rm d}\mathbf{r}|\epsilon_I(\mathbf{r},\omega)|d_{\alpha,i}^* G_{ik}^*(\mathbf{r}_{\beta},\mathbf{r},\omega) G_{jk}(\mathbf{r}_{\beta},\mathbf{r},\omega) d_{\beta,j},
    \label{eq:J_gain}
\end{align}
where we used
\begin{equation}
    \lim_{t\rightarrow \infty}\int_0^t{\rm d}\tau e^{-i(\omega_{0}-\omega)\tau}=\pi\delta(\omega-\omega_0)+i\frac{\mathcal{P}}{\omega-\omega_{0}}.\label{eq: MarkovLimitInt}
\end{equation}
Note, that we have also implicitly assumed that the detuning is smaller than an appreciable change in the bath/reservoir functions, so that $e^{i\Delta_{\alpha\beta}t}\rightarrow 1$ in the Markov limit.

Compared to the single emitter case the third (fourth) term of $\int_0^\infty{\rm d}\tau J_{\alpha\beta,{\rm loss}}$ as well as the first (second) term of $ \int_0^\infty{\rm d}\tau J_{\alpha\beta,{\rm gain}}$ are no longer purely real (imaginary), and contain both decay terms and frequency shift terms.  Separating both contributions (decay and frequency shift terms) then leads to 
\begin{align}
     \partial_t\tilde{\rho}_{\rm m}(t)=&\sum_{\alpha,\beta = a,b}\frac{\Gamma^{\downarrow}_{\alpha\beta}}{2}\left(\sigma^-_\beta\tilde{\rho}_{\rm m}\sigma^+_\alpha - \sigma^+_\alpha\sigma^-_\beta\tilde{\rho}_{\rm m} + \sigma^-_\alpha\tilde{\rho}_{\rm m}\sigma^+_\beta - \tilde{\rho}_{\rm m}\sigma^+_\beta\sigma^-_\alpha  \right)
     \nonumber\\
     &-i\sum_{\alpha,\beta = a,b}\delta_{\alpha\beta}^{\downarrow}\left(\sigma^-_\beta\tilde{\rho}_{\rm m}\sigma^+_\alpha - \sigma^+_\alpha\sigma^-_\beta\tilde{\rho}_{\rm m} - \sigma^-_\alpha\tilde{\rho}_{\rm m}\sigma^+_\beta + \tilde{\rho}_{\rm m}\sigma^+_\beta\sigma^-_\alpha  \right)\nonumber\\
     +&\sum_{\alpha,\beta = a,b}\frac{\Gamma^{\uparrow}_{\alpha\beta}}{2}\left(\sigma^+_\beta\tilde{\rho}_{\rm m}\sigma^-_\alpha - \sigma^-_\alpha\sigma^+_\beta\tilde{\rho}_{\rm m}  +\sigma^+_\alpha\tilde{\rho}_{\rm m}\sigma^-_\beta - \tilde{\rho}_{\rm m}\sigma^-_\beta\sigma^+_\alpha\right)\nonumber\\
     &-i\sum_{\alpha,\beta = a,b}\delta_{\alpha\beta}^{\uparrow}\left(\sigma^+_\beta\tilde{\rho}_{\rm m}\sigma^-_\alpha - \sigma^-_\alpha\sigma^+_\beta\tilde{\rho}_{\rm m}  -\sigma^+_\alpha\tilde{\rho}_{\rm m}\sigma^-_\beta + \tilde{\rho}_{\rm m}\sigma^-_\beta\sigma^+_\alpha\right)\label{eq: MasterInt8},
\end{align}
with 
\begin{align}
    \Gamma^{\downarrow}_{\alpha\beta}&=2{\rm Re}\int_0^\infty{\rm d}\tau J_{\alpha\beta,{\rm loss}}(\tau),~\Gamma^{\uparrow}_{\alpha\beta}=2{\rm Re}\int_0^\infty{\rm d}\tau J_{\alpha\beta,\rm gain}(\tau),\\
    \delta_{\alpha\beta}^{\downarrow}&=2{\rm Im}\int_0^\infty{\rm d}\tau J_{\alpha\beta,\rm loss}(\tau),~\delta_{\alpha\beta}^{\uparrow}=2{\rm Im}\int_0^\infty{\rm d}\tau J_{\alpha\beta,\rm gain}(\tau).
\end{align}

Finally, again we will
assume that the dipole parameters are very close to each other, which leads to $\Gamma^{\downarrow (\uparrow)}_{\alpha\beta}\approx \Gamma^{\downarrow (\uparrow)}_{\beta\alpha}$ as well as $\delta^{\downarrow (\uparrow)}_{\alpha\beta}\approx \delta^{\downarrow (\uparrow)}_{\beta\alpha}$, yielding to the simplified form
\begin{align}
     \partial_t\tilde{\rho}_{\rm m}(t)=&\sum_{\alpha,\beta = a,b}\frac{\Gamma^{\downarrow}_{\alpha\beta}}{2}\left(2\sigma^-_\alpha\tilde{\rho}_{\rm m}\sigma^+_\beta - \sigma^+_\alpha\sigma^-_\beta\tilde{\rho}_{\rm m} - \tilde{\rho}_{\rm m}\sigma^+_\beta\sigma^-_\alpha  \right)\nonumber\\
     &-i\sum_{\alpha,\beta = a,b}\delta_{\alpha\beta}^{\downarrow}\left[\sigma^+_\alpha\sigma^-_\beta,\tilde{\rho}_{\rm m}\right]-i\sum_{\alpha,\beta = a,b}\delta_{\alpha\beta}^{\uparrow}\left[\sigma^-_\alpha\sigma^+_\beta,\tilde{\rho}_{\rm m}\right]\nonumber\\
     +&\sum_{\alpha,\beta = a,b}\frac{\Gamma^{\uparrow}_{\alpha\beta}}{2}\left(2\sigma^+_\alpha\tilde{\rho}_{\rm m}\sigma^-_\beta  - \sigma^-_\alpha\sigma^+_\beta\tilde{\rho}_{\rm m} - \tilde{\rho}_{\rm m}\sigma^-_\beta\sigma^+_\alpha\right)\label{eq: MasterInt9}.
\end{align}
Transforming back from the interaction picture, then we  obtain the desired 
master equation for the coupled TLS
in the loss and gain medium:
\begin{align}
    \partial_t\rho_{\rm m}(t)=& -i\sum_{\alpha=a,b}\omega_\alpha\left[\sigma^+_{\alpha}\sigma^-_{\alpha},\rho_{\rm m}\right] -i\sum_{\alpha,\beta = a,b}\delta_{\alpha\beta}^{\uparrow}\left[\sigma^-_\alpha\sigma^+_\beta,\tilde{\rho}_{\rm m}\right]-i\sum_{\alpha,\beta = a,b}\delta_{\alpha\beta}^{\downarrow}\left[\sigma^+_\alpha\sigma^-_\beta,\tilde{\rho}_{\rm m}\right]\nonumber\\
    &+\sum_{\alpha,\beta = a,b}\frac{\Gamma^{\downarrow}_{\alpha\beta}}{2}\left(2\sigma^-_\alpha\tilde{\rho}_{\rm m}\sigma^+_\beta - \sigma^+_\alpha\sigma^-_\beta\tilde{\rho}_{\rm m} - \tilde{\rho}_{\rm m}\sigma^+_\beta\sigma^-_\alpha  \right)\nonumber\\
     &+\sum_{\alpha,\beta = a,b}\frac{\Gamma^{\uparrow}_{\alpha\beta}}{2}\left(2\sigma^+_\alpha\tilde{\rho}_{\rm m}\sigma^-_\beta  - \sigma^-_\alpha\sigma^+_\beta\tilde{\rho}_{\rm m} - \tilde{\rho}_{\rm m}\sigma^-_\beta\sigma^+_\alpha\right)\label{eq: MasterFinal}.
\end{align}

This result shows the impact of gain explicitly in the master equation. 
In particular, we have additional terms, $\delta^{\uparrow}_{ab}$ as well as $\Gamma^{\uparrow}_{\alpha \beta}$ for $\alpha, \beta = a,b$, which contribute to the reverse process compared to the usual lossy terms. These additional terms build off of our previous work with a single emitter~\cite{franke_fermis_2021,ren_classical_2023,VanDrunen:24}, where we now include multiple quantum emitters. In Eq.~\eqref{eq: MasterFinal}, the new gain terms contribute to pumping the excited states, which is the precursor to achieving lasing~\cite{vyshnevyy_gain-dependent_2022}.
It is important to also note that the usual loss decay terms are also modified with gain, e.g.,
$\Gamma_{aa}^{\downarrow}$,
which depends on $|\epsilon_I({\bf r},\omega)|$ (in the gain region)
as well as an LDOS term [see third and first terms of Eq.~\eqref{eq:J_loss}, respectively]. This then results in the lossy rate, $\Gamma^{\downarrow}_{\alpha \beta}$ having a contribution from the gain ($\Gamma^{\uparrow}_{\alpha \beta}$) as well as the usual lossy contribution, where as the gain term for the reverse process is just $\Gamma^{\uparrow}_{\alpha \beta}$. This is consistent with the known physics for a quantum 
thermal model for pumping and bath coupling~\cite{Tian1992}.

\subsection{Optical Bloch equations in the bare state basis}

With the master equation result in Eq.~\eqref{eq: MasterFinal} for two  coupled emitters, we can easily determine the equations of motion for the population densities and coherence, i.e., the reduced Bloch equations. To make this approach more realistic to experimental conditions and emitters, we also add a possible pure dephasing term, $\Gamma^{\prime}$, but will keep it small, such that $\Gamma^{\prime}\ll \Gamma_{\alpha\alpha}^{\downarrow}$ (in any simulations shown later, as we primarily want to explore the radiative dynamics). Although this will not affect the dominant decay processes and lineshapes, it is important to allow the equations to reach a well defined 
steady state for population trapped states (such as the superradiant state), also allowing us to compute observables such as the emission spectra (described later).

The 
modified master equation is then:
\begin{align}
    \dot{\rho}_{\rm m} = ~ -&i\sum_{\alpha=a,b}\omega_\alpha'\left[\sigma^+_{\alpha}\sigma^-_{\alpha},\rho_{\rm m}\right] 
    -i\left[\left(\delta_{ab}^{\downarrow}(\sigma^+_a\sigma^-_b 
    + \sigma^+_b\sigma^-_a)
    + \delta_{ab}^{\uparrow}(\sigma^-_b\sigma^+_a + \sigma^-_a\sigma^+_b)\right),\rho_{\rm m}\right] \nonumber \\
    +& \sum_{\alpha,\beta = 
    a,b}\frac{\Gamma_{\alpha\beta}^{\downarrow}}{2}\left[2\sigma^-_\alpha\rho_{\rm m}\sigma^+_\beta-\sigma^+_\alpha\sigma^-_\beta\rho_{\rm m}-\rho_{\rm m}\sigma^+_\alpha \sigma^-_\beta\right] 
    + \sum_{\alpha,\beta = 
    a,b}\frac{\Gamma_{\alpha\beta}^{\uparrow}}{2}\left[2\sigma^+_\alpha\rho_{\rm m}\sigma^-_\beta-\sigma^-_\alpha\sigma^+_\beta\rho_{\rm m}-\rho_{\rm m}\sigma^-_\alpha \sigma^+_\beta\right] \nonumber \\
    +& \sum_{\alpha=a,b}\frac{\Gamma^{\prime}}{2}\left[2\sigma^+_\alpha\sigma^-_\alpha\rho_{\rm m}\sigma^-_\alpha\sigma^+_\alpha - \sigma^-_\alpha\sigma^+_\alpha\sigma^+_\alpha\sigma^-_\alpha\rho_{\rm m} - \rho_{\rm m} \sigma^-_\alpha\sigma^+_\alpha\sigma^+_\alpha\sigma^-_\alpha\right],
    \label{eq: two atom master with gain}
\end{align}
which again assumes that  $\Gamma^{\downarrow(\uparrow)}_{ab} \approx \Gamma^{\downarrow(\uparrow)}_{ba}$ and $\delta^{\downarrow(\uparrow)}_{ab} \approx \delta^{\downarrow(\uparrow)}_{ba}$, 
and we have also defined $\omega_\alpha'=\omega_\alpha+\delta_{\alpha\alpha}$.
Below we will assume equal dipole resonances
and dipole moments, so that
$\omega_a' =\omega_b'=\omega_0$. 
Thus all the coupling rates above are implicitly evaluated
at $\omega_0$, for example,
$\Gamma^{\downarrow(\uparrow)}_{ab}
\equiv \Gamma^{\downarrow(\uparrow)}_{ab}(\omega_0)$
and
$\delta^{\downarrow(\uparrow)}_{ab}\equiv \delta^{\downarrow(\uparrow)}_{ab}(\omega_0)$.

First, focusing on a single emitter, we can express the coherent and incoherent decay rates in terms of the photonic Green function, where the typical (LDOS) SE rate is given by:
\begin{equation}
    \Gamma^{\rm LDOS}_{aa} (\mathbf{r}_a, \omega) = \frac{2}{\hbar \epsilon_0} \mathbf{d}_a\cdot {\rm Im}\{\mathbf{G}(\mathbf{r}_a,\mathbf{r}_a, \omega)\}\cdot \mathbf{d}_a.
    \label{eq: ldos Gamma_11}
\end{equation}
Similarly, the SE rate into a homogeneous background medium can be expressed as:
\begin{equation}
    \Gamma_{0} (\omega) = \frac{2}{\hbar \epsilon_0} \mathbf{d}_a\cdot {\rm Im}\{\mathbf{G}_{\rm B}(\mathbf{r}_a,\mathbf{r}_a, \omega)\}\cdot \mathbf{d}_a,
    \label{eq: gamma_0}
\end{equation}
where $\mathbf{G}_{\rm B}(\mathbf{r}_a,\mathbf{r}_a, \omega)$ is the Green function in a homogeneous medium, where ${\rm Im}\{\mathbf{G}_{\rm B}(\mathbf{r}_a,\mathbf{r}_a, \omega)\} = \left(\frac{\omega^3 n_{\rm B}}{6\pi c^3}\right)\mathbb{1}$. This latter
rate will enable us to define the Purcell factor (enhanced SE rate).

When a linear gain medium is introduced, the LDOS contribution no longer describes the response of the system alone. To appreciate this, it is important to note that  the well known LDOS-SE formula is linked to the Green function identity, $\int_{\mathbb{R}^3}{\rm d}\mathbf{s}\epsilon_I(\mathbf{s})\mathbf{G}(\mathbf{r},\mathbf{s})\cdot\mathbf{G}^*(\mathbf{s},\mathbf{r}')={\rm Im}[\mathbf{G}(\mathbf{r},\mathbf{r}')]$, which involves an integration over all space. However, to use this identity, one must subtract the contribution from the gain, and this results in \textit{adding} the separate gain contribution term~\cite{franke_fermis_2021, ren_classical_2023}. Thus the \textit{total} SE rate,
in the presence of loss and gain, can thus be defined as
\begin{equation}
    \Gamma^{\rm \downarrow}_{ aa} (\mathbf{r}_a,\omega) = \Gamma^{\rm LDOS}_{aa} (\mathbf{r}_a,\omega) + \Gamma^{\rm \uparrow}_{aa} (\mathbf{r}_a,\omega),
    \label{eq: gamma_11_tot}
\end{equation}
where the gain contribution is given by
\begin{equation}
    \Gamma^{\rm \uparrow}_{ aa} (\mathbf{r}_a,\omega) = \frac{2}{\hbar \epsilon_0} \mathbf{d}_a\cdot \mathbf{K}(\mathbf{r}_a,\mathbf{r}_a,\omega)\cdot \mathbf{d}_a,
\end{equation}
and the real tensor, $\mathbf{K}(\mathbf{r}_a,\mathbf{r}_a,\omega)$, is expressed by
\begin{equation}
    \mathbf{K}(\mathbf{r}_a,\mathbf{r}_a,\omega) = \int_{V_{\rm gain}} d\mathbf{r} \left|{\rm Im}[\epsilon^{\rm gain}(\mathbf{r},\omega)]\right|\mathbf{G}(\mathbf{r}_a,\mathbf{r},\omega)\cdot \mathbf{G}^{*}(\mathbf{r},\mathbf{r}_a,\omega).
\end{equation}

The incoherent decay rate between the two coupled emitters, with dipole moments $\mathbf{d}_a = d_a\mathbf{e}_a$ and $\mathbf{d}_b = d_b\mathbf{e}_b$, is given by~\cite{PhysRevB.99.085311}:
\begin{equation}
    \Gamma_{ab}^{\rm NLDOS}(\mathbf{r}_a, \mathbf{r}_b, \omega) = \frac{2 d_a d_b}{\hbar \epsilon_0}{\rm Im}[\mathbf{e}_a \cdot \mathbf{G}(\mathbf{r}_a, \mathbf{r}_b,\omega)\cdot \mathbf{e}_b],
    \label{eq: gamma no gain}
\end{equation}
which is  known as the real photon transfer rate. 
The coherent decay rate between two dipoles is~\cite{PhysRevB.99.085311}
\begin{equation}
    \delta_{ab}^{\rm NLDOS}(\mathbf{r}_a, \mathbf{r}_b,\omega) = \frac{-d_a d_b}{\hbar \epsilon_0}{\rm Re}[\mathbf{e}_a \cdot \mathbf{G}(\mathbf{r}_a, \mathbf{r}_b,\omega)\cdot \mathbf{e}_b],
    \label{eq: delta no gain}
\end{equation}
and this is often termed the virtual photon transfer rate. This latter process causes a frequency shift of the excited states, while forming
subradiant and superradiant states. 
We refer to this term as a 
non-local density of states term (NLDOS),
as it depends on two different spatial arguments. 
Note also that in order to evaluate the principal value integrals, we considered frequencies ranging from $-\infty$ to $\infty$, so that we can exploit
the relation, ${\cal P}\int_{-\infty}^{\infty}dx f(x)/(x-x_0) = i\pi f(x)$. Thus we can express $\delta_{ab}^{\rm NLDOS}$  in terms of the real part of the Green function.

In the presence of a linear gain medium, 
as shown earlier, we must consider an additional contribution from the gain.
Then, similar to the single emitter case, we can define the tensor $\mathbf{K}_{ab}$ such that:
\begin{equation}
\mathbf{K}(\mathbf{r}_a,\mathbf{r}_b,\omega)
=\int_{V_{\rm gain}}\!d\mathbf{r}\left|{\rm Im}[\epsilon^{\rm gain}(\mathbf{r},\omega)]\right|\mathbf{G}(\mathbf{r}_a,\mathbf{r},\omega)\cdot\mathbf{G}^{*}(\mathbf{r},\mathbf{r}_b,\omega).
\label{eq: K}
\end{equation}
Consequently, the gain contributions to the coherent and incoherent decay rates are
\begin{align}
    \Gamma^{\rm \uparrow}_{ ab} (\mathbf{r}_a, \mathbf{r}_b, \omega) &= \frac{2}{\hbar \epsilon_0}\mathbf{d}_a \cdot {\rm Re} \{ \mathbf{K}(\mathbf{r}_a,\mathbf{r}_b,\omega) \} \cdot \mathbf{d}_b,
    \label{eq: gamma_12_gain}\\
    \delta^{\rm \uparrow}_{ ab} (\mathbf{r}_a, \mathbf{r}_b, \omega) &= - 
    \frac{1}{\hbar \epsilon_0}\mathbf{d}_a \cdot {\rm Im}\{\mathbf{K}(\mathbf{r}_a,\mathbf{r}_b,\omega)\}\cdot \mathbf{d}_b.
    \label{eq: delta_12_gain}
\end{align}

Thus, we find  
that the total coherent and incoherent decay rates in the presence of gain are given by
\begin{align}
    \Gamma^{\rm \downarrow}_{ab} (\mathbf{r}_a, \mathbf{r}_b, \omega) &= \Gamma^{\rm NLDOS}_{ab} (\mathbf{r}_a, \mathbf{r}_b, \omega) + \Gamma^{\rm \uparrow}_{ab} (\mathbf{r}_a, \mathbf{r}_b, \omega),
    \label{eq: gamma_12_tot}\\
   \delta^{\rm \downarrow}_{ab} (\mathbf{r}_a, \mathbf{r}_b, \omega) &=  \delta^{\rm NLDOS}_{ab} (\mathbf{r}_a, \mathbf{r}_b, \omega) +  \delta^{\rm \uparrow}_{ab} (\mathbf{r}_a, \mathbf{r}_b, \omega). 
    \label{eq: delta_tot}
\end{align}
These emission terms explicitly include the gain terms as well, consistent with a thermal reservoir model~\cite{Tian1992}. We note here that in practice, $\delta_{ab}^{\uparrow}$ ends up being quite small, and can even be zero under certain conditions. Indeed, using the properties of Green functions, Eq.~\eqref{eq: K} has zero imaginary component in the case of spatially symmetric positions, thus resulting in $\delta_{ab}^{\uparrow} = 0$. However, note that $\delta_{ab}^{\rm NLDOS}$ is still modified due to gain,
which is then the dominant effect. 

When considering two (or more) quantum emitters, one can also study the collective effects which arise from radiative coupling between the emitters~\cite{PhysRevApplied.4.044018,vyshnevyy_gain-dependent_2022}.
In the case of two emitters, by first assuming that the dipole moments for the emitters are equivalent ($d_a = d_b$), and the two emitters are located at spatially symmetric locations (though these are not necessarily requirements), one can easily define the rates for superradiance (enhanced collective emission) and subradiance (suppressed collective emission), denoted by $\Gamma^+$ and $\Gamma^-$, respectively. These rates are usually given by~\cite{Gangaraj:15}:
\begin{align}
    \Gamma^+ &= \Gamma^{\rm LDOS}_{aa} + \Gamma^{\rm NLDOS}_{ab} = \Gamma^{\rm LDOS}_{bb} + \Gamma^{\rm NLDOS}_{ba},
    \label{eq: superradiance}\\
    \Gamma^- &= \Gamma^{\rm LDOS}_{aa} - \Gamma^{\rm NLDOS}_{ab} = \Gamma^{\rm LDOS}_{bb} - \Gamma^{\rm NLDOS}_{ba}.
    \label{eq: subradiance}
\end{align}

When gain is included in the system, we observe an additional contribution from the gain to the superradiant and subradiant rates. Thus, the \textit{total} superradiance and subradiance decay rates are now given by
\begin{align}
    \Gamma^{\rm +, \downarrow} &= \Gamma^{\rm LDOS}_{aa} + \Gamma^{\rm NLDOS}_{ab} + \Gamma^{\rm \uparrow}_{aa} + \Gamma^{\rm \uparrow}_{ab},
    \label{eq: total superradiance}\\
    \Gamma^{\rm -, \downarrow} &= \Gamma^{\rm LDOS}_{aa} - \Gamma^{\rm NLDOS}_{ab}+\Gamma^{\rm \uparrow}_{aa} - \Gamma^{\rm \uparrow}_{ab}.
    \label{eq: total subradiance}
\end{align}
When relaxing the assumptions to include the possibility that the dipole moments of the two emitters are not the same and 
the emitters are not at spatially symmetric locations, we will find that $\Gamma_{aa}^{\downarrow(\uparrow)} \neq \Gamma_{bb}^{\downarrow(\uparrow)}$, and thus a more rigorous theoretical derivation is required without using the assumption that the incoherent decay rates of the two emitters are equivalent, with or without gain. 
Our master equation, however, can easily incorporate such studies as well.


One can also determine the optical Bloch equations from the master equation for two emitters (with or without gain), using the expectation value of an operator, which is $\braket{\dot{O}} = \braket{\dot{\rho}_s O} = \rm{Tr}\{\dot{\rho}_s O\}$. For the two emitter Hilbert space,
the basis states are tensor products of the state of each atom, such that the elements of the bare-state density matrix are given by $\rho_{ij,lm} = \bra{ij}\rho\ket{lm}$, where indices $i$ and $l$ represent the state of atom $a$, and indices $j$ and $m$ represent the state of atom $b$. 

The basis we are using is made up of the single ket states $\ket{g_a}$, $\ket{g_b}$, $\ket{e_a}$, $\ket{e_b}$, where the coupling of the two emitters results in double ket states such that the states of both atoms are being considered, meaning that the four basis states in the bare basis are (1) $\ket{g_ag_b}$,
(2) $\ket{g_ae_b}$, 
(3) $\ket{e_ag_b}$, and 
(4) $\ket{e_ae_b}$.
The density matrix is then made up of 16 elements, where the trace sums to one and $\rho_{ij,lm} = \rho^*_{lm,ij}$. The density matrix is
\begin{equation}
    \rho_{ij,lm} = 
\begin{pmatrix}
    \rho_{11} & \rho_{12} & \rho_{13} & \rho_{14} \\
    \rho_{21} & \rho_{22} & \rho_{23} & \rho_{24} \\
    \rho_{31} & \rho_{32} & \rho_{33} & \rho_{34} \\
    \rho_{41} & \rho_{42} & \rho_{43} & \rho_{44}
    \label{eq: density matrix 2 TLS}
\end{pmatrix}.
\end{equation}
The diagonal elements represent populations, where $\rho_{33}$ is the excited state population of atom $a$ with atom $b$ being in the ground state, $\rho_{22}$ is the excited state population of atom $b$ with atom $a$ being in the ground state, $\rho_{44}$ is the population of both excited states together, and $\rho_{11}$ is the population of both atoms in the ground state. 
In the single atom basis, the Pauli operators are given by the following $2\times2$ matrices:
\begin{equation}
    \sigma_i^+ = 
    \begin{pmatrix}
        0 & 0 \\
        1 & 0
    \end{pmatrix},
    ~~\sigma_i^- = 
    \begin{pmatrix}
        0 & 1 \\
        0 & 0
    \end{pmatrix}.
\end{equation}
In the two atom basis, which has been described above, the raising/lowering operators, $\sigma^{+/-}_i$, where $i = a,b$, are given by $4\times4$ matrices, obtained through an outer product with the identity matrix.

From the density matrix, the optical Bloch equations for these populations with gain,
in a rotating frame at
frequency $\omega_0$,
 where $\omega_a = \omega_b = \omega_0$, are
\begin{align}
    \dot{\rho}_{22} =& -\left(\Gamma_{bb}^{\downarrow}(\mathbf{r}_b)+\Gamma_{aa}^{\uparrow}(\mathbf{r}_a)\right)\rho_{22} 
    + \Gamma_{aa}^{\downarrow}(\mathbf{r}_a)\rho_{44}  + \Gamma_{bb}^{\uparrow}(\mathbf{r}_b)\rho_{11} 
    - \left(\Gamma_{ab}^{\downarrow}(\mathbf{r}_a, \mathbf{r}_b)+\Gamma_{ab}^{\uparrow}(\mathbf{r}_a, \mathbf{r}_b)\right)\rho_{23}^{\rm R} 
\nonumber \\ &- 2\left(\delta_{ab}^{\downarrow}(\mathbf{r}_a, \mathbf{r}_b)-\delta_{ab}^{\uparrow}(\mathbf{r}_a, \mathbf{r}_b)\right)\rho_{23}^{\rm I}, \label{eq: rho_22} \\
    \dot{\rho}_{33} =& -\left(\Gamma_{aa}^{\downarrow}(\mathbf{r}_a)+\Gamma_{bb}^{\uparrow}(\mathbf{r}_b)\right)\rho_{33} 
    + \Gamma_{bb}^{\downarrow}(\mathbf{r}_b)\rho_{44} + \Gamma_{aa}^{\uparrow}(\mathbf{r}_a)\rho_{11}  
    - \left(\Gamma_{ab}^{\downarrow}(\mathbf{r}_a, \mathbf{r}_b)+\Gamma_{ab}^{\uparrow}(\mathbf{r}_a, \mathbf{r}_b)\right)\rho_{23}^{\rm R} 
    \nonumber \\ &+ 2\left(\delta_{ab}^{\downarrow}(\mathbf{r}_a, \mathbf{r}_b)-\delta_{ab}^{\uparrow}(\mathbf{r}_a, \mathbf{r}_b)\right)\rho_{23}^{\rm I}, 
    \label{eq: rho_33} \\
    \dot{\rho}_{44} =& - \left(\Gamma_{aa}^{\downarrow}(\mathbf{r}_a) + \Gamma_{bb}^{\downarrow}(\mathbf{r}_b)\right)\rho_{44} 
    + \Gamma_{aa}^{\uparrow}(\mathbf{r}_a)\rho_{22} 
    + \Gamma_{bb}^{\uparrow}(\mathbf{r}_b)\rho_{33} 
    + 2\Gamma_{ab}^{\uparrow}(\mathbf{r}_a,\mathbf{r}_b)\rho_{23}^{\rm R} ,  \\
    \dot{\rho}_{23} =& -\left(\frac{\Gamma_{aa}^{\downarrow}(\mathbf{r}_a) + \Gamma_{bb}^{\downarrow}(\mathbf{r}_b)+\Gamma_{aa}^{\uparrow}(\mathbf{r}_a) + \Gamma_{bb}^{\uparrow}(\mathbf{r}_b)}{2}\right)\rho_{23}
    + \frac{\Gamma_{ab}^{\downarrow}(\mathbf{r}_a,\mathbf{r}_b)}{2}\left(2\rho_{44} -\rho_{33} - \rho_{22}\right)  
     + \frac{\Gamma_{ab}^{\uparrow}(\mathbf{r}_a,\mathbf{r}_b)}{2} 
     \left(2\rho_{11} -\rho_{33} - \rho_{22}\right) \nonumber \\
     +& \frac{\Gamma^{'}}{2}\rho_{23}- i\left( \delta_{ab}^{\downarrow}(\mathbf{r}_a,\mathbf{r}_b) - \delta_{ab}^{\uparrow}(\mathbf{r}_a,\mathbf{r}_b)\right)\left(\rho_{33} - \rho_{22}\right),\label{eq: bare coherence tot} 
\end{align}
which are derived from the master equation in Eq.~\eqref{eq: two atom master with gain}, where $\rho_{23}^{\rm R/I}$ represents the real/imaginary part of the coherence $\rho_{23}$. 
All decay rates and coupling parameters
are (implicitly) determined at frequency $\omega_0$, and we have added in the explicit dependence on dipole position for clarity. 
Again it is important to note that terms such as
$\Gamma_{aa}^{\downarrow}(\mathbf{r}_a)$
contain an LDOS contribution
as well as a term that requires an integration over the gain region, which modifies the 
SE rate at that spatial location
with a non-local gain contribution. This allows the two quanta state to become populated, which can then make transitions to both the superradiant and subradiant state.


These Bloch equations  form a closed set, so we can use any standard ODE numerical solver to solve them,
or solve them analytically if they admit a simple solution. Alternatively one can easily solve
the master equation in developed open-source software such as
QuTiP~\cite{JOHANSSON_qutip1,JOHANSSON_qutip2}.
From this set of equations, we can easily
determine the  excited-state populations  of atom $a$ and atom $b$, which are given by $\braket{\sigma^+_a \sigma^-_a}$ and $\braket{\sigma^+_b \sigma^-_b}$, respectively. Expressions for these populations can be determine based on the Bloch equations above, since for $i,j = a,b$ we have the following relationship:
\begin{equation}
    \sigma^+_i \sigma^-_i =  \sigma^+_i \sigma^-_i (\sigma^-_j \sigma^+_j + \sigma^+_j \sigma^-_j),
\end{equation}
where $(\sigma^-_j \sigma^+_j + \sigma^+_j \sigma^-_j)$ acts as the identity operator when applied to a state. Thus we 
obtain the populations of atom $a$ and $b$, $\rho_{aa}$ and $\rho_{bb}$, from
\begin{align}
\rho_{aa} &= \braket{\sigma^+_a\sigma^-_a}
=  \rho_{33}
+ \rho_{44},\label{eq: atom a pop full}\\
    \rho_{ bb} &= \braket{\sigma^+_b\sigma^-_b}
= \rho_{22}
+ \rho_{44},\label{eq: atom b pop full}
\end{align}
which now describes the excited state populations of the two emitters in the bare basis, including the dynamics from all of the basis states that contribute to populating the emitters.

For many studies
with coupled atoms in lossy media,
the  WEA
is invoked, so that effectively the coupled system resembles a three-level system, and one can solve the system under {\it linear response}. Such a system can also be solved classically because of this correspondence. For example, with no gain, if we 
start with a single excitation, the $\ket{e_ae_b}$ state is not populated, so that $\braket{\sigma^+_a\sigma^-_a}
=  \rho_{33}$ and $\braket{\sigma^+_b\sigma^-_b}
= \rho_{22}$ under these conditions.
However, with any finite gain value, this WEA no longer applies, and the system is inherently nonlinear, even with no initial excitations, as the gain will act as an incoherent pumping term.

\subsection{Optical Bloch equations in the dressed-state basis}

When considering an emitter interacting with fields or with another emitter, it is often useful to present the optical Bloch equations in the dressed basis~\cite{Gangaraj:15,Karanikolas:20,PhysRevB.93.165305,10.1063/5.0119264}. This basis provides additional physical insights and makes the solution more compact and efficient, and allows us to more naturally investigate superradiant and subradiant decay rates.

We first diagonalize the system Hamiltonian, which allows us to identify the system eigenstates, which become the basis states for the dressed basis. To find the system eigenstates, we must first determine the eigenvalues by solving the eigenvalue equation, where an eigenvalue $\lambda$ can be solved from the fact that ${\rm det}(H - \lambda I)=0$, where $I$ is the identity. First, to further simplify, we assume that the two emitters are located at spatially symmetric positions, such that $\Gamma_{aa}^{\downarrow(\uparrow)} = \Gamma_{bb}^{\downarrow(\uparrow)}$. Then, it can be found that the four eigenvalues are given by $[0, \omega_0 - \delta_{12},  \omega_0 + \delta_{12},2\omega_0,]$. 

\begin{figure}
    \centering
    \includegraphics[width=0.6\linewidth]{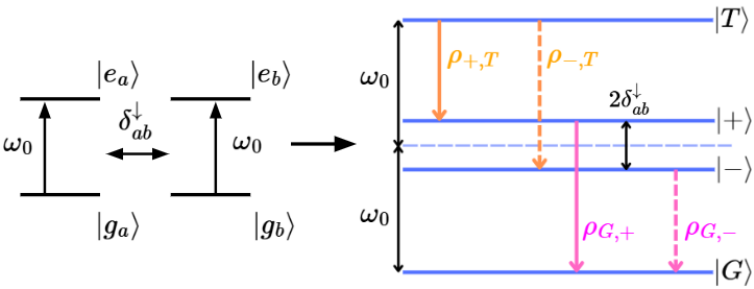}
    \caption{A schematic energy level diagram to show  how the bare state energy levels hybridize to form the dressed energy basis. On the left, we see the two TLSs which are coupled via $\delta_{ab}^{\downarrow}$, where we assume that the energy difference between the ground and excited states in both cases is $\hbar \omega_0$. On the right, we show the energy levels in the dressed basis, where the ground $\ket{G}$ and two-quanta $\ket{T}$ states are separated by $2\omega_0$, and the superradiant $\ket{+}$ and subradiant $\ket{-}$ states are separated from each other by $2\delta_{ab}^{\downarrow}$, where $\delta_{ab}^{\downarrow} = \delta_{ab}^{\rm NLDOS}+\delta^{\uparrow}_{ab}$. We note that $\delta^{\uparrow}_{ab}$ is zero when the emitters at located at spatially symmetric positions, and otherwise is quite small. However, the gain also impacts $\delta_{ab}^{\rm NLDOS}$, as discussed in the main text,
    which can then have a significant impact on the splitting.
    The solid lines show the dominant transitions 
    with finite gain, which renders the problem nonlinear. 
   }
    \label{fig: energy levels}
\end{figure}

From the eigenenergies, the eigenstates can be determined, as the eigenenergies and eigenstates are related through $H_{\rm eff}\mathbf{v}=\lambda\mathbf{v}$, where $\mathbf{v}$ are the eigenstates and $H_{\rm eff}$ is the effective system Hamiltonian, where we include the coherent interactions but do not include losses in Liouville space.
We find the following four eigenstates:
\begin{equation}
    \mathbf{v}_1 = 
    \begin{bmatrix}
        1 \\
        0 \\
        0  \\
        0
    \end{bmatrix},
    \mathbf{v}_2 = 
    \begin{bmatrix}
        0 \\
        -1/\sqrt{2} \\
        1/\sqrt{2}  \\
        0
    \end{bmatrix},
    \mathbf{v}_3 = 
    \begin{bmatrix}
        0 \\
        1/\sqrt{2} \\
        1/\sqrt{2}  \\
        0
    \end{bmatrix},
    \mathbf{v}_4 = 
    \begin{bmatrix}
        0 \\
        0 \\
        0  \\
        1
    \end{bmatrix}.
\end{equation}
Thus, this new dressed basis consists of the ground state $\ket{G} = \ket{g_ag_b}$, the biexciton (two-quanta) state $\ket{T} = \ket{e_a e_b}$, and the super ($\ket{+}$) and subradiant 
($\ket{-}$) states $\ket{\pm} = \frac{1}{\sqrt{2}}(\ket{e_ag_b}\pm \ket{g_ae_b})$. A visualization of these states can be found in Fig.~\ref{fig: energy levels}, where we show the states of the two individual emitters in the bare basis and then show how the basis hybridizes to form the dressed basis. In the dressed basis we show the different energy levels with
total (with gain) contributions to the coherent decay rates, which correspond to the eigenvalues. As previously mentioned, when there is no gain in the system and we are working within the dressed basis, the two-quanta state decouples from the basis, leaving only three basis states remaining. However, since gain acts as a pumping mechanism itself, it populates the two-quanta state, meaning that it is crucial to not decouple this state from the dressed basis. The solid transition lines in Fig.~\ref{fig: energy levels} will contribute more dominantly when considering spectral emission compared to the dotted transition lines. This results in different spectral linewidths than what one would expect from simple linear dipole emitters, which can be seen later in the spectral results, as well as in \cite{PhysRevB.99.085311}.

The dressed-state optical Bloch equations can be written as
\begin{align}
    \dot{\rho}_{++} &= (\Gamma^{\downarrow}_{aa} + \Gamma^{\downarrow}_{ab})\rho_{TT} + (\Gamma^{\uparrow}_{aa} + \Gamma^{\uparrow}_{ab})\rho_{GG} 
    - (\Gamma^{\downarrow}_{aa} + \Gamma^{\downarrow}_{ab})\rho_{++} - (\Gamma^{\uparrow}_{aa} + \Gamma^{\uparrow}_{ab})\rho_{++} 
    - \frac{\Gamma^{'}}{2}(\rho_{++}-\rho_{--}), \label{eq: rho++}
    \\
    \dot{\rho}_{--} &= (\Gamma^{\downarrow}_{aa} - \Gamma^{\downarrow}_{ab})\rho_{TT} + (\Gamma^{\uparrow}_{aa} - \Gamma^{\uparrow}_{ab})\rho_{GG} 
    - (\Gamma^{\downarrow}_{aa} - \Gamma^{\downarrow}_{ab})\rho_{--} - (\Gamma^{\uparrow}_{aa} - \Gamma^{\uparrow}_{ab})\rho_{--}
    + \frac{\Gamma^{'}}{2}(\rho_{++}-\rho_{--}), \label{eq: rho--}
    \\
    \dot{\rho}_{TT} &= -2\Gamma^{\downarrow}_{aa}\rho_{TT} + \Gamma^{\uparrow}_{aa}(\rho_{++}+\rho_{--}) + \Gamma^{\uparrow}_{ab}(\rho_{++}-\rho_{--}),
    \label{eq: rho_TT}
    \\
    \dot{\rho}_{+-} &= -(\Gamma^{\downarrow}_{aa} + 2i\delta^{\downarrow}_{ab} + 2i\delta^{\uparrow}_{ab})\rho_{+-} - \Gamma_{aa}^{\uparrow}(\rho_{++}+\rho_{--})
    + \frac{\Gamma^{\prime}}{2}(\rho_{+-}-\rho_{-+}).\label{eq: rho+-}
\end{align}

In the steady-state limit, where the 
density matrix elements no longer change with time (i.e., the derivatives are equal to 0), one can obtain the following analytical solutions, where again we maintain that $\Gamma_{aa}^{\downarrow(\uparrow)} = \Gamma_{bb}^{\downarrow(\uparrow)}$, but $\Gamma_{ab}^{\downarrow(\uparrow)}$ can be different:
\begin{align}
    \rho_{++}(t_{ss}) &= \frac{(\Gamma_{aa}^{\uparrow}+\Gamma_{ab}^{\uparrow}) - (\Gamma_{aa}^{\uparrow} + \Gamma_{ab}^{\uparrow} - \Gamma^{\prime}/2)\rho_{--}(t_{ss}) + (\Gamma_{aa}^{\downarrow}+\Gamma_{ab}^{\downarrow} - \Gamma_{aa}^{\uparrow} - \Gamma_{ab}^{\uparrow})\rho_{TT}(t_{ss})}{(\Gamma_{aa}^{\downarrow}+\Gamma_{ab}^{\downarrow}) + 2(\Gamma_{aa}^{\uparrow} + \Gamma_{ab}^{\uparrow}) + \Gamma^{\prime}/2}, \label{eq: rho++b}
    \\
   \rho_{--}(t_{ss}) &= \frac{(\Gamma_{aa}^{\uparrow} - \Gamma_{ab}^{\uparrow}) - (\Gamma_{aa}^{\uparrow} - \Gamma_{ab}^{\uparrow} - \Gamma^{\prime}/2)\rho_{++}(t_{ss}) + (\Gamma_{aa}^{\downarrow}-\Gamma_{ab}^{\downarrow} - \Gamma_{aa}^{\uparrow} + \Gamma_{ab}^{\uparrow})\rho_{TT}(t_{ss})}{(\Gamma_{aa}^{\downarrow}-\Gamma_{ab}^{\downarrow}) + 2(\Gamma_{aa}^{\uparrow} - \Gamma_{ab}^{\uparrow}) + \Gamma^{\prime}/2}, \label{eq: rho--b}
    \\
    \rho_{TT}(t_{ss}) &= \frac{(\Gamma_{aa}^{\uparrow}+\Gamma_{ab}^{\uparrow})\rho_{++}(t_{ss}) + (\Gamma_{aa}^{\uparrow}-\Gamma_{ab}^{\uparrow})\rho_{--}(t_{ss})}{2\Gamma_{aa}^{\downarrow}},
    \label{eq: rhoTTb}
    \\
    {\rm Re}[\rho_{+-}(t_{ss})] &= \frac{\Gamma_{aa}^{\uparrow}(\rho_{++}(t_{ss})+\rho_{--}(t_{ss}))-2{\rm Im}[\rho_{+-}(t_{ss})](\delta_{ab}^{\downarrow}+\delta_{ab}^{\uparrow})}{\Gamma_{aa}^{\downarrow}},
    \\
    {\rm Im}[\rho_{+-}(t_{ss})] &= \frac{2{\rm Re}[\rho_{+-}(t_{ss})](\delta_{ab}^{\downarrow}+\delta_{ab}^{\uparrow})}{\Gamma^{\prime} - \Gamma_{aa}^{\downarrow}}.\label{eq: rho+-b}
\end{align}
For these solutions to be valid, the gain must be finite, in which case these solutions will yield the same steady-state values regardless of the initial condition. The advantage of having solutions which has $\Gamma_{aa}^{\downarrow(\uparrow)} = \Gamma_{bb}^{\downarrow(\uparrow)}$, but not necessarily equal to $\Gamma_{ab}^{\downarrow(\uparrow)}$ is for a more robust understanding of the physics before the lasing regime. We note here that without including pure dephasing, one could obtain steady-state solution dependent on initial condition, if $\Gamma_{aa}^{\downarrow (\uparrow)} = \Gamma_{ab}^{\downarrow (\uparrow)}$. Thus, pure dephasing is not only a valuable inclusion for making the system more realistic, but also for ensuring that the subradiant state experiences decay for symmetric cases, which (as also mentioned earlier) is required to properly observe the spectrum, which will be defined and studied later.

In the case of a single excitation with no gain and no dephasing, the two-quanta state decouples from the basis, which simplifies the set of dressed-state optical Bloch equations even further. Here, the initial state of the system only has non-zero density matrix elements in $\rho_{++}, \rho_{--}, \rho_{+-}, \rho_{-+}$, and thus the Bloch equations can be rewritten as:
\begin{align}
    \dot{\rho}_{++} &= - (\Gamma^{\downarrow}_{aa} + \Gamma^{\downarrow}_{ab})\rho_{++} = -\Gamma^+\rho_{++}, \label{eq: no_gain_superradiant_pop}
    \\
    \dot{\rho}_{--} &= - (\Gamma^{\downarrow}_{aa} - \Gamma^{\downarrow}_{ab})\rho_{--} = -\Gamma^-\rho_{--}, \label{eq: no_gain_subradiant_pop}
    \\
    \dot{\rho}_{+-} &= -(\Gamma^{\downarrow}_{aa}+ 2i\delta^{\downarrow}_{ab})\rho_{+-},
\end{align}
which recovers the usual superradiant and subradiant decay rates, and are independent of the initial condition of the system. These differential equations yield the following analytical results,
\begin{align}
    \rho_{++} &= C_1e^{-\Gamma^{+}t}, \\
    \rho_{--} &= C_2 e^{-\Gamma^{-}t}, \\
    \rho_{+-} &= C_3 e^{-(\Gamma_{aa}^{\downarrow}+2i\delta^{\downarrow}_{ab})t},
\end{align}
where $C_1, C_2, C_3$ are constants of integration which can be solved depending on the initial condition of the system.
However, when gain is present in the system, the gain is able to excite both atoms incoherently, and
the $\ket{T}$ state does not decouple from the basis. Indeed,  this is the case even with a single atom -- the gain pumping breaks the WEA, and thus the classical solution fails.

\subsection{Measures of entanglement for coupled 
emitter systems}

For many studies with coupled emitters,
one is interested in exploring the
entanglement dynamics. 
For our purposes,  we will look at the logarithmic entanglement negativity (EN), which is defined by~\cite{Plenio_2005_log_EN,Sang_2021_EN,Vidal_2002_entanglement}:
\begin{equation}\label{eq: log EN}
    E_N(\rho) = {\rm log}_2({\cal N}+1),
\end{equation}
where
${\cal N} = 2 \sum_{i}
{\rm max}(0,-\lambda_{i})$,
with $\lambda_i<0$.
Here $\lambda_i$ are the negative eigenvalues of the density matrix $\rho^{T_b}$,
 where the superscript $T_b$ denotes the partial transpose of the density matrix with respect to emitter $b$.
 This can readily be extended to systems with multiple
 quantum emitters. 

Another common measure of entanglement  is concurrence, which ranges from 0 to 1, where a concurrence of 0 indicates  no entanglement, and a concurrence of 1 indicates complete entanglement~\cite{Frank_Verstraete_2001,Horst_2013_two_qubit,Adam_Miranowicz_2004}. However, in this work we will not investigate this measure of entanglement, and instead will focus solely on logarithmic EN.
Note also that the
(logarithmic) EN provides a more robust definition of entanglement as it can better handle mixed states and systems with more than two quantum emitters~\cite{Vidal_2002_entanglement}.

\subsection{Spectral calculations}
In a rotating frame, the stationary (long time limit) spectrum is obtained from the two-time first-order correlation functions calculated from,
\begin{equation}
	S(\omega) = \text{Re}\left[ \int_0^{\infty} dt' \int_0^{\infty} \!\! d\tau \langle 
    {\bf E}^-(t') \cdot {\bf E}^+(t'+\tau)\rangle e^{i\Delta_{\omega }\tau} \right], 
 \label{eq:Sw}
\end{equation}
where $\Delta_{\omega } = \omega - \omega_{\rm 0}$. This definition is valid for any initial condition and time-dependent excitation.

For steady steady solutions, 
where the one-time density matrix have a ``steady state'' solution, then we can also use
\begin{equation}
	S_{\rm }(\omega) = \text{Re}\left[ \int_0^{\infty} \!\! d\tau \langle {\bf E}^-(t_{\rm ss})
    \cdot {\bf E}^+(t_{\rm ss}+\tau)\rangle e^{i\Delta_{\omega }\tau} \right], 
 \label{eq:SwSS}
\end{equation}
evaluated in steady state. 
This is more appropriate and convenient to use in the presence of any finite gain.
The quantum correlation functions are computed 
using the quantum regressions theorem, and
for convenience, we will use QuTiP to compute 
our spectral calculations 
below~\cite{JOHANSSON_qutip1,JOHANSSON_qutip2}. 

The electric field operator,
at some point detector position
${\bf r}_{\rm D}$
in a rotating wave approximation, is given by~\cite{PhysRevA.91.051803}
\begin{equation}
{\bf E}^-(t) \propto 
\sum_i \mathbf{G}(\mathbf{r}_{\rm D},\mathbf{r}_{i};\omega)\cdot \mathbf{d}_{i} \sigma_i^+(t),
\end{equation}
which is not the same as a classical
system  or a bosonic emitter system. 
For coupled TLSs that have incoherent coupling rates comparable to the SE rates (where $\Gamma^{\downarrow(\uparrow)}_{aa} \approx \Gamma^{\downarrow(\uparrow)}_{ab}$), and in the presence gain (or incoherent pumping), the spectral calculations can easily show nonlinear effects, and the system breaks the WEA approximation. Commonly, an incoherent
pump term can be phenomenologically inserted into the master equation to compute observables; however the inclusion of gain media to the system already provides a pumping mechanism without having to add in an additional pumping term ``by hand''. 
To observe these effects in a rotating frame, the emission spectrum can be calculated from~\cite{PhysRevA.91.051803}:
 \begin{equation}
S(\omega) = \sum_{n,n'}\text{Re}\Big\{g_{n,n'}(\omega)S_{n,n'}^0(\omega)\Big\},
\end{equation}
for $n,n'=a,b$, where $g_{n,n'}(\omega) = \frac{1}{\epsilon_0^2}\mathbf{d}_n\cdot\mathbf{G}^*(\mathbf{r}_n,\mathbf{r}_{\rm D};\omega)\cdot \mathbf{G}(\mathbf{r}_{\rm D},\mathbf{r}_{n'};\omega)\cdot \mathbf{d}_{n'}$ is associated with the propagation of the emitted fields to the detector at position $\mathbf{r}_D$, and
\begin{equation}
S_{n,n'}^0(\omega) = \lim_{t\rightarrow \infty}\Bigg[\int_0^{\infty}d\tau \langle \sigma_{n}^+(t+\tau)\sigma_{n'}^-(t)\rangle e^{-i(\omega-\omega_0)\tau}\Bigg],
\end{equation}
where in the case of  identical emitters, we have that $S_{a,a}^0(\omega) = S^0_{b,b}(\omega)$, and $S_{a,b}^0(\omega) = (S^0_{b,a}(\omega))^*$. 
In the spirit of a Markov approximation, 
we can neglect the frequency dependence
of $g(\omega)$, and thus 
 \begin{equation}
S(\omega) \approx \sum_{n,n'}\text{Re}
\Big\{S_{n,n'}^0(\omega)
\Big\}.
\end{equation}

\subsection{Quasinormal mode theory for obtaining the Green's functions and coupling rates}
\label{sec: theory QNM}

Finally, we discuss QNM theory
as a practical way to compute the desired response functions and coupling rates for photonic cavity systems. We will then use this theory in our results section, showing an example for a metal dimer cavity system, whose Green functions can be quantitatively well described in terms of just a single QNM, with and without gain in the system. This then forms a very convenient approach for calculating the coherent and incoherent decay rates which are needed for the master equation.

The QNMs, $\bf{\Tilde{f}}_\mu$, 
are the mode solutions 
to the (classical) Helmholtz equation, 
with open boundary conditions~\cite{ren_quasinormal_2021}:
\begin{equation}
    {\bm \nabla} \cross {\bm \nabla} \cross \Tilde{{\bf f}}_\mu({\bf r}) - \left(\frac{\Tilde{\omega}_{\mu}}{c} \right)^2 \epsilon({\bf r}, \Tilde{\omega}_\mu)\Tilde{{\bf f}}_\mu({\bf r}) = 0,
    \label{eq1}
\end{equation}
where $c$ is the speed of light in a vacuum, $\Tilde{\omega}_\mu$ is the QNM complex eigenfrequency $\Tilde{\omega}_\mu = \omega_\mu - i\gamma_\mu$, and $\epsilon({\bf r}, \Tilde{\omega}_\mu)$ is the complex dielectric function.
The inhomogeneous 
Helmholtz equation for an arbitrary polarization source,  
can be used to define the
Green function, from
$
c^2\mathbf{\nabla}\times\mathbf{\nabla}\times{\mathbf{G}}(\mathbf{r},\mathbf{r}^\prime,\omega) -\omega^2\epsilon(\mathbf{r},\omega){\mathbf{G}}(\mathbf{r},\mathbf{r}^\prime,\omega) \nonumber 
=  \omega^2 {\mathbb{1}}\delta(\mathbf{r} - \mathbf{r}'),
$
where 
the electric field solution is at $\mathbf{r}$, when a source field dipole is at $\mathbf{r}^\prime$.
Within or near the cavity region, 
the Green function can be expressed as a sum of normalized QNMs~\cite{ren_near-field_2020, leung_completeness_1994, ge_quasinormal_2014}:
\begin{equation}
    {\bf G}_{\rm }({\bf r}_a, {\bf r}_b, \omega) = \sum_{\mu} A_{\mu}(\omega) \Tilde{{\bf f}}_{\mu}({\bf r}_a) \Tilde{{\bf f}}_{\mu}({\bf r}_b) + \mathbf{G}_{\rm others},
    \label{eq2}
\end{equation}
where $A_{\mu}(\omega)= {\omega}/{[2(\Tilde{\omega}_\mu - \omega)]}$~\cite{bai_efficient_2013-1},
and we note the vector product is unconjugated (i.e., not $\tilde {\bf f}_\mu \tilde {\bf f}_\mu^*$). The term $\mathbf{G}_{\rm others}$ is a potential correction term which includes additional contributions to the Green function;
for our examples later, this
could account for non-modal
 Coulomb screening from the metal nanorods. 
 
For single mode cavity systems where
a single QNM dominates,
$\mu = c$, then the  Green function
is simply
\begin{equation}
    {\bf G}_c({\bf r}_a, {\bf r}_b, \omega) \approx A_c(\omega) \Tilde{{\bf f}}_c({\bf r}_a) \Tilde{{\bf f}}_c ({\bf r}_b). 
    \label{eq: QNM GF}
\end{equation}
when $\mathbf{r}_b = \mathbf{r}_a$, where $\mathbf{r}_a$ is the location of a dipole with dipole moment $\mathbf{d}_a = d_a\mathbf{e}_a$, we can calculate the SE rate from a single emitter within the 
cavity structure. Using the photonic Green function $\mathbf{G}_c(\mathbf{r}_a, \mathbf{r}_a,\omega)$, calculated with Eq.~\eqref{eq: QNM GF}, we can substitute this in for the Green function in Eq.~\eqref{eq: ldos Gamma_11}, in order to analytically determine the SE rate. In principle, we can also add in a background contribution 
from ``other'' contributions if desired.

By implementing the single mode limit for the Green function as shown in Eq.~\eqref{eq: QNM GF}, we obtain
the analytical form of Eq,~\eqref{eq: K}
\begin{align}
\mathbf{K}(\mathbf{r}_a,\mathbf{r}_a,\omega)
&=\Big|A_{c}(\omega)\Big|^2(\mathbf{e}_{a}\cdot \Tilde{{\bf f}_c}(\mathbf{r}_a))(\mathbf{e}_{a}\cdot \Tilde{{\bf f}_c}^*(\mathbf{r}_a)) 
\int_{V_{\rm gain}}d\mathbf{r}\Big|{\rm Im}[\epsilon^{\rm gain}(\mathbf{r},\omega)]\Big|\Big|\Tilde{{\bf f}_c}({\bf r})\Big|^2,
\label{eq: K with G_c}
\end{align}
which is explicitly real, meaning that the gain contribution term to the SE rate is
\begin{align}
    \Gamma^{\rm \uparrow}_{aa,\rm QNM}({\bf r}_a, \omega)
=\frac{2|\mathbf{d}_a|^2}{\hbar \epsilon_0}\Big|A_{c}(\omega)\Big|^2\Big|\mathbf{e}_{a}\cdot \Tilde{{\bf f}}_c(\mathbf{r}_a)\Big|^2 
\int_{V_{\rm gain}}d\mathbf{r}\Big|{\rm Im}[\epsilon^{\rm gain}(\mathbf{r},\omega)]\Big|\Big|\Tilde{{\bf f}}_c({\bf r})\Big|^2.
\label{eq: gamma_11_gain}
\end{align}

For NLDOS component the incoherent and coherent coupling rates between the two coupled emitters, given generally by Eqs.~\eqref{eq: gamma no gain} and \eqref{eq: delta no gain}, we employ the same method as above, where we substitute the QNM expansion for the Green function as given by Eq.~\eqref{eq: QNM GF} in order to analytically determine the coupling rates. 
For the gain contribution to coherent coupling rate, in the case of a single dominant QNM, 
 $\mu = c$, then
 \begin{equation}\label{eq: delta_ab_up_singleQNM}
     \delta^{\uparrow}_{ab, \rm QNM}(\mathbf{r}_a,\mathbf{r}_b, \omega) = -\frac{1}{\hbar \epsilon_0}\mathbf{d}_a\cdot{\rm Im}\left\{ \Big|A_{c}(\omega)\Big|^2(\mathbf{e}_{a}\cdot \Tilde{{\bf f}_c}(\mathbf{r}_a))(\mathbf{e}_{a}\cdot \Tilde{{\bf f}_c}^*(\mathbf{r}_b)) 
\int_{V_{\rm gain}}d\mathbf{r}\Big|{\rm Im}[\epsilon^{\rm gain}(\mathbf{r},\omega)]\Big|\Big|\Tilde{{\bf f}_c}({\bf r})\Big|^2\right\}\cdot \mathbf{d}_b,
 \end{equation}
which is zero for the case of spatially symmetric positions with equivalent QNM phases, because $(\mathbf{e}_{a}\cdot \Tilde{{\bf f}_c}(\mathbf{r}_a))(\mathbf{e}_{a}\cdot \Tilde{{\bf f}_c}^*(\mathbf{r}_b))$ will give an explicitly real result, similarly for $\delta_{\alpha \alpha}$. This has been discussed more generally using Green function properties previously, however this result remains the same in the case of a single QNM. We note here that this theory is not restricted to having a dispersionless gain medium~\cite{VanDrunen:24}; however for simplicity, that is what we investigate in this work.

\subsection{Classical full-dipole Maxwell simulations for the loss and gain rates}\label{sec: full_dipole}

To confirm that the QNM method accurately describes the photonic Green function and thus the decay and coupling rates (with and without gain), one can use a classical numerical full-dipole method, with no mode approximations.
For our work below, we use
COMSOL Multiphysics to calculate the Green function and integrated Poynting vectors, and power flow arguments to determine the rates as described in~\cite{ren_classical_2023}.
However, other numerical Maxwell solvers could also compute
the same functions, such as finite-difference
time-domain (FDTD) approaches. In our case, we prefer to use COMSOL as this is the approach we use to compute the QNMs, so we can compare with exactly the same numerical grids and simulation software. 

The four contributions to the total power flow that are of interest, from a dipole at
location ${\bf r}_a$ are 
\begin{align}
P_{\rm LDOS}(\mathbf{r}_{a},\omega)&=\int_{\Sigma_{\rm d}} \hat{{\bf n}} \cdot {\bf S}_{\rm }^{\rm poyn}({\bf r}, \omega)d\mathbf{r},\label{eq: PLDOS_num}\\
P_{\rm nloss/gain}(\mathbf{r}_{a},\omega)&= -{\rm sgn}(\epsilon_{\rm Im}^{\rm loss/gain}(\omega))
\int_{\Sigma_{\rm loss/gain}} \hat{{\bf n}} \cdot {\bf S}_{\rm }^{\rm poyn}({\bf r}, \omega)d\mathbf{r},\label{eq: Pnlossgain_surf}\\
P_{\rm rloss}(\mathbf{r}_{a},\omega)&=\int_{\Sigma_{\rm far}} \hat{{\bf n}} \cdot {\bf S}_{\rm }^{\rm poyn}({\bf r}, \omega)d\mathbf{r},\label{eq: Prloss}
\end{align}
where, respectively, these power contributions represent the power flowing away from a local region around the emitter, the non-radiative loss or gain, and the radiative losses (due to the metal). 
We use the sign function in Eq.~\eqref{eq: Pnlossgain_surf} (${\rm sgn}[\epsilon_{\rm Im}^{\rm loss/gain}]={\rm sgn}[{\rm Im}(\epsilon^{\rm loss/gain})]=\pm1$) to define net positive powers for $P_{\rm nloss/gain}(\mathbf{r}_{a},\omega)$.
We have also used $\Sigma_{\rm d/loss/gain/far}$ for a closed surface enclosing the dipole, lossy region, gain region, and the closed far-field surface, respectively; in addition, ${\bf S}_{\rm }^{\rm poyn}({\bf r}, \omega)$ is the Poynting vector.


The conventional LDOS decay rate (in  Purcell factor units) for a dipole at $\mathbf{r}_a$ is given by (`FD' subscripts refer to full-dipole calculations)
\begin{equation}\label{eq: FP_num_LDOS}
\frac{\Gamma_{aa, \rm FD}^{\rm LDOS}(\mathbf{r}_a,\omega)}
{\Gamma_0(\omega)}
= 
\frac{P_{\rm LDOS}(\mathbf{r}_a,\omega)}{P_0({\omega})},
\end{equation}
where $P_0(\omega)$ is the background power contribution from the same emitter in a homogeneous medium (i.e. without the gain and the resonator structure). However, this rate
is not the correct rate for SE decay
 when gain is included in the medium. 
 Instead, this expression can be corrected by adding the contribution from the gain region, as follows~\cite{ren_classical_2023}:
\begin{align}\label{eq: FP_num_LDOSpGain}
\frac{\Gamma_{aa,\rm FD}^{\downarrow}(\mathbf{r}_a,\omega)}
{\Gamma_0(\omega)}
=
     \frac{P_{\rm LDOS}(\mathbf{r}_a,\omega) + P_{\rm gain}(\mathbf{r}_a,\omega)}{P_0({\omega})}.
\end{align}
 We can express Eq.~\eqref{eq: FP_num_LDOSpGain}
 as 
\begin{align}
 \Gamma_{aa,\rm FD}^{\downarrow}(\mathbf{r}_a,\omega)= \Gamma_{aa, \rm FD}^{\rm LDOS}(\mathbf{r}_a,\omega) +\Gamma_{aa,\rm FD}^{\uparrow}(\mathbf{r}_a,\omega),\label{eq: Gamma_aa_down_num}
 \end{align}
 with
 \begin{align}
 \frac{\Gamma_{aa,\rm FD}^{\uparrow}(\mathbf{r}_a,\omega)}{\Gamma_0(\omega)}=\frac{ P_{\rm gain}(\mathbf{r}_a,\omega)}{P_0({\omega})}\label{eq: Gamma_aa_up_num}.
\end{align}

For coupled dipole emitters with real dipole moments $\mathbf{d}_{a/b} =d_{a/b}\mathbf{e}_{\rm a/b}$, at the spatial points ${\bf r}_{a/b}$, one can obtain the numerical Green's function (accounting for the NLDOS contribution) from the scattered field $\mathbf{E}^{\rm scatt}(\mathbf{r}_{b},\omega)$ at $\mathbf{r}_{b}$, originating from a dipole source at $\mathbf{r}_{a}$:
\begin{align}
\mathbf{E}^{\rm scatt}(\mathbf{r}_{b},\omega)=\frac{\mathbf{d}_{a}\cdot\mathbf{G}_{\rm num}(\mathbf{r}_{a},\mathbf{r}_{b},\omega)}{\epsilon_{0}}.
\end{align}
Thus,  the numerical photon transfer rate and coherent exchange rate  are 
\begin{align}
\Gamma_{ab, \rm FD}^{\rm NLDOS}({\bf r}_a,{\bf r}_b,\omega)
    &= \frac{2}{\hbar\epsilon_{0}} {\bf d}_a \cdot {\rm Im} \{{\bf G}_{\rm num}^{\rm }({\bf r}_a,{\bf r}_b,\omega)\} \cdot {\bf d}_b
    ,\label{eq: Gamma12_num_LDOS}\\
\delta_{ab, \rm FD}^{\rm NLDOS}({\bf r}_a,{\bf r}_b,\omega)
    &= \left(-\frac{1}{2}\right)\frac{2}{\hbar\epsilon_{0}} {\bf d}_a \cdot {\rm Re} \{{\bf G}_{\rm num}^{\rm }({\bf r}_a,{\bf r}_b,\omega)\} \cdot {\bf d}_b.\label{eq: Delta12_num_LDOS}
\end{align}

In addition to above NLDOS contributions, we also introduce a classical gain correction
to the inter-emitter rate, as~\cite{Novotny2007Principles,ren_classical_2023}
\begin{align}
\label{eq: P12_num_gain}
P_{ab,\rm FD}^{\uparrow}(\mathbf{r}_{a},\mathbf{r}_{b},\omega)
&=-\frac{1}{2} \int_{V_{\rm gain}} {\rm }\{\mathbf{J}^{*}_{b}(\mathbf{r},\omega)\cdot\mathbf{E}_{a}(\mathbf{r},\omega)\}d\mathbf{r}\\
&=+\frac{1}{2} \int_{V_{\rm gain}} {\rm }\{\omega\epsilon_{0}|\epsilon^{\rm gain}_{\rm Im}|\mathbf{E}^{*}_{b}(\mathbf{r},\omega)\cdot\mathbf{E}_{a}(\mathbf{r},\omega)\}d\mathbf{r},
\end{align}
where the current source $\mathbf{J}_{b}(\mathbf{r},\omega)$($=\omega\epsilon_{0}\epsilon_{\rm Im}^{\rm gain}\mathbf{E}_{b}(\mathbf{r},\omega)$)  originates from the dipole at $\mathbf{r}_{b}$, while the electric field $\mathbf{E}_{a}(\mathbf{r},\omega)$ originates from the dipole at $\mathbf{r}_{a}$.
Subsequently,  the gain contributions to the photon transfer rate and exchange rate (assuming $d_a = d_b$) are 
\begin{align}
\frac{\Gamma_{{ab},{\rm FD}}^{\uparrow}({\bf r}_a,{\bf r}_b,\omega)}{\Gamma_0(\omega)}
   & = \frac{{\rm Re}[P_{ab, \rm FD}^{\uparrow}({\bf r}_a,{\bf r}_b,\omega)]}{P_{0}(\omega)},\label{eq: Gamma_ab_up_num}\\
\frac{\delta_{{ab},{\rm FD}}^{\uparrow}({\bf r}_a,{\bf r}_b,\omega)}{\Gamma_0(\omega)}
   & = \left(-\frac{1}{2}\right)\frac{{\rm Im}[P_{ab, \rm FD}^{\uparrow}({\bf r}_a,{\bf r}_b,\omega)]}{P_{0}(\omega)},\label{eq: delta_ab_up_num}    
\end{align}
which in fact have the same analytical expression as the quantum corrections shown in Eqs.~\eqref{eq: gamma_12_gain} and \eqref{eq: delta_12_gain}, when the Green function solution is used to expand the electric field and current source.

The total numerical inter-dipole rates are then 
\begin{align}
{\Gamma_{{ab},{\rm FD}}^{\downarrow}({\bf r}_a,{\bf r}_b,\omega)}
    =\Gamma_{{ab},{\rm FD}}^{\rm NLDOS}({\bf r}_a,{\bf r}_b,\omega)+\Gamma_{{ab},{\rm FD}}^{\uparrow}({\bf r}_a,{\bf r}_b,\omega),\label{eq: Gamma_ab_down_num} \\
{\delta_{{ab},{\rm FD}}^{\downarrow}({\bf r}_a,{\bf r}_b,\omega)}
    = \delta_{{ab},{\rm FD}}^{\rm NLDOS}({\bf r}_a,{\bf r}_b,\omega)+\delta_{{ab},{\rm FD}}^{\uparrow}({\bf r}_a,{\bf r}_b,\omega).\label{eq: delta_ab_down_num} 
\end{align}

\section{Application: Population dynamics and spectra from a 
gain-modified plasmonic dimer system with two quantum emitters}
\label{sec: results}

\subsection{Cavity geometry and simulation details}

\begin{figure}[hbt]
    \subfloat[3D Schematic]{\includegraphics[width=0.28\linewidth]{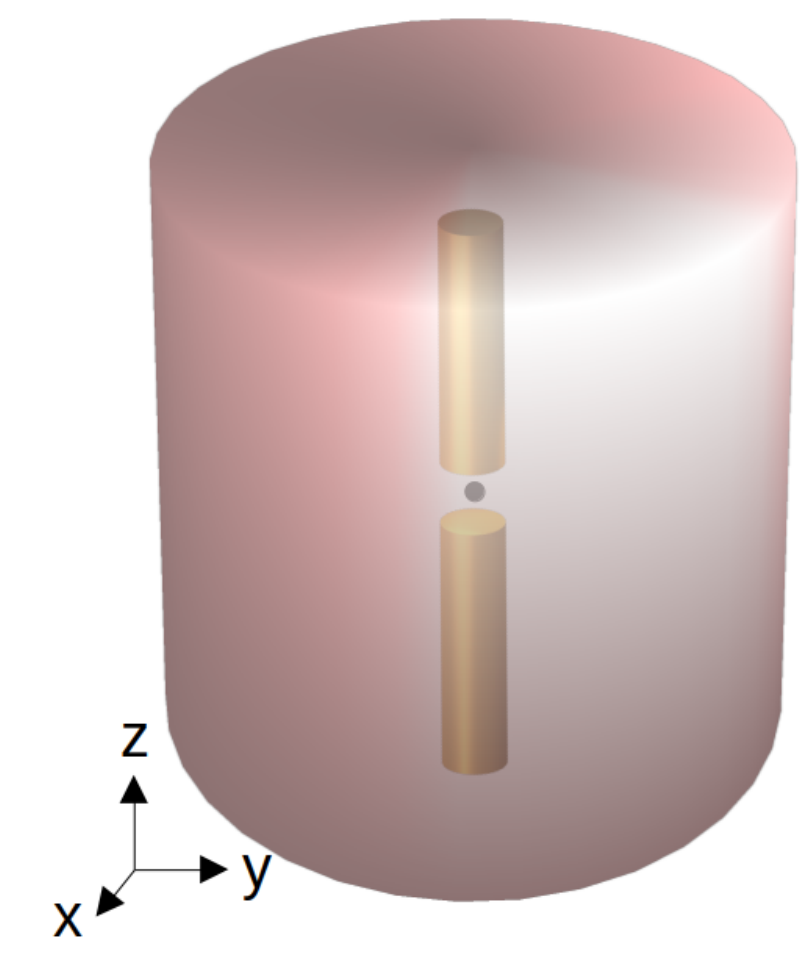}}
    \hspace{0.35cm}
    \subfloat[2D Schematic]{\includegraphics[width=0.52\linewidth]{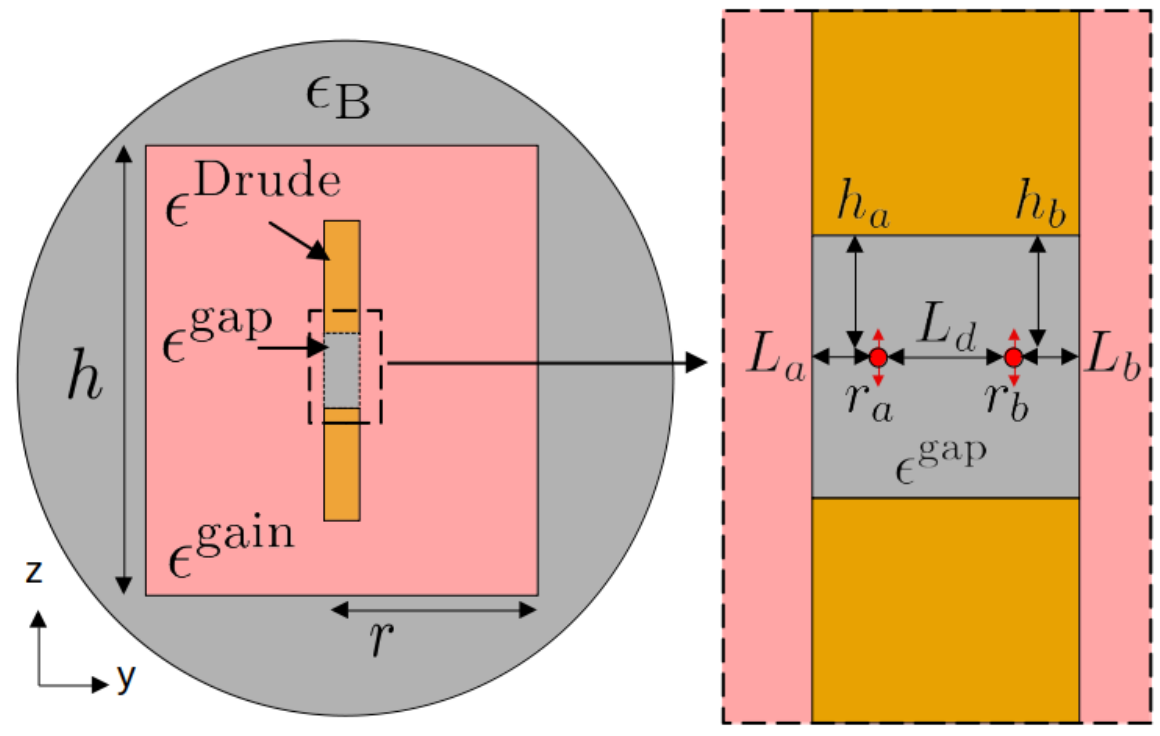}}
    \caption{{\bf Cavity structure of interest containing loss and gain parts.} (a) A  3D schematic of the plasmonic resonator system, used in all our numerical calculations. (b) A 2D diagram of the same system. All the dielectric functions for the materials are labeled, where the background medium has $\epsilon_{\rm B} = 2.25$ ($\epsilon_{\rm B} = n_{\rm B}^2$), the gain region has $\epsilon^{\rm gain} = 2.25 - i\alpha_g$ (with $\alpha_g$ the gain parameter), the gold nanorods have $\epsilon^{\rm Drude}(\omega)$ which is governed by the Drude model (see main text), and the gap region has $\epsilon^{\rm gap}=2.25$. The gold nanorods have a length of $80~$nm, a radius of $10~$nm, and the gap distance is $20~$nm; $h$ represents the height of the gain region, and is $400~$nm, and $r$ represents the radius of the gain region, which is $200~$nm. On the right, we present a zoom-in of the gap region, where two $\mathbf{\hat{z}}$ polarized emitters are located at symmetric locations $\mathbf{r}_{a} = (0,-5~{\rm nm},0)$ and $\mathbf{r}_{b} = (0,5~{\rm nm},0)$,  with a distance of $L_d=10~$nm between them (the origin of the coordinate system is at the gap center). Each emitter is located at a distance $L_{a/b}$ away from the edge of the gap region where the gain begins, where $L_{a/b} = 5~$nm in this schematic. Finally, each emitter is a distance $h_{a/b}$ away from the (upper) nanorod, which is $h_{a/b} = 10~$nm in this schematic, and must be at least $5~$nm away from each nanorod at all times to avoid additional screening effects from the metal structure. Note that for all following results on two coupled emitters, we use spatially symmetric locations, but this is not a model restriction. 
    }
    \label{fig: resonator schematic}
\end{figure}

In this model application section, we present semi-analytical and numerical results for the temporal dynamics and spectra of two coupled quantum emitters within an example 
gain-lossy cavity system. First, in Sec.~\ref{sec: QNM and coupling rates} we justify the use of the QNM theory for calculating the coupling rates by comparing the rates found with the Green function from QNMs and a numerical full dipole method. Additionally, we show the coupling rates for various gain values which will be used for subsequent calculations. Next, we study the near resonant dynamics in the bare state basis in Sec.~\ref{sec: near res dynamics}, and compare these results to the off-resonant dynamics in the bare state basis in Sec.~\ref{sec: off res dynamics}. We then change bases to observe the temporal dynamics in the dressed state basis in Sec.~\ref{sec: dressed state basis}, to better understand how gain impacts the superradiant, subradiant, and two-quanta state populations. Finally, in Sec.~\ref{sec: power spectrum}, we study the emitted power spectrum for a phenomenologically added pump as well as our full gain model, and discuss the implications of the incoherent cross-coupling pumping term
that emerges from our model.

For the  example cavity structure, we consider  a 
fully three-dimensional plasmonic resonator with gain media region, as shown in Fig.~\ref{fig: resonator schematic}. In this system, we wish to study the quantum population dynamics of two coupled emitters located within the gap region of the plasmonic resonator, which is composed of two gold nanorods. The permittivity of the gold dimer is described by the Drude model, where 
  $  \epsilon^{\rm Drude}(\omega)=1-\frac{\omega_{\rm p}^{2}}{\omega^{2}+i\omega\gamma_{\rm p}}$,
with $\hbar\omega_{\rm p}=8.2934$ eV
and $\hbar\gamma_{\rm p}=0.0928$ eV. The gain region, which encompasses the metal dimer and quantum emitters, has permittivity given by $\epsilon^{\rm gain} = 2.25 - i\alpha_g$, where $\alpha_g$ is the gain parameter which governs the amount of gain in the system. In this work, we assume that the gain region is dispersionless for simplicity, however this does not necessarily need to be the case, and gain enhancements can still be observed even in the presence of a dispersive gain medium~\cite{VanDrunen:24}.
In the general theory above,
we stress that there are no restrictions on the models for
$\epsilon^{\rm gain}(\mathbf{r},\omega)$
or 
$\epsilon^{\rm loss}(\mathbf{r},\omega)$. The only restriction is that the combined system remains linear, which can easily be checked with our QNM approach (i.e., $\gamma_c>0$).

For the gain-loss cavity system considered, previous work has shown that
 significant Purcell enhancements
 are possible, especially around the QNM resonance; however this was demonstrated with only one emitter, and without considering any coupled-emitter
 transitions, Bloch equation saturation effects or gain excitation beyond the WEA~\cite{VanDrunen:24}.  Moreover, the
 response system for the gain-compensated metal system is known to be accurately represented with QNM theory with only one cavity mode, and is thus
 a good testbed for looking at collective emitter systems as well.

For the calculation of the QNMs or the full-dipole numerical simulation as a check for the QNM semi-analytical solution, COMSOL Multiphysics is used, with the radio frequency module. The QNM calculation is done in complex frequency space, where the QNM pole(s) are found using the Pad\'e approximation, which uses the electric field to iteratively determine the pole frequency~\cite{bai_efficient_2013-1}.
To identify the resonance response of the emitter with $\mathbf{\hat{z}}$ polarization, the frequency is chosen to be close to the QNM frequency (such that $\Tilde{\omega}_p = \Tilde{\omega}_c (1 - 10^{-5})$). The field QNMs can then be found using a scattered electric field, where the (normalized) QNMs are given by~\cite{ren_quasinormal_2021}
\begin{equation}
\tilde{\mathbf{f}}_c(\mathbf{r}) = \sqrt{\frac{\epsilon_0}{A_c(\Tilde{\omega}_p) \mathbf{d}\cdot\mathbf{E}^{\rm scatt}(\mathbf{r}_0,\Tilde{\omega}_p)}}\mathbf{E}^{\rm scatt}(\mathbf{r},\Tilde{\omega}_p),
\end{equation}
where the emitter is located at $\mathbf{r}_0$, the scattered field from the dipole is the difference of the total electric field and the background electric field (without the resonant structure),
$\mathbf{E}^{\rm scatt}(\mathbf{r},\Tilde{\omega}_p) = \mathbf{E}^{\rm tot}(\mathbf{r},\Tilde{\omega}_p) - \mathbf{E}^{\rm background}(\mathbf{r},\Tilde{\omega}_p)$, and $A_c(\Tilde{\omega}_p) = \Tilde{\omega}_p/2(\Tilde{\omega}_c - \Tilde{\omega}_p)$~\cite{kamandar_dezfouli_regularized_2018}. 
Here, we note that the mode itself is not frequency dependent, and this expression is calculated using discrete, complex frequencies.
For further information, see Refs.~\cite{bai_efficient_2013-1, ren_near-field_2020}.

As shown in Fig.~\ref{fig: resonator schematic}, the finite sized gain region surrounds the metal dimer structure, and is modeled to sit within a
background region. The computational domain is $600~$nm in radius, surrounded by $300~$nm of perfectly matched layers (PMLs) which act as absorbing boundary conditions such that no radiation bounces off of the edge of the computational domain and back into the region with the structure. The maximum
mesh element size for the small sphere of radius $1$~nm surrounding the dipole is $0.1~$nm, inside the metal is $2~$nm, and
outside the metal is $80~$nm.

\subsection{Quasinormal mode and coupling rates
for finite gain values in a plasmonic cavity system}
\label{sec: QNM and coupling rates}

\begin{figure}[ht]
    \subfloat[]{\includegraphics[width=0.52\linewidth]{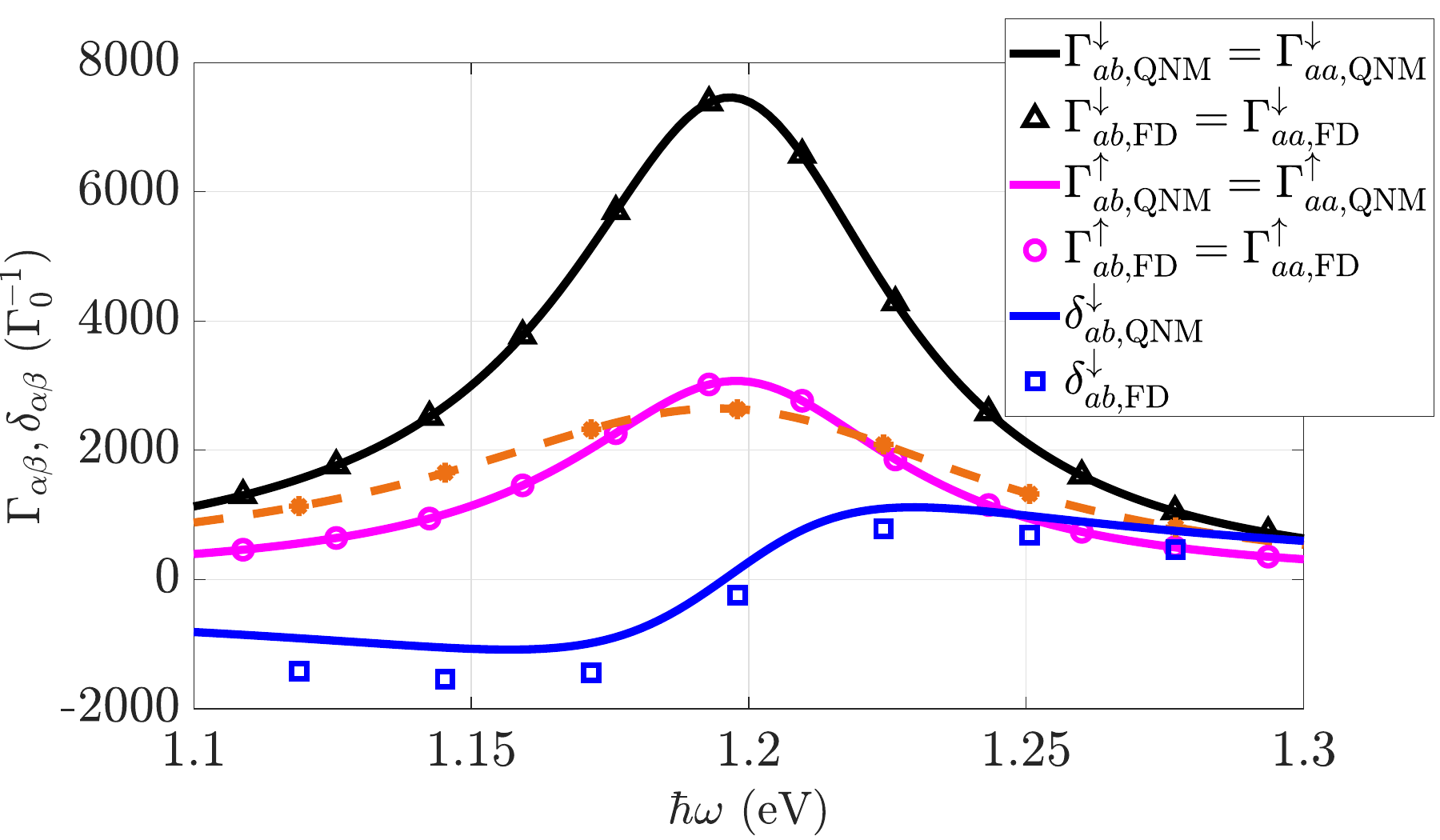}}
    \hspace{0.8cm}
    \subfloat[]{\includegraphics[width=0.25\linewidth]{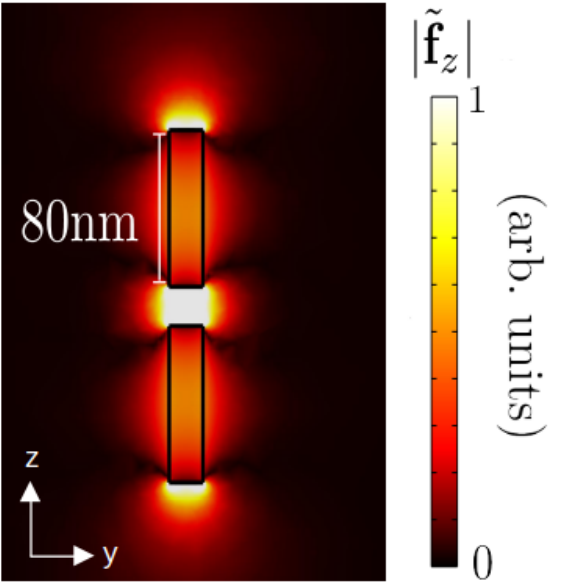}}
    \caption{{\bf Example emitter rates (and Purcell factors) and mode profile in the presence of gain.} (a) The coherent and incoherent decay rates for two coupled emitters located at symmetric locations (see Fig.~\ref{fig: resonator schematic}).
    This corresponds to the case where $\alpha_g = 0.1$. The black curve is calculated with Eq.~\eqref{eq: gamma_12_tot} or Eq.~\eqref{eq: gamma_11_tot} (where the latter can only be used in the single emitter case), the magenta curve represents $\Gamma_{ab}^{\uparrow}$, calculated with Eq.~\eqref{eq: gamma_12_gain}, and the blue curve represents $\delta_{ab}^{\downarrow}$, calculated with Eq.~\eqref{eq: delta_tot}. Due to the symmetric nature, $\Gamma_{aa}^{\downarrow(\uparrow)} = \Gamma_{ab}^{\downarrow(\uparrow)}$. The symbols represent the full dipole numerical solution for the respective rates, as a check for the QNM solution, calculated with Eqs.~\eqref{eq: Gamma_ab_down_num}, \eqref{eq: Gamma_aa_down_num}, \eqref{eq: Gamma_ab_up_num}, \eqref{eq: Gamma_aa_up_num}, and \eqref{eq: delta_ab_down_num}. As a reference, LDOS decay rates with no gain are shown as the dashed orange curve (QNM results, Eq.~\eqref{eq: ldos Gamma_11}, Eq.~\eqref{eq: QNM GF}) and symbols (full dipole calculations, Eq.~\eqref{eq: FP_num_LDOS}). 
    In this case, due to the symmetric nature, $\Gamma^{\rm LDOS}_{aa} = \Gamma_{ab}^{\rm NLDOS}$. (b) 2D surface plot of the spatial QNM profile, using only the dominant ($z$) component, for the case when $\alpha_g = 0.1$.
    }
    \label{fig: QNM vs FD}
\end{figure}
\begin{figure}[ht]
    \subfloat[$\alpha_g = 0.001$]{\includegraphics[width=0.33\linewidth]{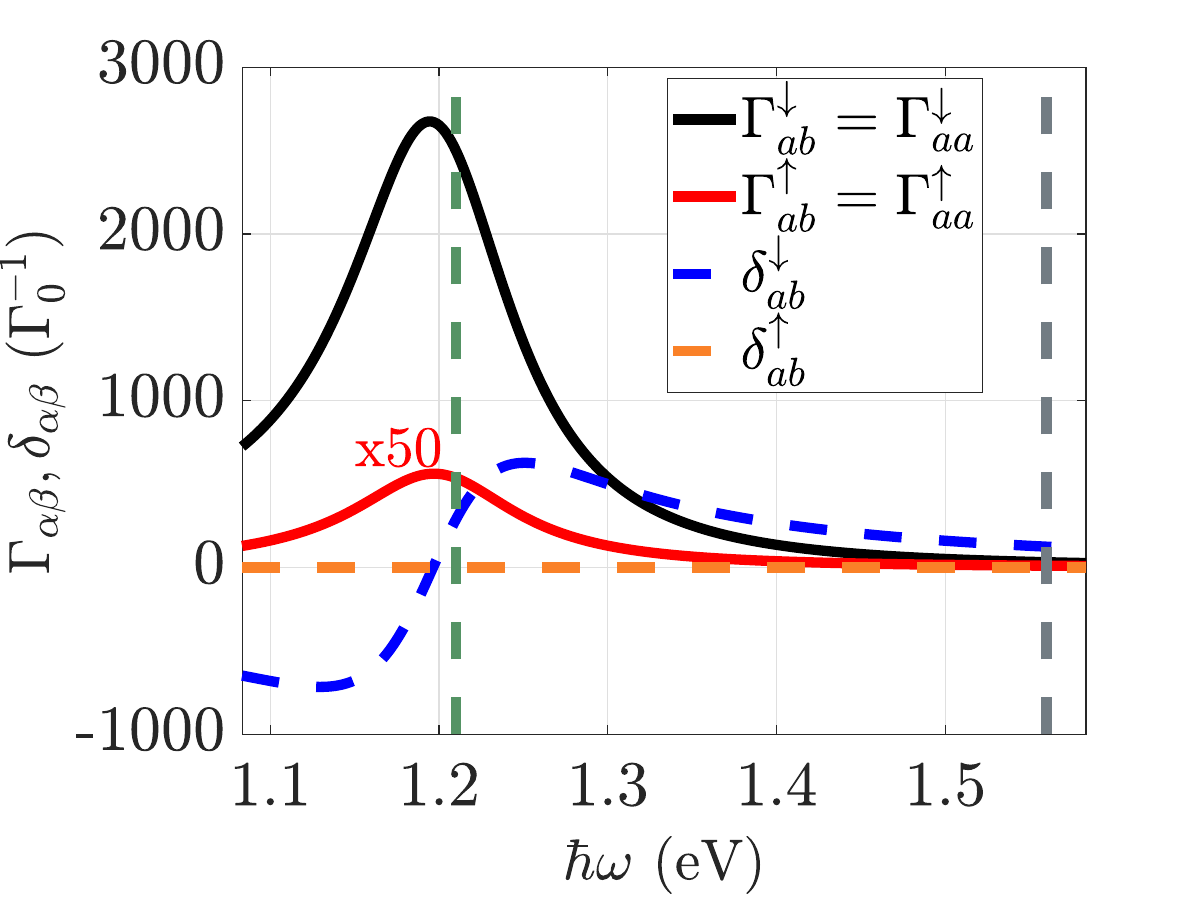}}
    \subfloat[$\alpha_g = 0.1$]{\includegraphics[width=0.33\linewidth]{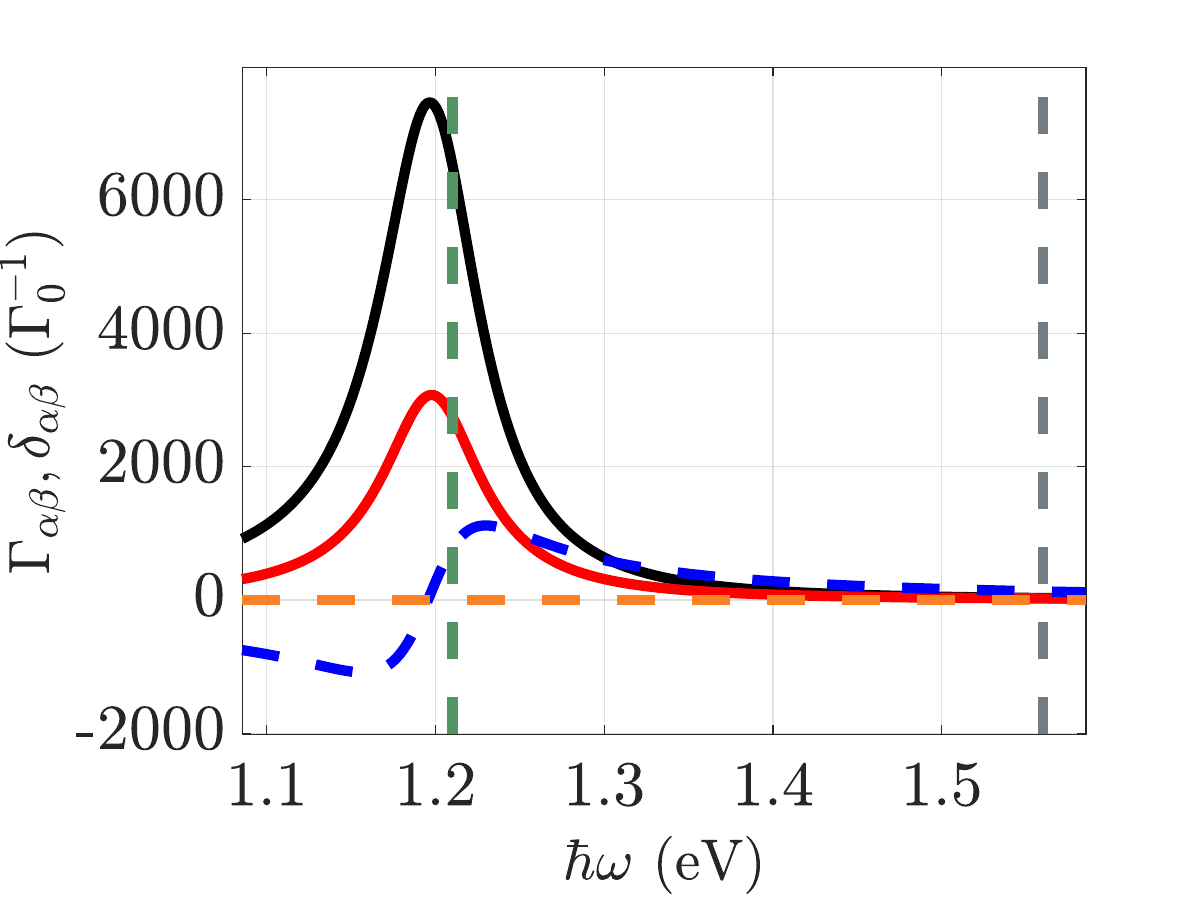}}
    \subfloat[$\alpha_g = 0.22$]{\includegraphics[width=0.33\linewidth]{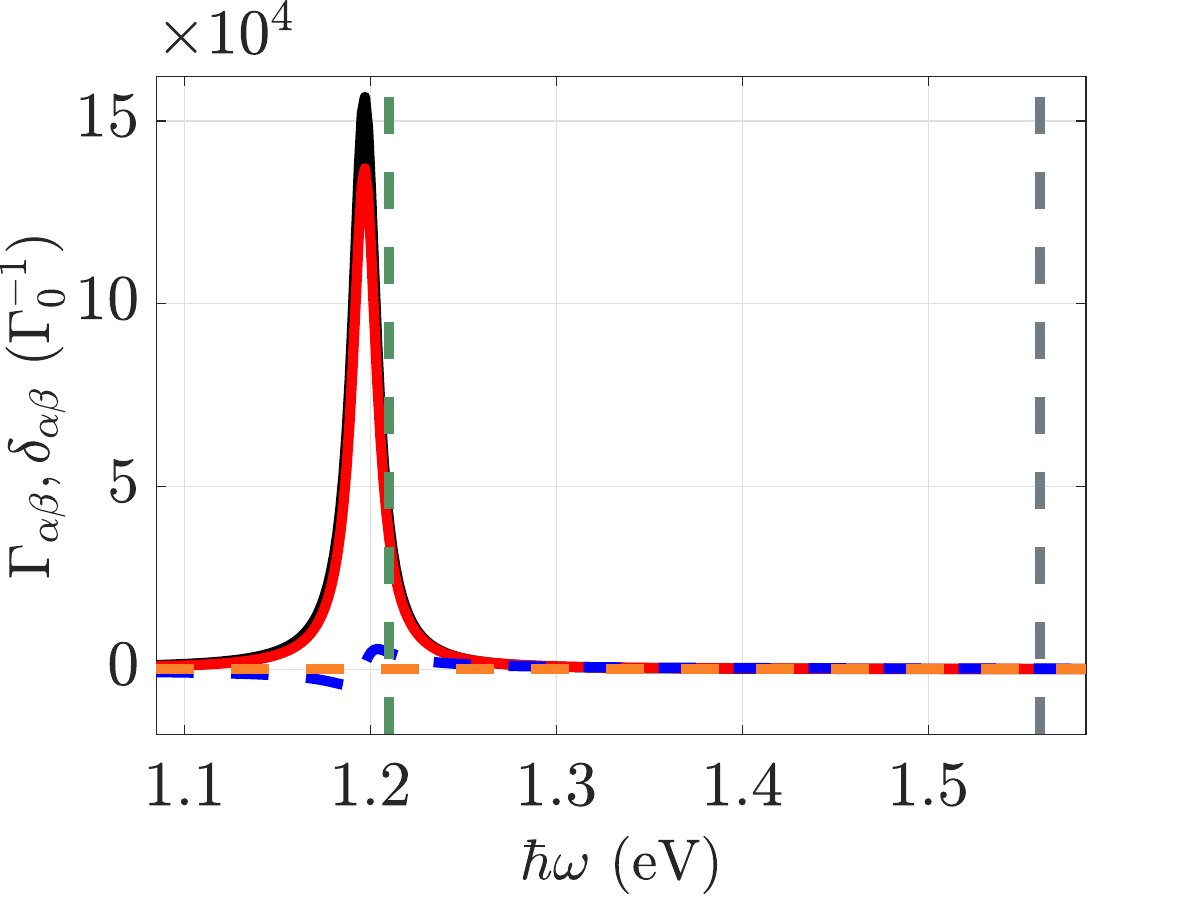}} \\
    \subfloat[$\alpha_g = 0.001$]{\includegraphics[width=0.33\linewidth]{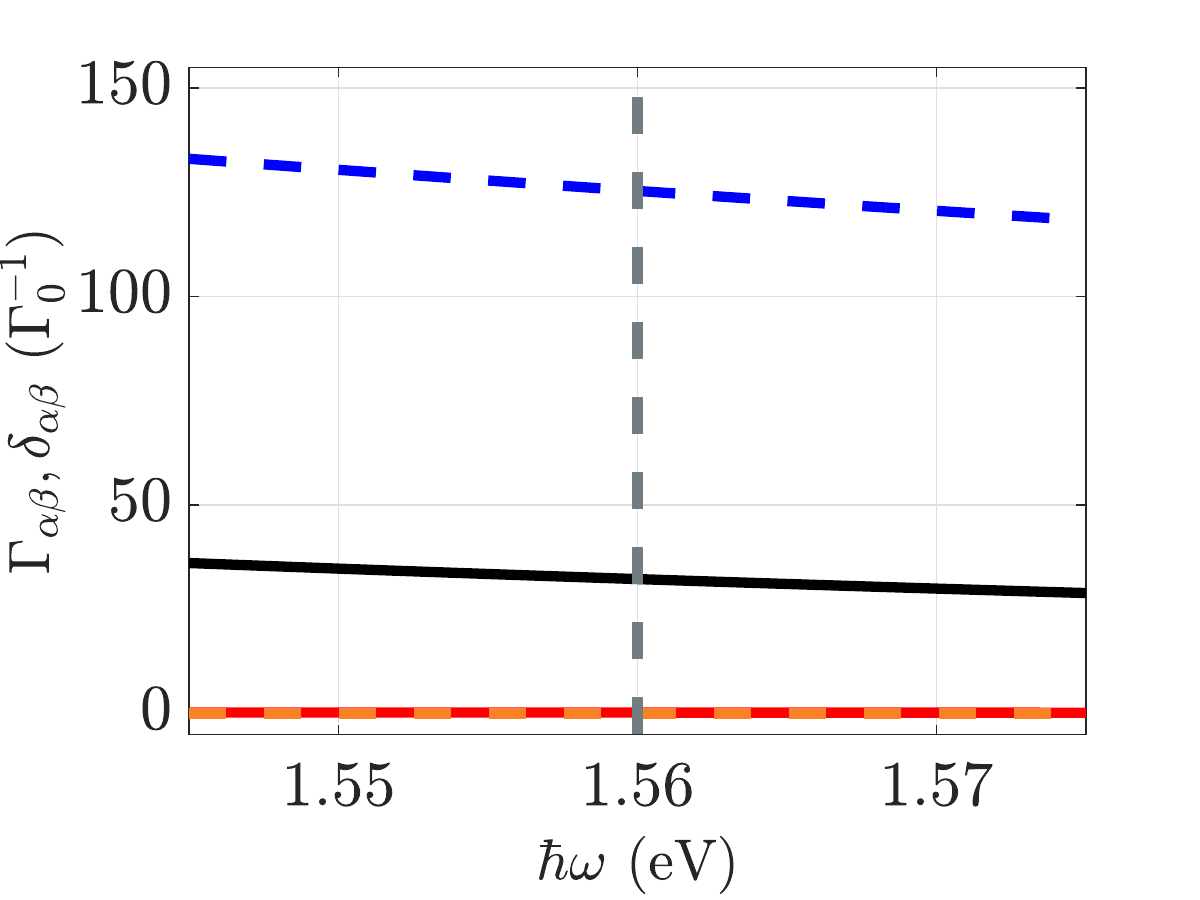}}
    \subfloat[$\alpha_g = 0.1$]{\includegraphics[width=0.33\linewidth]{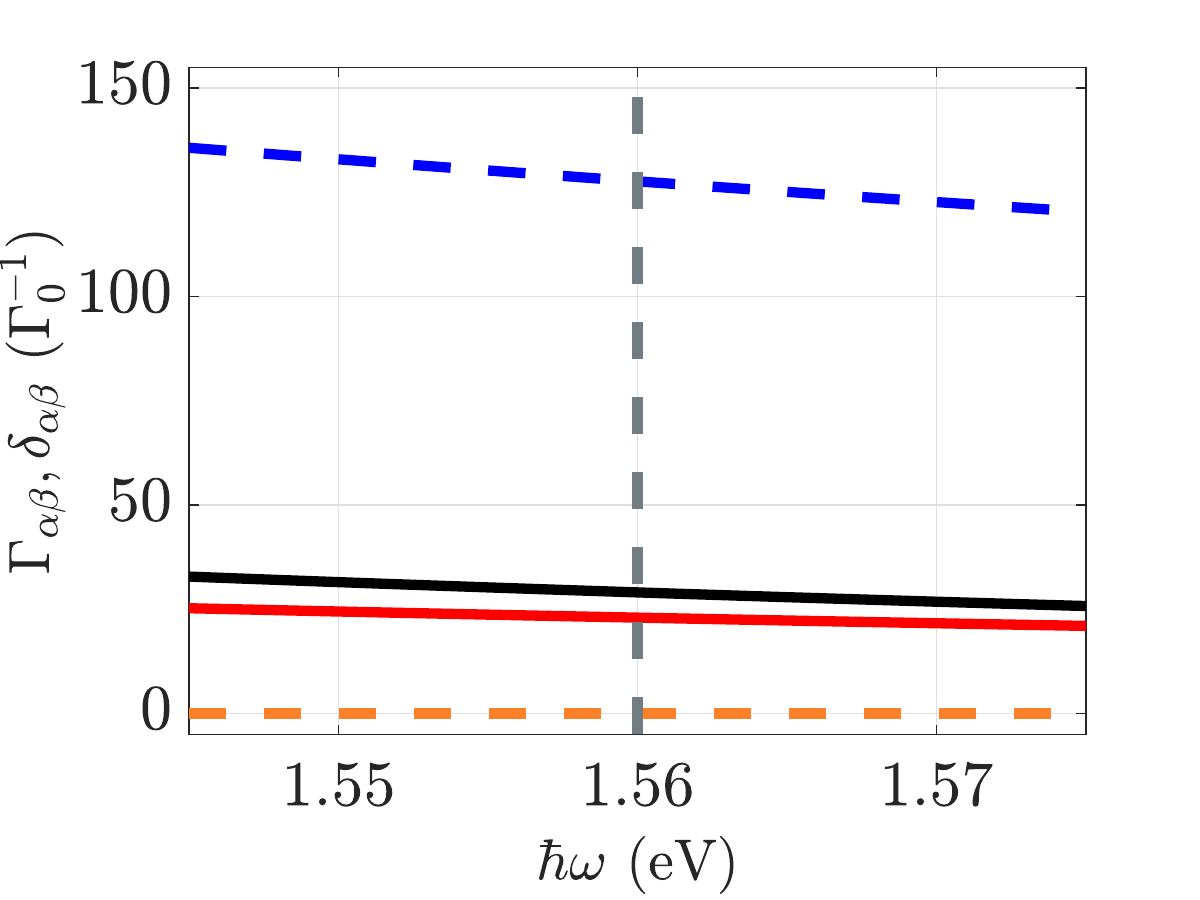}}
    \subfloat[$\alpha_g = 0.22$]{\includegraphics[width=0.33\linewidth]{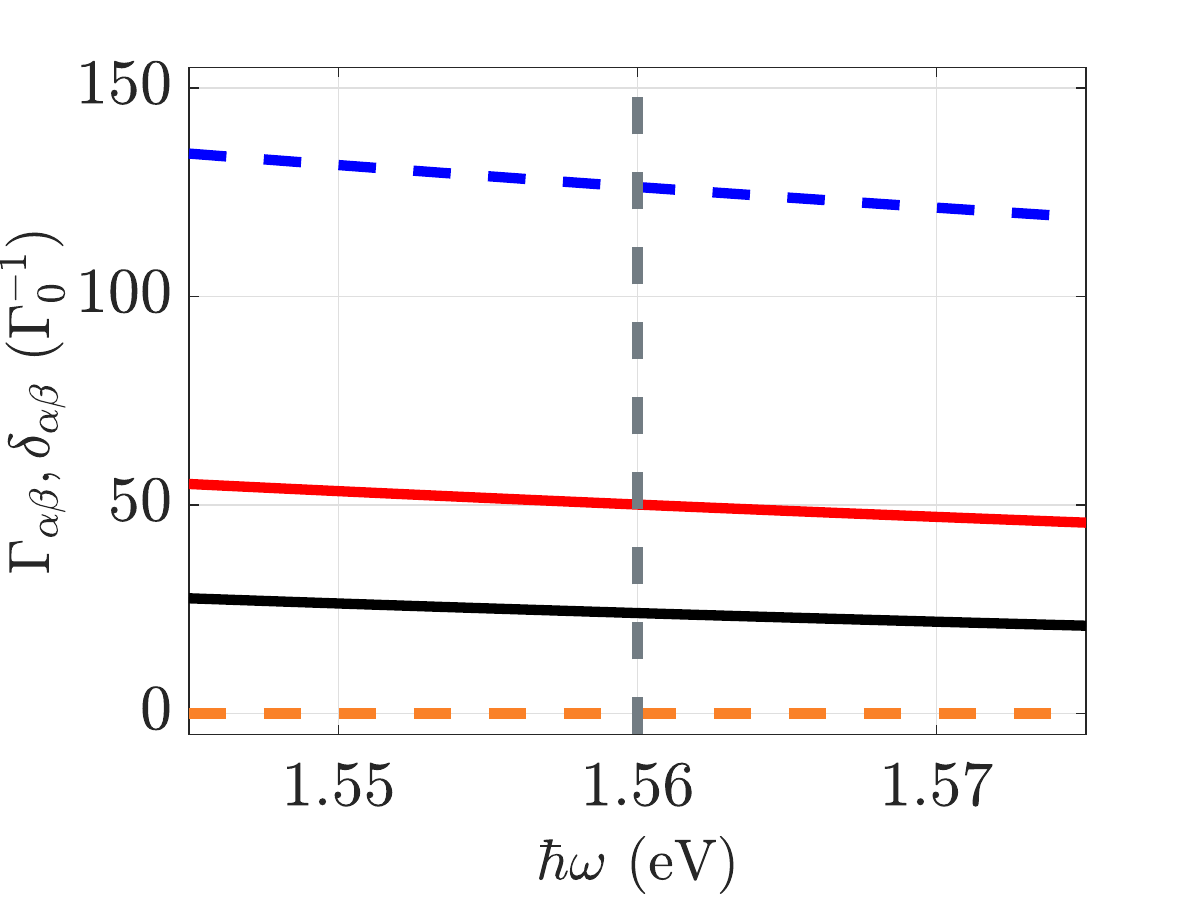}}
    \caption{\textbf{Various emitter rates needed for the master equation, with different values of gain.}
(a) Gain-compensated coherent and incoherent decay rates for one of two (identical) coupled quantum emitters.
For (a),  there is a very small amount of gain, $\alpha_g = 0.001$, but this allows one to reach steady-state by being non zero. The solid black line is the total lossy incoherent decay rate (SE rate), calculated by Eq.~\eqref{eq: gamma_12_tot}; the red solid line is the additional gain contribution to the incoherent decay rate (which has been multiplied by a factor of 50 for clarity), calculated with Eq.~\eqref{eq: gamma_11_gain}; the dashed blue line is the total lossy coherent inter-emitter decay rate, calculated with Eq.~\eqref{eq: delta_tot}; and finally, the dashed orange line is the gain contribution to the inter-emitter coherent decay rate, given by Eq.~\eqref{eq: delta_12_gain}. The dashed vertical lines at $\hbar \omega = 1.21, ~1.56$~eV represent the frequencies at which we will be considering later. (b) Decay rates for one of two  coupled emitters, with the same spatial configuration and equations as in subplot (a), but now with  $\alpha_g = 0.1$. (c) Decay rates for one of two  coupled emitters, with the same spatial configuration and equations as in subplot (a), but now with  $\alpha_g = 0.22$. (d) A zoom in of subplot (a) around the $\hbar \omega = 1.56~$eV region. (e)  A zoom in of subplot (b) around $\hbar \omega = 1.56~$eV. (f) A zoom in of subplot (c) around $\hbar \omega = 1.56~$eV.
   }
    \label{fig: decay rates}
\end{figure}

To demonstrate the validity of the QNM approach to calculating the coupling rates of the two quantum emitters, we show an example of the \textit{total} coupling and decay rates for two coupled quantum emitters in the cavity structure with gain, comparing the solutions calculated with the analytical QNM method as described in Sec.~\ref{sec: theory QNM} to a numerical full dipole method (Sec.~\ref{sec: full_dipole}), as well as a spatial map of the profile of the single QNM which dominates in this system. For this example, we use $\alpha_g = 0.1$. In Fig.~\ref{fig: QNM vs FD} (a), we first show $\delta_{ab}^{\downarrow}$ and $\Gamma_{ab}^{\downarrow}$ (which is equivalent to $\Gamma_{aa}^{\downarrow}$ in this case because the emitters are located at spatially symmetric locations with equivalent dipole moments) as a function of frequency for both the QNM method and a numerical check. Notably, no fitting parameters are used. First, we observe that there is clearly only a single mode dominating, and additionally we observe excellent agreement between the two methods for $\Gamma_{ab}^{\downarrow}$. 
This is extremely convenient as computing the full Green function at various spatial locations is computationally intensive, and such calculations do not clearly highlight the underlying cavity-mode physics. We also obtain excellent agreement for $\Gamma_{ab}^{\uparrow}$, highlighting how crucial this additional gain term is, and how it is accurately described within the QNM method. 

While we do not observe as good an agreement for $\delta_{ab}^{\downarrow}$, this is due to non-modal contributions from the metal nanorods, as a result of Coulomb screening. This contribution is contained in $\mathbf{G}_{\rm others}$, which is basically a screened version of $\mathbf{G}_{\rm B}$, which has a longitudinal and transverse component~\cite{Scheel}. It has been shown that this additional screening term plays a bigger role in the real part of the Green function compared to the imaginary part~\cite{PhysRevA.108.043502,PhysRevA.105.023702}, but a quasi-static Green function can work well if required~\cite{ge_quasinormal_2014}.
Below we will simply include the single QNM component, since any additional contributions to $\delta_{ab}$ simply cause
a small frequency shift in a Markov approximation.
Moreover, since we have concluded that the QNM method for calculating the photonic Green function is accurate, we will no longer include the subscript QNM when considering the various rates, as all calculations from now on will be with the QNM method.

In the presence of gain,
we also show a 2D spatial map of the QNM mode in Fig.~\ref{fig: QNM vs FD} (b), where one can observe how the field strength changes with respect to position around the resonator. Although not shown, this spatial map is virtually identical to the case with no gain~\cite{VanDrunen:24}.

In Fig.~\ref{fig: decay rates}, we show the coherent and incoherent decay rates for three different gain values, namely $\alpha_g = 0.001$, $\alpha_g = 0.1$, and $\alpha_g = 0.22$, where we highlight the region of interest that we will be focusing on when considering the temporal dynamics and emission spectra. In particular, we consider two regions of interest;  $\hbar \omega = 1.56$~eV, which is away from the QNM resonance which peaks around $\hbar \omega = 1.2$~eV, and $\hbar \omega = 1.21$~eV, which lies much closer to the QNM resonance. Looking at the magnitudes of the various rates at these two different locations, one can see stark differences between the enhancements due to gain. Additionally in Fig.~\ref{fig: decay rates} we show a zoom-in around the frequency of $\hbar \omega = 1.56~$eV to highlight the magnitudes of the decay rates at that frequency, as they are not impacted by the gain enhancement as severely as we observe around the QNM resonance. We note here that we do not consider the rate for $\delta^{\uparrow}_{aa}$ and similarly for $\delta^{\uparrow}_{bb}$, as by referring to Eq.~\eqref{eq: K}, there is no imaginary part of this function when the two position variables are the same, thus this contribution is effectively zero for our purposes. Similarly, for spatially symmetric locations, we observe that $\delta_{ab}^{\uparrow}$ is zero for all cases, as Eq.~\eqref{eq: K} would again have no imaginary component. 


\subsection{Near resonant emitter dynamics}
\label{sec: near res dynamics}

\begin{figure}[th]
    \subfloat{\includegraphics[width=0.325\linewidth]{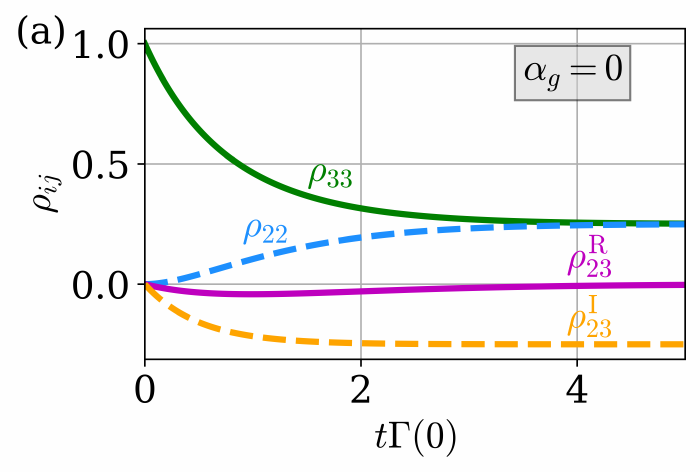}}
    \subfloat{\includegraphics[width=0.34\linewidth]{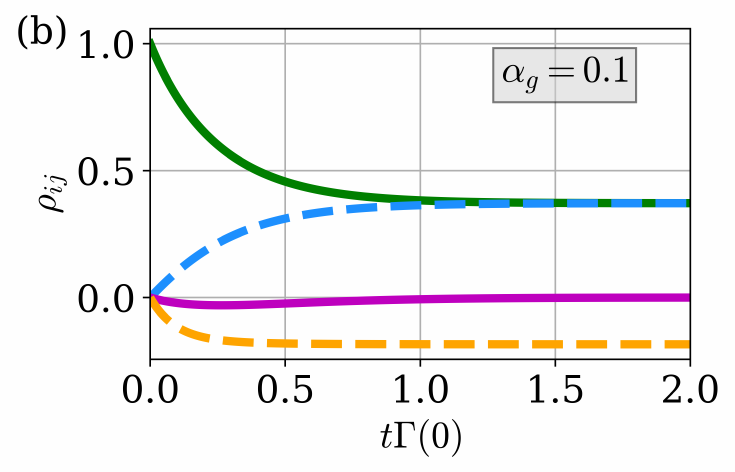}}
    \subfloat{\includegraphics[width=0.325\linewidth]{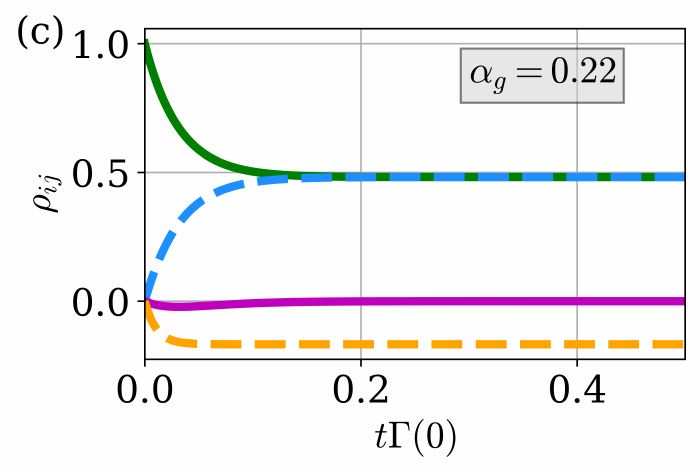}}\\
    \subfloat{\includegraphics[width=0.325\linewidth]{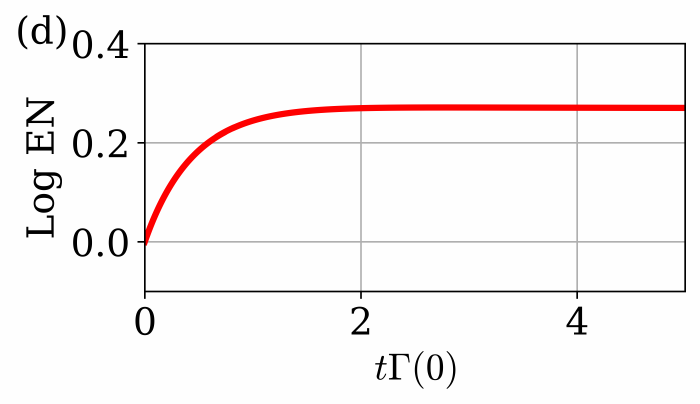}}
    \subfloat{\includegraphics[width=0.34\linewidth]{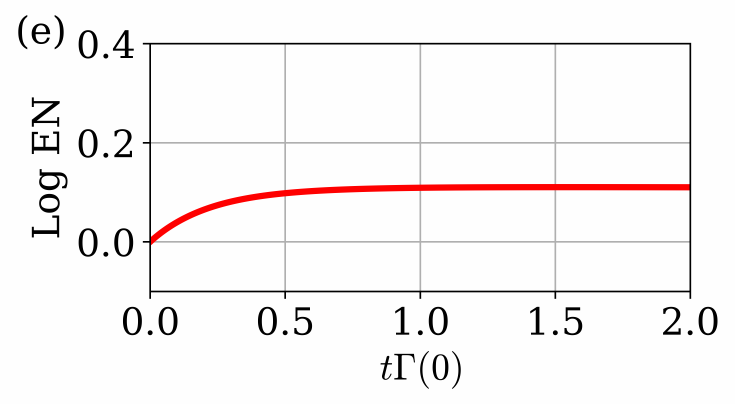}}
    \subfloat{\includegraphics[width=0.325\linewidth]{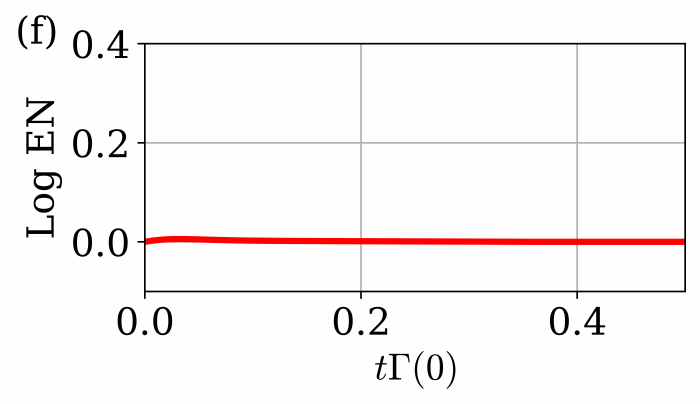}} 
    \caption{\textbf{Near resonant population dynamics.} (a) Population dynamics of the two coupled dipoles at $\hbar \omega = 1.21~$eV.
    Initially, the system starts with atom $a$ being excited and atom $b$ being in the ground state. In this case, there is no gain present in the system, and we are working within the bare state basis. The green and blue curves correspond to the atom populations using the analytical expressions found with the optical Bloch equations, as given in Eqs.~\eqref{eq: atom a pop full} and \eqref{eq: atom b pop full}, respectively, under the WEA. The solid magenta curve and the dashed orange curve represent the real and imaginary parts of the coherence, respectively, as given by Eq.~\eqref{eq: bare coherence tot}. The time units are normalized to
    $\Gamma(\alpha_g=0) = \Gamma^{\downarrow}_{aa}$, for the case with no gain, where $\Gamma(0) = 2473.84\Gamma_0$, where $\Gamma_0$ is the background SE rate, given by Eq.~\eqref{eq: gamma_0}, at $\hbar \omega = 1.21~$eV. In all cases shown, we include a small dephasing term, such that $\Gamma' = 0.001\Gamma(0)$.
(b) The population dynamics of the two coupled dipoles at $\hbar \omega = 1.21~$eV, with the same spatial configuration and initial condition as in subplot (a), but this time with some gain in the system, where $\alpha_g = 0.1$.
(c) The population dynamics of the two coupled dipoles at $\hbar \omega = 1.21~$eV, with the same spatial configuration and initial condition as in subplot (a), but this time with a larger amount of gain in the system, namely $\alpha_g = 0.22$. 
(d) Entanglement dynamics using results of subplot (a), where the logarithmic EN is calculated by Eq.~\eqref{eq: log EN}.
(e) Entanglement dynamics using results of subplot (b).
(f) Entanglement dynamics using results of subplot (c).
}
    \label{fig: pops and EN with single excitation 1.21eV renorm}
\end{figure}

In this section, we examine the population and entanglement dynamics in the bare state basis of the coupled quantum emitters within the plasmonic resonator structure, with the symmetrical condition such that $\Gamma^{\downarrow(\uparrow)}_{aa} = \Gamma^{\downarrow(\uparrow)}_{ab}$. Here, we specifically look at the near resonant dynamics of the emitters, namely at $\hbar \omega = 1.21~$eV, which is denoted by the vertical green dotted line in Fig.~\ref{fig: decay rates}. In Fig.~\ref{fig: pops and EN with single excitation 1.21eV renorm}, we show the emitter populations and coherence dynamics for three different gain cases: $\alpha_g = 0$ (no gain), $\alpha_g = 0.1$, and $\alpha_g = 0.22$, where the system initially starts with one emitter excited and the other in the ground state. To calculate these temporal dynamics, we use the optical Bloch equations in the bare basis, as given by Eqs.~\eqref{eq: rho_22}-\eqref{eq: bare coherence tot}, where the decay rates used in the equations are graphically demonstrated in Fig.~\ref{fig: decay rates}. 

First, in the case with no gain, the emitters reach a steady-state population of 0.25. As we add a moderate amount of gain ($\alpha_g = 0.1$), the steady-state populations increase to 0.37, and we see a further increase to 0.48 when we consider a larger amount of gain ($\alpha_g = 0.22$). Notably, there is little coherent interaction evident in the dynamics - each emitter reaches its steady-state value without oscillations in the evolution of the density matrix elements. This is also shown in the real part of the coherence, $\rho_{23}^{\rm R}$, which is shown in magenta, and is effectively zero for all times when there is gain. Furthermore, as more gain is added, steady-state is reached faster, as the inclusion of gain results in greater decay rates, as shown in Fig.~\ref{fig: decay rates}.

Additionally in Fig.~\ref{fig: pops and EN with single excitation 1.21eV renorm}(d-f), we show the entanglement dynamics of the coupled emitter system via logarithmic EN, calculated by Eq.~\eqref{eq: log EN}. Without gain, long-lived entanglement is possible to realize, however as gain is added to this system, the amount increases in the incoherent decay rates compared to the coherent decay rates which mitigates the coherence has the same effect on entanglement. When a large amount of gain is included ($\alpha_g = 0.22$), the logarithmic EN is zero over time, indicating no entanglement between the two emitters. These results demonstrate that while adding gain can drastically increase the incoherent decay rates in the region of the QNM resonance, the coherent interactions are weakened as a result.


\subsection{Off-resonant emitter dynamics}
\label{sec: off res dynamics}

\begin{figure}[ht]
    \subfloat{\includegraphics[width=0.33\linewidth]{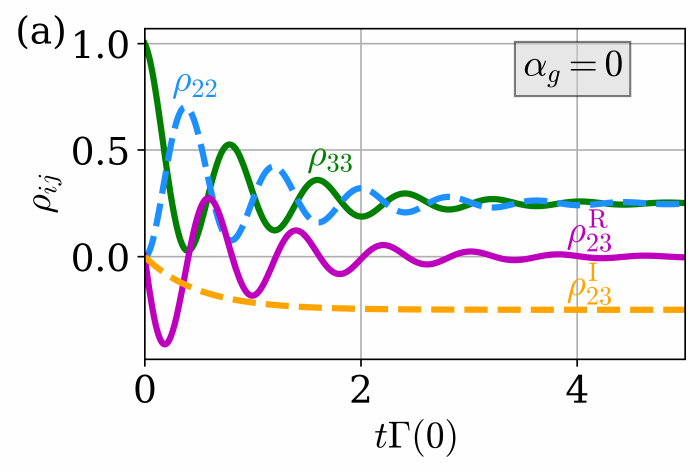}}
    \subfloat{\includegraphics[width=0.33\linewidth]{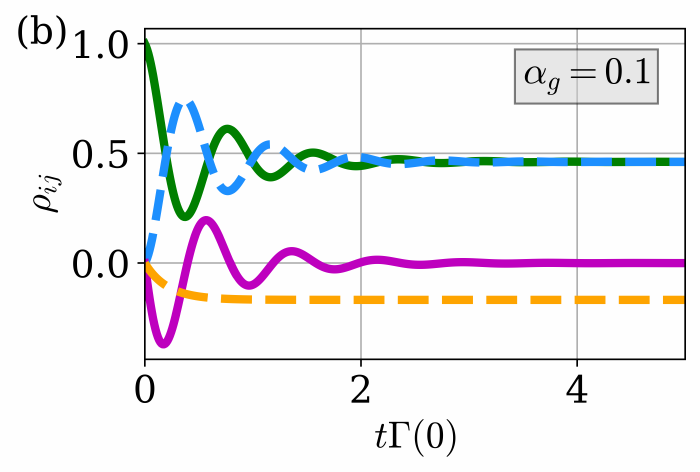}}
    \subfloat{\includegraphics[width=0.33\linewidth]{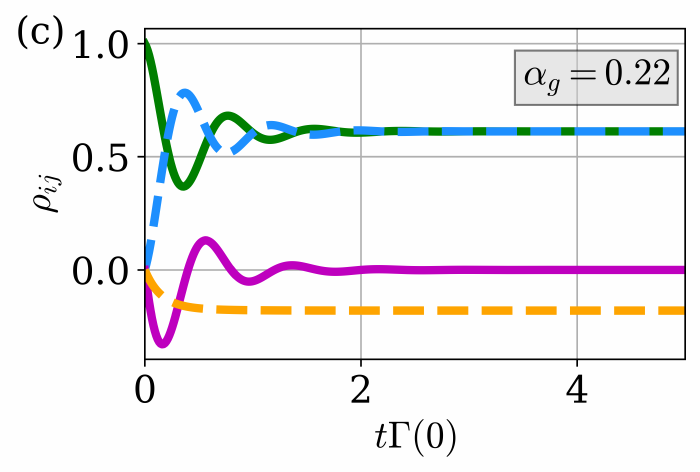}}\\
    \subfloat{\includegraphics[width=0.33\linewidth]{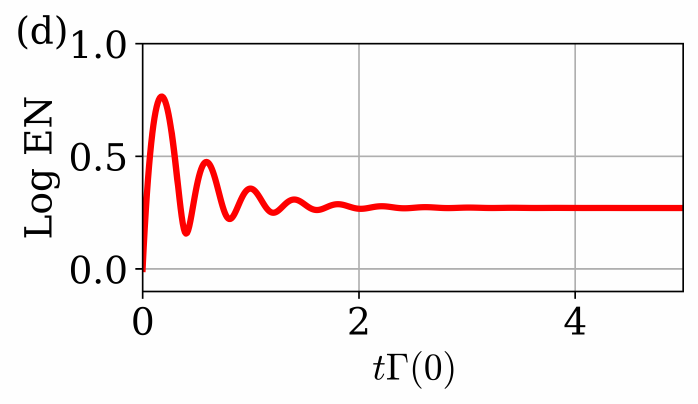}}
    \subfloat{\includegraphics[width=0.33\linewidth]{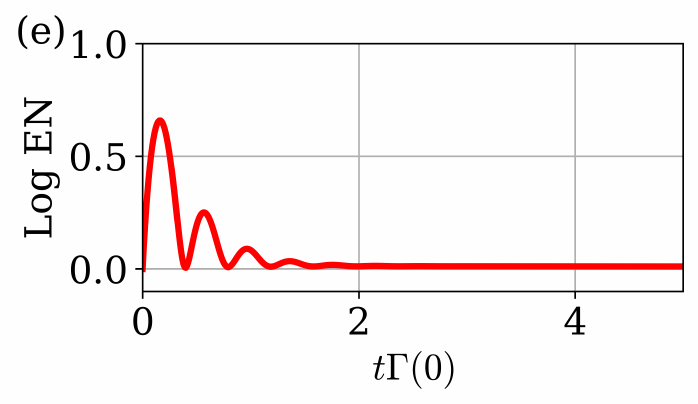}}
    \subfloat{\includegraphics[width=0.33\linewidth]{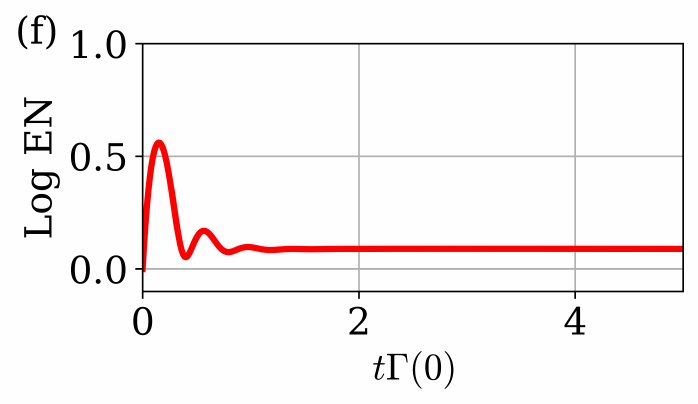}} 
    \caption{\textbf{Off-resonant 
    population dynamics.} 
    (a) Population dynamics of the two coupled dipoles at $\hbar \omega = 1.56~$eV.
    Initially, the system starts with atom a being excited and atom b being in the ground state. In this case, there is no gain present in the system, and we are working within the bare state basis. The green and blue curves correspond to the atom populations using the analytical expressions found with the Optical Bloch equations, as given in Eqs.~\eqref{eq: atom a pop full} and \eqref{eq: atom b pop full}, respectively, under the WEA. The solid magenta curve and the dashed orange curve represent the real and imaginary parts of the coherence, respectively, as given by Eq.~\eqref{eq: bare coherence tot}. On the x-axis, $\Gamma(0)$ refers to the $\Gamma^{\downarrow}_{aa}$ rate with no gain, where $\Gamma^{\downarrow}_{aa} = 32.1\Gamma_0$, where $\Gamma_0$ is the background SE rate, given by Eq.~\eqref{eq: gamma_0}. In all cases shown, we include a small dephasing term, such that $\Gamma' = 0.001\Gamma(0)$.
(b) The population dynamics of the two coupled dipoles at $\hbar \omega = 1.56~$eV, with the same spatial configuration and initial condition as in subplot (a), but this time with some gain in the system, where $\alpha_g = 0.1$. 
(c) The population dynamics of the two coupled dipoles at $\hbar \omega = 1.56~$eV, with the same spatial configuration and initial condition as in subplot (a), but this time with a larger amount of gain in the system, namely $\alpha_g = 0.22$. 
(d) Entanglement dynamics with respect to the population dynamics as shown in subplot (a), where the logarithmic EN is calculated by Eq.~\eqref{eq: log EN}.
(e) Entanglement dynamics with respect to the population dynamics as shown in subplot (b).
(f) Entanglement dynamics with respect to the population dynamics as shown in subplot (c).
}
    \label{fig: pops and EN with single excitation 1.56eV renorm}
\end{figure}

Similar to the previous section, we again show the population and entanglement dynamics of two coupled quantum emitters in a plasmonic resonator in the bare state basis with the same symmetrical condition that we used previously, but this time we consider the off-resonant dynamics. Referring back to Fig.~\ref{fig: decay rates}, we are now considering the dynamics at $\hbar \omega = 1.56~$eV, which is denoted by the vertical gray dotted line, and is far from the QNM resonance at $\hbar \omega = 1.2~$eV. In Fig.~\ref{fig: pops and EN with single excitation 1.56eV renorm}, we demonstrate the population dynamics with and without gain, where we again consider the three following cases: $\alpha_g = 0$, $\alpha_g = 0.1$, $\alpha_g = 0.22$. The system initially starts with a single excitation, one emitter starts in the excited state, and the other starts in the ground, and we use the optical Bloch equations as given by Eq.~\eqref{eq: rho_22}-\eqref{eq: bare coherence tot}. 

Beginning with the case with no gain, the emitter populations $\rho_{22}$ and $\rho_{33}$ reach a steady-state value of 0.25. By adding gain, we obtain steady-state populations of 0.46 when $\alpha_g = 0.1$, and further to 0.61 when $\alpha_g = 0.22$. Additionally, we observe clear oscillations in $\rho_{22}$ and $\rho_{33}$ as well as in the real part of the coherence ($\rho_{23}^{\rm R}$), which showcases that the coherent interactions are still being maintain with gain when looking off-resonance. However, as $\alpha_g$ is increased, steady-state is reached faster, which agrees with what was found in the near resonance dynamics in Fig.~\ref{fig: pops and EN with single excitation 1.21eV renorm}. 

Moreover, we also show the entanglement dynamics in Fig.~\ref{fig: pops and EN with single excitation 1.56eV renorm} by plotting the logarithmic EN from Eq.~\eqref{eq: log EN} over time. Similarly to the near resonance case, when there is no gain in the system, we also observe the best long-lived entanglement. However, when $\alpha_g = 0.22$, we do obtain a non-zero steady-state logarithmic EN, showcasing that long-lived entanglement can be realized with gain when far from the QNM resonance. Additionally, we observe oscillations in the logarithmic EN, which makes sense now that coherent decay rates outweigh the incoherent rates.
For the remainder of this work, we will focus solely at the frequency of $\hbar \omega = 1.56~$eV to observe the more interesting effects with the two coupled quantum emitters.



\subsection{Off-resonant emitter dynamics in a dressed state basis }
\label{sec: dressed state basis}

\begin{figure}[ht]
    \subfloat[Superradiant initial condition, dressed basis, $\alpha_g = 0$]{\includegraphics[width=0.32\linewidth]{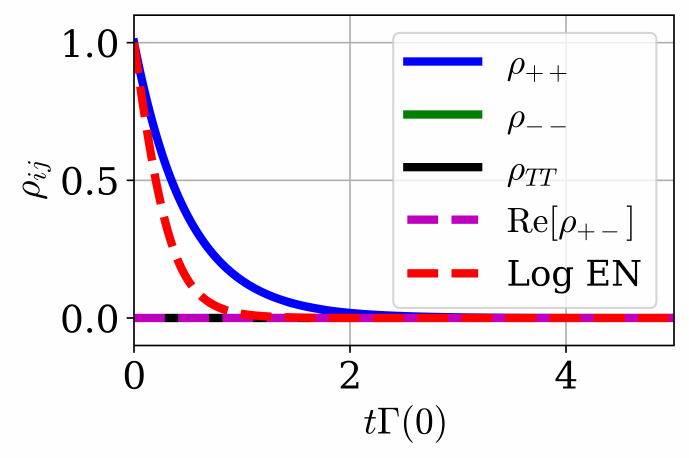}}
    \subfloat[Subradiant initial condition, dressed basis, $\alpha_g = 0$]{\includegraphics[width=0.335\linewidth]{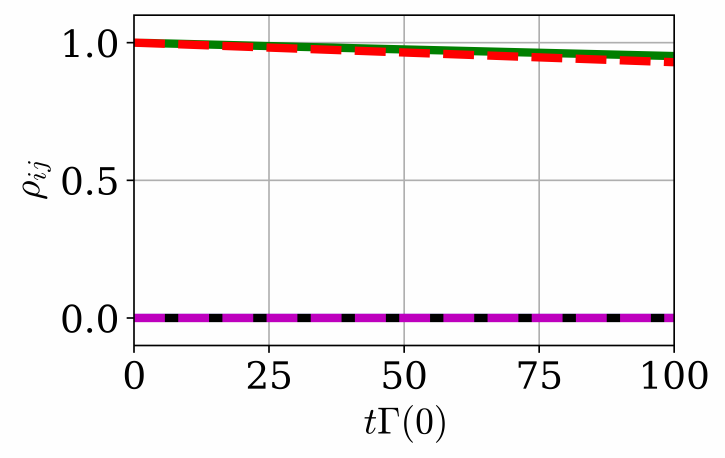}}
    \subfloat[Ground initial condition, dressed basis, $\alpha_g = 0$]{\includegraphics[width=0.32\linewidth]{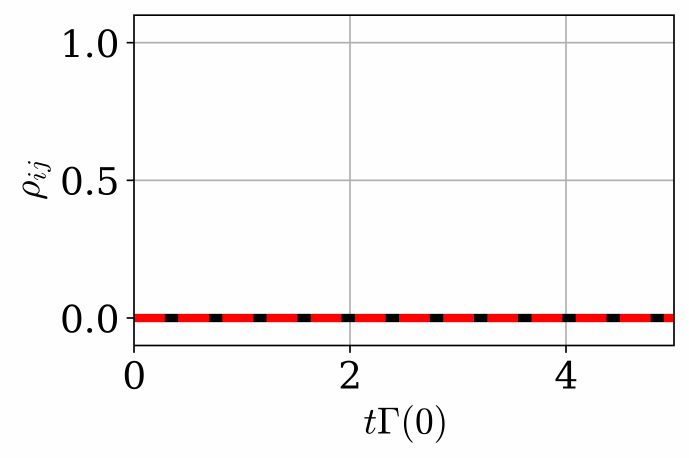}}\\
    \subfloat[Superradiant initial condition, dressed basis, $\alpha_g = 0.22$]{\includegraphics[width=0.32\linewidth]{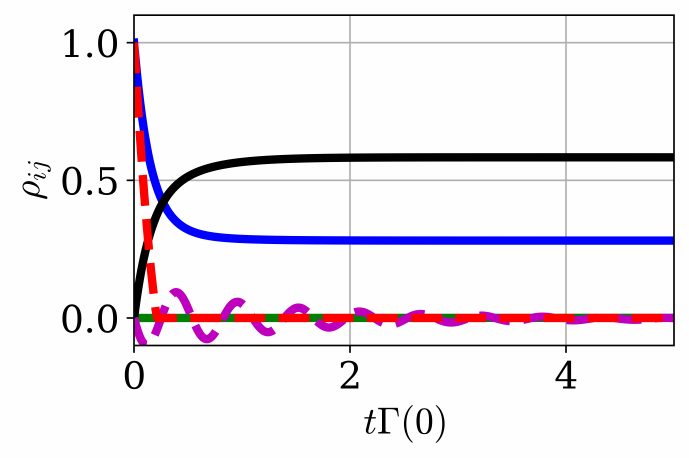}}
    \subfloat[Subradiant initial condition, dressed basis, $\alpha_g = 0.22$]{\includegraphics[width=0.335\linewidth]{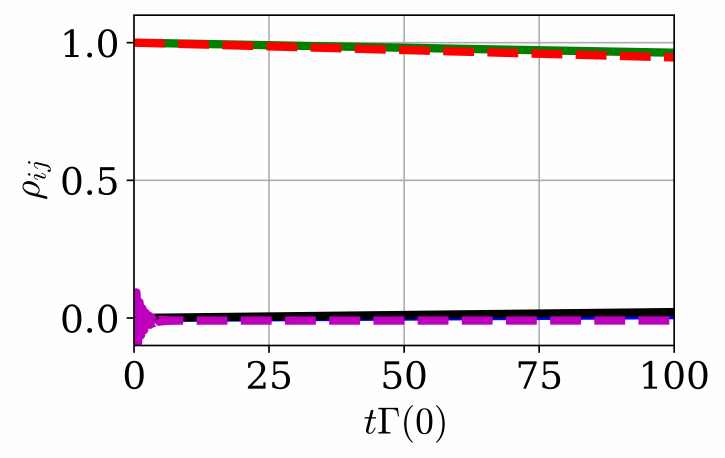}}
    \subfloat[Ground initial condition, dressed basis, $\alpha_g = 0.22$]{\includegraphics[width=0.32\linewidth]{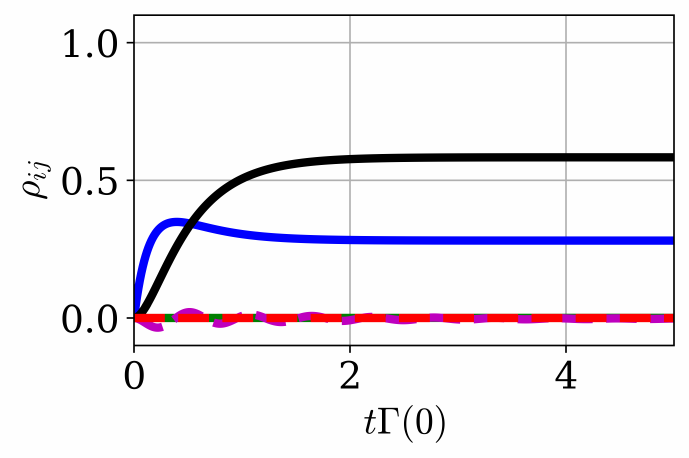}} 
    \caption{\textbf{Off-resonant dynamics in a  dressed-state basis.} (a) Population dynamics of the two coupled dipoles at $\hbar \omega = 1.56~$eV.
    Initially, the system starts in the superradiant state, $\ket{+} = \frac{1}{\sqrt{2}}(\ket{e_1g_2}+\ket{g_1e_2})$. In this case, there is no gain in the system, where $\alpha_g = 0$, and we are working within the dressed state basis. The dark blue and green lines [note green lines remain at zero in panels (a,c,d,f)] represent the populations of the superradiant and subradiant dressed states, respectively, calculated with Eqs.~\eqref{eq: rho++} and \eqref{eq: rho--}. The black line represents the population of the two-quanta state, as calculated by Eq.~\eqref{eq: rho_TT}. The dotted magenta line corresponds to the real coherence between the superradiant and subradiant state, as given by Eq.~\eqref{eq: rho+-}. The dashed red line shows the logarithmic EN, calculated with Eq.~\eqref{eq: log EN}. On the x-axis, $\Gamma(0)$ refers to the $\Gamma^{\downarrow}_{aa}$ rate with no gain, where $\Gamma^{\downarrow}_{aa} = 32.1\Gamma_0$, where $\Gamma_0$ is the background SE rate, calculated with Eq.~\eqref{eq: gamma_0}. In all cases shown, we include a small dephasing term, such that $\Gamma' = 0.001\Gamma(0)$.
    Note the curve for $\rho_{--}$ remains as zero.
    (b) The population dynamics of the two coupled dipoles at $\hbar \omega = 1.56~$eV, with the same spatial configuration as in subplot (a), but this time with the system initially starting the subradiant initial condition, $\ket{-} = \frac{1}{\sqrt{2}}(\ket{e_1g_2}-\ket{g_1e_2})$. 
(c) The population dynamics of the two coupled dipoles at $\hbar \omega = 1.56~$eV, with the same spatial configuration as in subplot (a), but this time with the initial state starting in the ground state. 
(d) The population dynamics of the two coupled dipoles at $\hbar \omega = 1.56~$eV, with the same spatial configuration and initial condition as in subplot (a), but with gain present in the system, where $\alpha_g = 0.22$. 
(e) The population dynamics of the two coupled dipoles at $\hbar \omega = 1.56~$eV, for the same spatial configuration and initial condition as in subplot (b), but with gain present in the system, where $\alpha_g = 0.22$.
(f) The population dynamics of the two coupled dipoles at $\hbar \omega = 1.56~$eV, with the same spatial configuration and initial condition as in subplot (c), but with gain present in the system, where $\alpha_g = 0.22$.
}
    \label{fig: dressed pops with gain norm}
\end{figure}

Next, we investigate the temporal dynamics in the dressed state basis, to observe how gain impacts the superradiant and subradiant states, as shown in Fig.~\ref{fig: dressed pops with gain norm}. Without gain, first we note that the two-quanta state, $\rho_{TT}$, which is represented by the black line, is never populated, hence why this state can be decoupled from the basis when there is no gain or pumping mechanism present. Additionally, all states eventually decay to zero, though the subradiant state would take a long time to decay since $\Gamma^{\downarrow(\uparrow)}_{aa} = \Gamma^{\downarrow(\uparrow)}_{ab} \gg \Gamma^{\prime}$. When looking at the case with gain, where $\alpha_g = 0.22$, there are now non-zero steady-state populations. Notably, the two-quanta state, which was previously decoupled from the basis, now reaches a steady-state population of 0.58 when $\alpha_g = 0.22$. This is due to the gain, which acts as a pump to populate this state, which typically is not populated unless there is a pump present. Furthermore, in the case with no gain, when the system starts in the ground state, it remains in the ground state over time. However, when gain is included, the two-quanta, superradiant, and subradiant states become populated over time (where the population of the subradiant state is directly related to the dephasing decay rate, which is defined as being very small relative to the other incoherent decay rates). This result showcases the advantage of using gain, as we can obtain non-zero steady-state solutions in the dressed basis without having to include additional pump terms due to the gain providing the pumping mechanism.


\subsection{Emitted spectra from the coupled emitter system}
\label{sec: power spectrum}

\begin{figure}[th]
    \subfloat[$\Gamma_{\rm pump}=0.001\Gamma(0)$]{\includegraphics[width=0.32\linewidth]{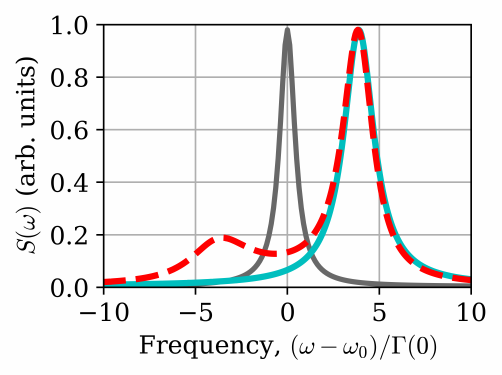}}
    \subfloat[$\Gamma_{\rm pump}=0.01\Gamma(0)$]{\includegraphics[width=0.32\linewidth]{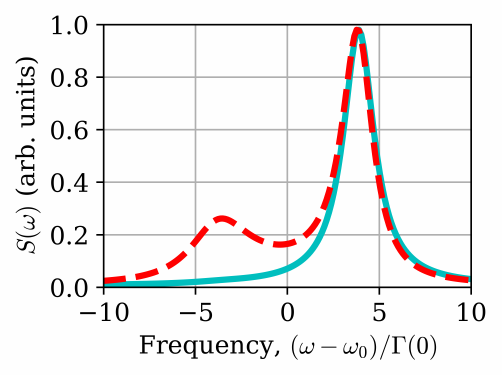}}
    \subfloat[$\Gamma_{\rm pump}=0.1\Gamma(0)$]{\includegraphics[width=0.32\linewidth]{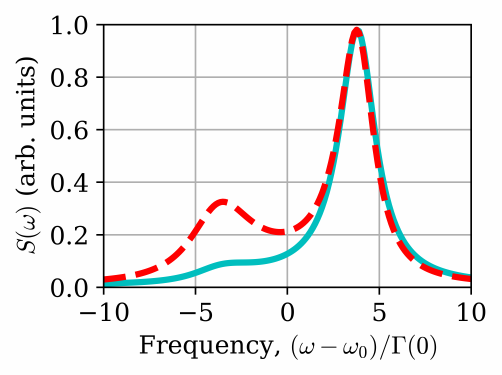}} \\
    \caption{{\bf Emission spectrum with no gain, but with an increasing heuristic (incoherent) pump rate}. (a) Emission spectrum for two coupled emitters.
    There is no gain in the system, but there is a small pure dephasing term, such that $\Gamma^{\prime} = 0.001\Gamma(0)$, and there is also a small pump term, where $\Gamma_{\rm pump} = 0.001\Gamma(0)$. The solid grey line represents the single atom spectrum, as a reference.  The dashed red curve represents the spectrum with no incoherent cross-pumping term (i.e., no $\Gamma_{ab}^{\uparrow}$), where as the solid cyan curve represents the spectrum which does include this term. Both curves are calculated with the steady-state spectral solution, given by Eq.~\eqref{eq:SwSS}. On the $x$-axis, $\Gamma(0)$
    refers to the $\Gamma^{\downarrow}_{aa}$ rate with no gain, where $\Gamma^{\downarrow}_{aa} = 32.1\Gamma_0$, with $\Gamma_0$ being the background SE rate from Eq.~\eqref{eq: gamma_0}.  
    (b) Spectrum for two coupled emitters with the same spatial configuration as in subplot (a), but this time with a larger pump, where $\Gamma_{\rm pump} = 0.01 \Gamma(0)$.  
(c) Spectrum for two coupled emitters with the same spatial configuration as in subplot (b), but this time with a larger pump term again, where $\Gamma_{\rm pump} = 0.1 \Gamma(0)$. 
}
    \label{fig: spectra no gain}
\end{figure}
\begin{figure}[ht]
    \subfloat[$\alpha_g = 0.001$]{\includegraphics[width=0.32\linewidth]{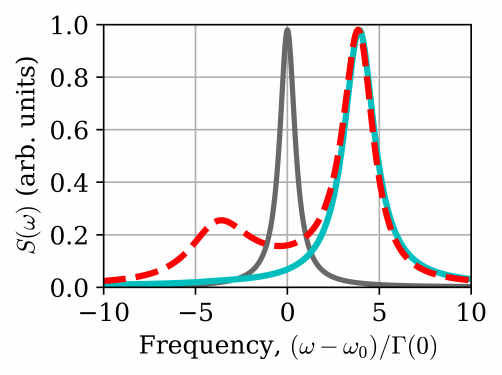}}
    \subfloat[$\alpha_g = 0.1$]{\includegraphics[width=0.32\linewidth]{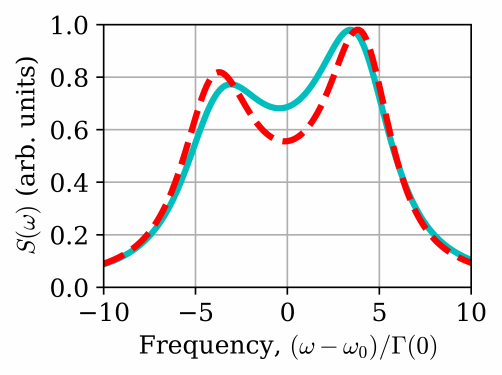}}
    \subfloat[$\alpha_g = 0.22$]{\includegraphics[width=0.32\linewidth]{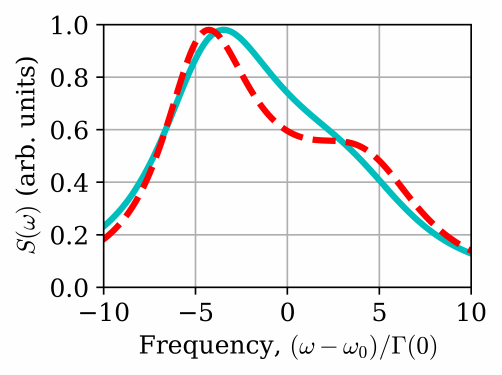}}
    \caption{{\bf Emission spectrum with full gain model with no heuristic gain or pump parameters.} (a) Emission 
    spectrum for two coupled emitters. 
    In this case, there is a very small amount of gain in the system, namely $\alpha_g = 0.001$. The solid gray line represents the single atom spectrum, as a reference. 
    The dashed red curve represents the case where $\Gamma_{ab}^{\uparrow} = 0$, and the solid cyan line represents the spectrum when $\Gamma_{ab}^{\uparrow} = \Gamma_{aa}^{\uparrow}$, where both are calculated with the steady-state spectral equation given by Eq.~\eqref{eq:SwSS}. On the $x$-axis in all cases, $\Gamma(0)$
    refers to the $\Gamma^{\downarrow}_{aa}$ rate with no gain, where $\Gamma^{\downarrow}_{aa} = 32.1\Gamma_0$, where $\Gamma_0$ is the background SE rate from Eq.~\eqref{eq: gamma_0}. In all cases shown, we include a small dephasing term, such that $\Gamma' = 0.001\Gamma(0)$.
    (b) Spectrum for two coupled emitters with the same spatial configuration as in subplot (a), but this time with gain, where $\alpha_g = 0.1$.
(c) Spectrum for two coupled emitters with the same spatial configuration as in subplot (a), but this time with gain, where $\alpha_g = 0.22$. 
}
    \label{fig: spectra with gain}
\end{figure}

\begin{figure}[th]
    \centering
    \includegraphics[width=\columnwidth]{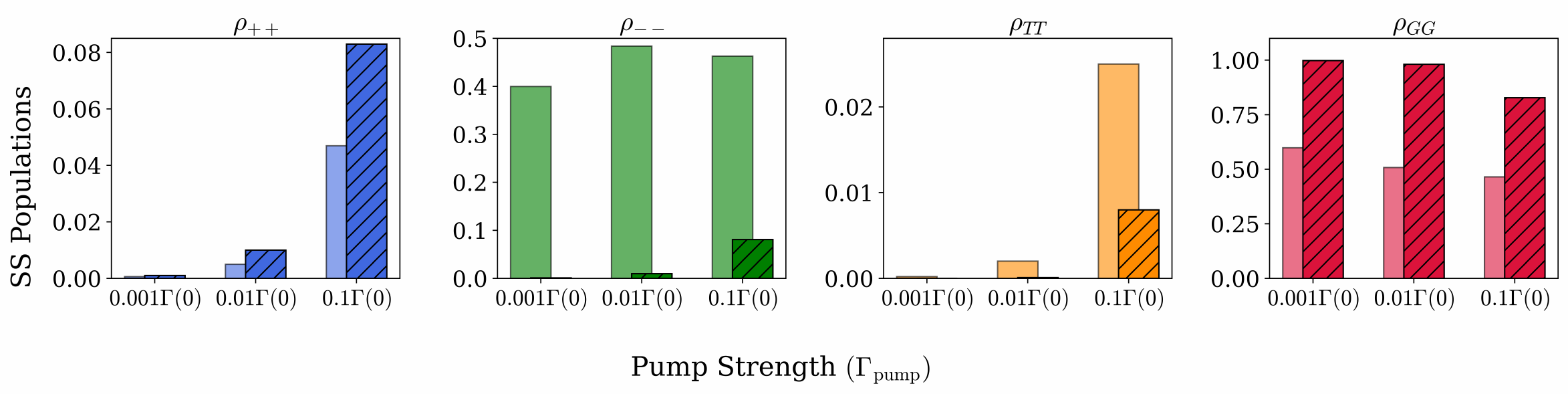}
    \caption{\textbf{Steady state populations with a heuristic pump model.} Steady-state populations in the dressed state basis for the four basis states, where the solid colored bars represent the model without considering incoherent coupling, and the bars with the diagonal black lines represent the model with the incoherent coupling included.
    }
    \label{fig: ss vals pump model}
\end{figure}
\begin{figure}[ht]
    \centering
    \includegraphics[width=\columnwidth]{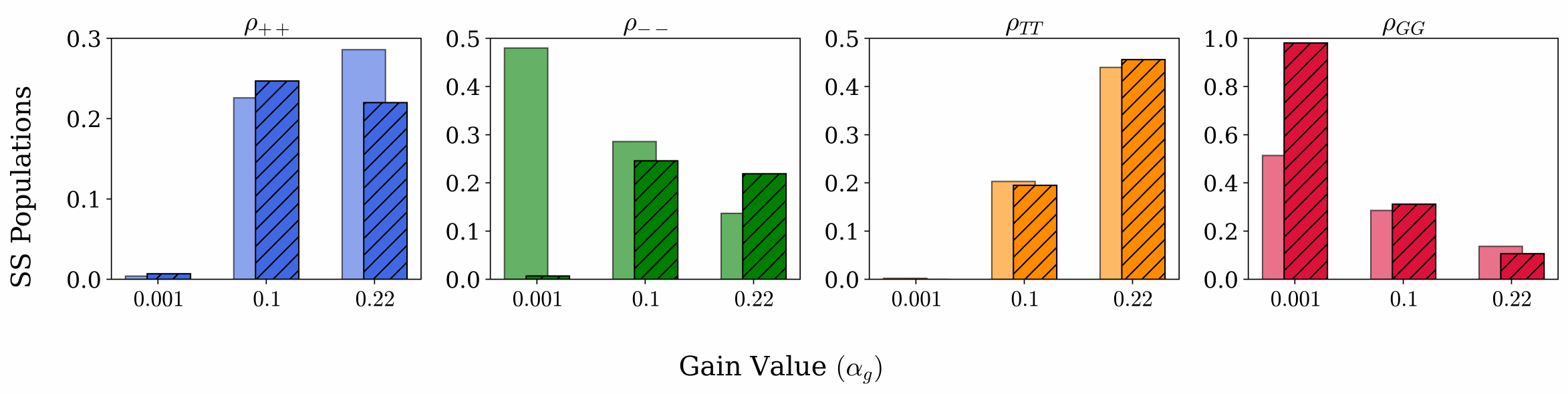}
    \caption{\textbf{Steady state populations with our full gain model.} Steady-state populations in the dressed state basis for the four basis states, where the solid colored bars represent the model without considering incoherent coupling, and the bars with the diagonal black lines represent the model with the incoherent coupling included.}
    \label{fig: ss vals gain model}
\end{figure}

Finally, we calculate the emitted spectrum using the steady-state spectrum method given in Eq.~\eqref{eq:SwSS}. First, to highlight the role of a phenomenological pump term when observing the emitted spectrum, we show a few examples of the spectra with no gain for the coupled quantum emitters in the plasmonic resonator, which can be found in Fig.~\ref{fig: spectra no gain}. In these cases with no gain, we phenomenologically add in a pumping term, such that the pump excites each atom individually, as shown in the master equation solution in Ref.~\cite{PhysRevB.99.085311}. In particular, here we want to highlight the role of the inter-emitter incoherent pump term due to $\Gamma_{ab}^{\uparrow}$, so we plot the spectrum with and without this term included. We stress this term is not known in the literature, as far as we are aware, and this is another new aspect of our theory.
We also stress that the linewidths are substantially different
to what is expected from a linear dipole theory, for which there would be no subradiant emission without pumping; with pumping, then we have a transition from 
$\ket{E}$ to $\ket{+}$ (lower energy peak), as well
as from $\ket{+}$ to $\ket{G}$ (higher energy peak),
as shown in the dressed-state picture of Fig.~\ref{fig: resonator schematic}.

We show three examples in Fig.~\ref{fig: spectra no gain}; all of these have a small amount of emitter pure dephasing included (where $\Gamma^{\prime} = 0.001\Gamma(0)$), and the pump strength increases as follows: $\Gamma^{\rm pump} = 0.001 \Gamma(0)$, $\Gamma^{\rm pump} = 0.01 \Gamma(0)$, and $\Gamma^{\rm pump} = 0.1 \Gamma(0)$, where again $\Gamma(0)$ is the $\Gamma^{\downarrow}_{aa}$ rate with no gain, which is $\Gamma^{\downarrow}_{aa} = 32.1\Gamma_0$, where $\Gamma_0$ is the background SE rate, given by Eq.~\eqref{eq: gamma_0}. When including the pump term which incoherently pumps the coupling between the emitters, we let this pump term be equivalent to what was quoted above. From these figures, we can clearly see that the $\Gamma_{ab}^{\uparrow}$ has a notable impact on the spectrum, particularly around the subradiant peak. Without including this term, the subradiant peak is visible even for very small pump values; however once this term is included, the subradiant peak is only visible once the pump strength increases sufficiently. In the superradiant peak, while it is minimal in this case, the pump terms result in broadening of the peak. 
The subradiant peak being visible, even when a dark state, is consistent with the results
in \cite{PhysRevB.99.085311}, which also explains the general broadening of the linewidths for this nonlinear spectra.

A deeper physical understanding of how $\Gamma_{ab}^{\uparrow}$ pumps the system can be investigated through considering the steady-state populations of the ground state, superradiant state, subradiant state, and two-quanta state, as the pump strength changes.
Based on the steady-state populations as given in Fig.~\ref{fig: ss vals pump model}, it is clear why the subradiant peak is always visible when one does not consider pumping with the inter-emitter term, $\Gamma_{ab}^{\uparrow}$; since without this term, one realizes larger $\rho_{TT}(t_{ss})$ values, and then transitions occur from the biexciton state to the subradiant state, a transition that is not allowed in a WEA. When the inter-emitter incoherent coupling pump term is considered, the subradiant state is not as visible in the spectrum until the pump reaches a sufficient fraction of the loss rate -- in Fig.~\ref{fig: spectra no gain}, the subradiant peak  starts to become visible (see the cyan curve) when $\Gamma_{\rm pump} = 0.1\Gamma(0)$.

Next we study the effects of gain from the plasmonic cavity 
(as shown in Fig.~\ref{fig: resonator schematic}) in a fully 
self-consistent way, and use the derived master equation as given by Eq.~\eqref{eq: two atom master with gain}; we repeat a similar process as above where we compare the spectra with and without the process, $\Gamma_{ab}^{\uparrow}$. In Fig.~\ref{fig: spectra with gain}, we first compute the spectrum with a very small amount of gain, where $\alpha_g = 0.001$. The decay and coupling rates used in the master equation for all cases are shown in Fig.~\ref{fig: decay rates}. With this small amount of gain, only the superradiant peak is visible in the spectrum when $\Gamma_{ab}^{\uparrow}$ is present in the system, which is similar to the spectral results when using smaller phenomenological pump values as investigated previously. 

When more gain is added to the system, namely when $\alpha_g = 0.1$, the two peaks can now be seen (i.e., the subradiant superradiant states), even without including the $\Gamma_{ab}^{\uparrow}$ term, as the gain is now acting as a sufficiently large pumping mechanism in the system. When a larger amount of gain, $\alpha_g = 0.22$, is added to the system, we see that the linewidths and spectral weights flipped between the peaks -- this is an indication of much strong pumping and gain-induced broadening~\cite{PhysRevB.99.085311}.

The impact of $\Gamma_{ab}^{\uparrow}$ is also visible with larger values of $\alpha_g=0.22$, which we can most clearly see by looking at the steady-state populations of the four basis states in the dressed state basis, which can be found in Fig.~\ref{fig: ss vals gain model}. 
Similar to what we saw with the phenomenological pumping, when $\alpha_g = 0.001$ and $0.1$, we find that $\rho_{TT}(t_{ss})$ is greater than when we set $\Gamma_{ab}^{\uparrow} = 0$, which then results in more distinct subradiant peaks in the spectra. When the gain parameter is increased to $\alpha_g = 0.22$, we find that $\rho_{TT}(t_{ss})$ actually has a greater value than when $\Gamma_{ab}^{\uparrow}$ is included, which highlights the importance of having this term in the master equation for large amount of gain, which is useful for understanding the emission dynamics before reaching a potential lasing regime (not considered in this work). Without properly considering how gain pumps the incoherent coupling between emitters, we can lose important information regarding the long-time limits of the population densities. 

Note that for a heuristic pumping model, one should
really use
$(\Gamma\rightarrow \Gamma+\Gamma_p) [\sigma^-]$
as well as $\Gamma_p L[\sigma^+]$ (detailed balanced for thermal energy~\cite{Tian1992,PhysRevB.81.033309}), in the Lindblad SE decay and pump terms.
However, since $\Gamma_{\rm pump} \ll \Gamma(0)$, there is very little impact for the additional SE in such a model. 



\section{Conclusions}
\label{sec: conclusions}

In summary, we have introduced  a general theory of the emission dynamics from multiple quantum emitters
in a gain-loss cavity system,
to 
show how gain impacts the theory of photon-coupled dipoles, treated as TLSs. 
This work 
extends previous work 
developed for a 
single quantum emitter~\cite{PhysRevA.105.023702,franke_fermis_2021} in a gain-loss linear medium.
We first derived a Born-Markov master equation, explicitly treating the gain and loss parts of the medium,
showing the impact on
 $n$ active emitter coupling. We then restricted the theory to have only two coupled quantum emitters,
 to analytically derive the optical Bloch equations in both the bare state basis and the dressed state basis, where a breakdown in the WEA 
 occurs  because of gain (causing nonlinear excitation of both emitters, beyond a linear response). We also demonstrated that a full treatment of the gain medium at the master equation level yields gain terms which act as pumping terms not only to the individual emitters, but also to the incoherent coupling between the emitters.

We  subsequently applied this theory to 
study two coupled quantum emitters in a finite-size gain-surrounded plasmonic resonator, taking advantage of a well-defined cavity system and QNM theory to analytically calculate decay rates from the Green functions with ease. By incorporating gain into this system and studying the temporal dynamics of the density matrix, we found that increasing the gain yields an increase in the steady-state populations for each of the emitters in the bare basis, especially when considering the emitters at a frequency far detuned from the QNM cavity resonance. Through an investigation of the near resonant dynamics versus off-resonant dynamics, we found that despite observing large enhancements in the incoherent decay rates around the QNM resonance, more interesting population and entanglement dynamics occur when $\delta_{ab} > \Gamma_{aa}^{\downarrow}, \Gamma_{aa}^{\uparrow}$.
Even though the gain mitigates some of the coherent interactions, long-lived entanglement is still accessible with large amounts of gain. Furthermore, a breakdown of the WEA with gain in the dressed basis allows for the gain to pump the two-quanta state, which typically is assumed to decouple from the basis (lossy case), and non-zero steady-state values for the two-quanta state can be achieved with gain in the system, notably without having to add any pump terms by hand. 

Finally, we showed how the full treatment of gain at the master equation level provides a deeper understanding for the emitted  spectrum, a key experimental observable. We showed that by phenomenologically adding an incoherent pump term, which is a common model choice, one misses crucial incoherent interactions between the emitters which are still mediated by the gain part of the system. 

These findings are important 
for a fundamental understanding of radiative coupling and emission in loss-gain systems,
and for having a better understanding of multi-emitter laser theories, as the regime of linear amplification is a precursor to the lasing regime. The advantage of our approach is that all rates are derived from first principles, and work for and loss-gain media, which opens many doors for future work, with all input coming from the medium Green functions. While this work largely focused on having just two coupled quantum emitters in a cavity system, our findings lay the foundation for studying the dynamics of many coupled quantum emitters in a nanophotonic system with lossy and gain media, and could also find use for applications such as plasmon-assisted lasing and spasing. Lastly, we highlight that our approach can be extended to treat the QNMs at the level of quantized modes~\cite{PhysRevA.105.023702}, allowing for a full quantum field treatment of the general loss-gain system.

\acknowledgements
This work was supported by the Natural Sciences and Engineering Research Council of Canada (NSERC), 
the Canadian Foundation for Innovation (CFI), Queen's University, Canada, and the
Alexander von Humboldt Foundation through a Humboldt Research Award.
We also thank 
CMC Microsystems for the provision of COMSOL Multiphysics.

\vspace{1cm}

\bibliography{references}

\end{document}